\documentclass{article}

\usepackage{arxiv}

\usepackage[utf8]{inputenc} 
\usepackage[T1]{fontenc}    
\usepackage{hyperref}       
\usepackage{url}            
\usepackage{booktabs}       
\usepackage{amsmath,amssymb,amsfonts}       
\usepackage{nicefrac}       
\usepackage{microtype}      
\usepackage{lipsum}
\usepackage{graphicx}
\graphicspath{ {./images/} }
\usepackage{algorithmic}
\usepackage{xcolor}
\usepackage{stfloats}
\usepackage{url}
\usepackage{adjustbox}
\usepackage{enumerate}
\usepackage[numbers]{natbib} 
\usepackage{enumitem}
\usepackage{mathrsfs}
\usepackage{textcomp}
\usepackage{multirow}
\usepackage{booktabs}
\usepackage{siunitx}
\usepackage{chngcntr}
\usepackage[english]{babel} 
\usepackage{blindtext}

\newcommand{\acemod}{\textsc{Ace}Mod\xspace}
\newcommand{\phasetrace}{\textsc{Ph}ASE\,\textsc{Tra}CE\xspace}

\title{Multi-scale phylodynamic modelling of rapid punctuated pathogen evolution}

\author{
 Quang Dang Nguyen\textsuperscript{*} \\
  Centre for Complex Systems\\
  Faculty of Engineering\\
  The University of Sydney, Australia\\
  \And
 Sheryl L. Chang  \textsuperscript{*}\\
  Centre for Complex Systems\\
  Faculty of Engineering\\
  The University of Sydney, Australia \\
  \texttt{sheryl.chang@sydney.edu.au} \\
   \And
 Carl J. E. Suster \\
  Centre for Infectious Diseases and 
  \\Microbiology–Public Health\\
  Westmead Hospital\\
  NSW, Australia\\
  \And 
  Rebecca J. Rockett \\ 
  Centre for Infectious Diseases and \\
  Microbiology–Public Health\\
  Westmead Hospital\\
  NSW, Australia\\
  \And 
  Vitali Sintchenko\\
  Sydney Infectious Diseases Institute \\
  The University of Sydney\\
  NSW, Australia\\
  \And 
  Tania C. Sorrell \\
  Sydney Infectious Diseases Institute \\
  The University of Sydney\\
  NSW, Australia\\
  \And
  Mikhail Prokopenko\\
  Centre for Complex Systems\\
  Faculty of Engineering\\
  The University of Sydney, Australia\\
}

\begin{document}
\maketitle

\begin{abstract}
Computational multi-scale pandemic modelling remains a major and timely challenge. Here we identify specific requirements for a new class of models simulating pandemics across three scales: (1) pathogen evolution, often punctuated by the rapid emergence of new variants, (2) human interactions within a heterogeneous population, and (3) public health responses which constrain individual actions to control the disease transmission.
We then present a pandemic modelling framework satisfying these requirements and capable of simulating feedback loops between dynamics unfolding at these different scales. 
The developed framework comprises a stochastic agent-based model of pandemic spread, coupled with a phylodynamic model that incorporates within-host pathogen evolution. It is validated with a case study, modelling the punctuated evolution of SARS-CoV-2, based on global and contemporary genomic surveillance data, which captures a large heterogeneous population. We demonstrate that the model replicates the essential features of the COVID-19 pandemic and virus evolution, while retaining computational tractability and scalability.
\end{abstract}

\textsuperscript{*}: \textit{These authors contributed equally}\\

\section{Introduction}
\label{intro}
Digital epidemiology is an emerging field, rapidly capitalising on the increasing availability of high-resolution genomic, immuno-epidemiological, demographic and human mobility data, social media analytics, high-performance computing power, as well as innovations in simulation methods and data science. In principle, these diverse sources of data should allow modellers to develop informative multi-scale pandemic models with a better capacity (i) to simulate realistic epidemiological, immunological and evolutionary dynamics, and (ii) to anticipate longer-term epidemiological and evolutionary dynamics as the pandemic unfolds.
Yet, pandemic modelling continues to face significant challenges in concisely capturing relevant characteristics of pandemic pathogens, such as their pathogenesis, transmissibility, antigenicity, etc., as well as predicting long-term phylodynamic trajectories and future public health risks. These challenges arise due to (i) the inherent complexity of rapidly evolving pathogens, (ii) population heterogeneity (demographic, immunological and behavioural), and (iii) multi-objective public health interventions carried out under severe pressure and non-trivial social dynamics. 

The modelling complexity is exacerbated by (iv) different time scales needed to model pandemics on multiple levels, ranging from the range of evolutionary drivers and rates of the implicated pathogens to natural infections in individuals and their social behaviour and interactions, (v) fragmentation of data across heterogeneous sources, and (vi) computational complexity of multiple simulations over a sufficiently long horizon, required to examine the distribution of outcomes in many stochastic realisations of the model and across ranges of uncertainties in parameters such as substitution rate, fitness, accumulated mutations, and genomic diversity. Multi-scale models often suffer from the ``curse of dimensionality'', when computational costs increase exponentially with the number of degrees of freedom~\cite{fabiani_task_oriented_2024}. 

A principal modelling problem is the presence of feedback loops: for example, pandemic mitigation measures may indirectly affect the pathogen evolution, leading to the emergence of more transmissible lineages. Higher transmissibility may increase the need for more vigorous interventions, which in turn may cause changes in how populations respond and behave, and constrain the pathogen evolution in a specific way. This feedback loop contributes to the formation of recurrent waves of infection, fluctuating genomic diversity, non-linear increases in fitness levels, and a potentially delayed transition to endemicity. Consequently, multi-scale modelling of a major pandemic crisis, such as COVID-19, quickly becomes intractable. 

Over the last decades, stochastic agent-based modelling (ABM) has been established as a robust tool for tracing fine-grained effects of complex intervention policies in diverse epidemic and pandemic settings~\cite{Eubank2024,germann2006mitigation,nsoesie2012sensitivity,rockett2020revealing}. Most recently, these studies produced policy recommendations developed for COVID-19 control, which were adopted in Australia \cite{chang_modelling_2020,scott2020modelling,blakely2020probability}, the USA \cite{Kerr2021}, the UK \cite{ferguson2020impact}, and broadly by the WHO \cite{world2020combined}. In these ABMs, each agent represents an individual human host with a set of demographic, epidemiological, and immunological attributes.

A largely unexplored avenue to leverage the precision and fidelity of ABMs is to extend them with comprehensive phylodynamic modelling of evolving pathogens, going beyond existing phylogenetic models which define simplified evolutionary landscapes~\cite{Ferguson, Nielsen}. This necessitates a new class of multi-scale phylodynamic ABMs.

An effective framework for multi-scale phylodynamic agent-based modelling should include the following distinct capabilities that produce quantifiable outcomes: 

\begin{enumerate}[label={{Capability} \arabic*:},leftmargin=*]
    \item Model and examine epidemic or pandemic patterns over a mid- to long-term timeframe, with respect to complex transmission and immunological profiles, affected by varying pharmaceutical and non-pharmaceutical interventions:
        \begin{enumerate}
        \item Reproduce and predict salient peaks and recurrent waves of incidence, prevalence, and other epidemic dynamics. 
        \item Explore possible transitions and pathways to endemicity or elimination.
        \end{enumerate}
    \item Examine the pathogen fitness with respect to its phylodynamics: 
    \begin{enumerate}
        \item Trace changes in transmissibility with respect to pathogen mutations. This includes analysis of the average reproductive number within the population, and the accumulated mutations measured by the average genomic distance between a circulating genome and the ancestral genome, e.g., in terms of the nucleotide substitution rate. 
        \item Examine the functional dynamics (i.e., the relationship between the genome sequence and the pathogen's epidemic behaviour). This analysis quantifies the individual contributions of amino acids to changes in the pathogen's fitness, and traces the dynamics of these contributions to transmissibility and antigenicity over time. 
    \end{enumerate}

    \item Detect and evaluate the emergence and dominance of variants of concern:
    \begin{enumerate}
        \item Explore concordance between phylodynamics and disease dynamics, e.g., relating accumulated mutations, genomic diversity, and saltations in pathogen fitness to incidence peaks. 
        \item Detect abrupt changes in genomic diversity, and evaluate emergence of variants of concern, by using appropriate quantitative techniques (such as Augmented Dickey-Fuller stationarity test, CUSUM, etc.), supported by suitable data visualisation methods (such as phylogenetic trees).\\
    \end{enumerate}
\end{enumerate}

\begin{figure}
    \centering
    \includegraphics[width=\columnwidth,trim={0 20cm 0 0},clip]{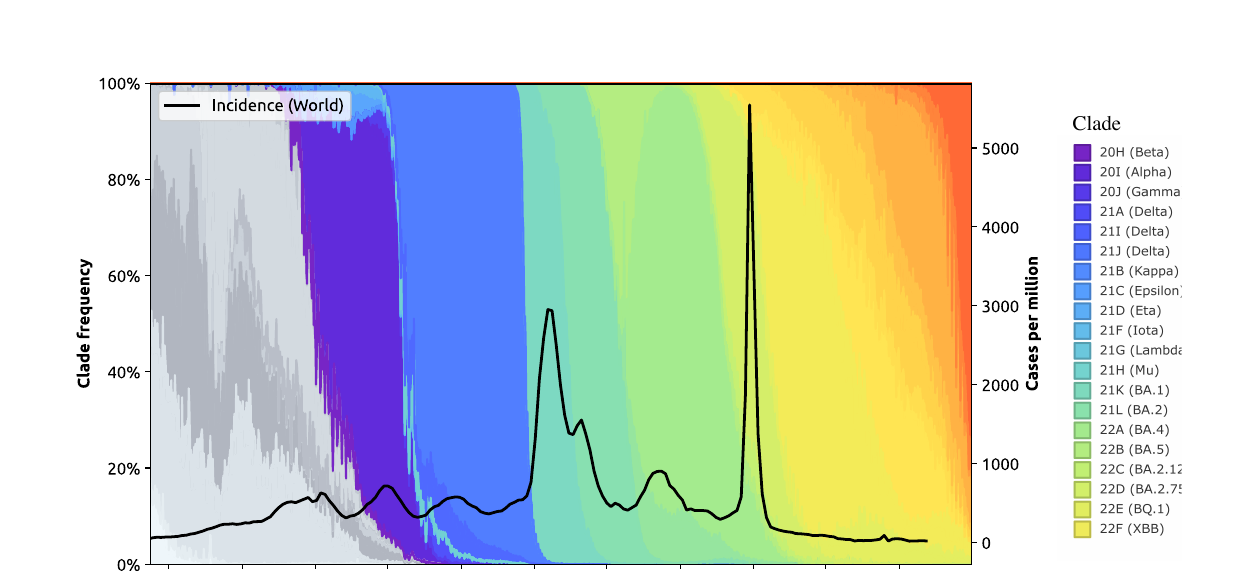}
    \caption{Capability 1. Pandemic patterns in terms of the worldwide incidence (solid, black line),  measured as new weekly cases per million \cite{ourworldindata}, overlaid with the frequency of circulating variants between 2020 and 2024, plotted using open SARS-CoV-2 sequence data (7,075,645 samples) from GenBank and the Robert Koch Institute, processed by Nextstrain \cite{nextstrain_data}.
    }
    \label{fig:obj1}
\end{figure}

\begin{figure}
    \centering
    \includegraphics[width=\columnwidth]{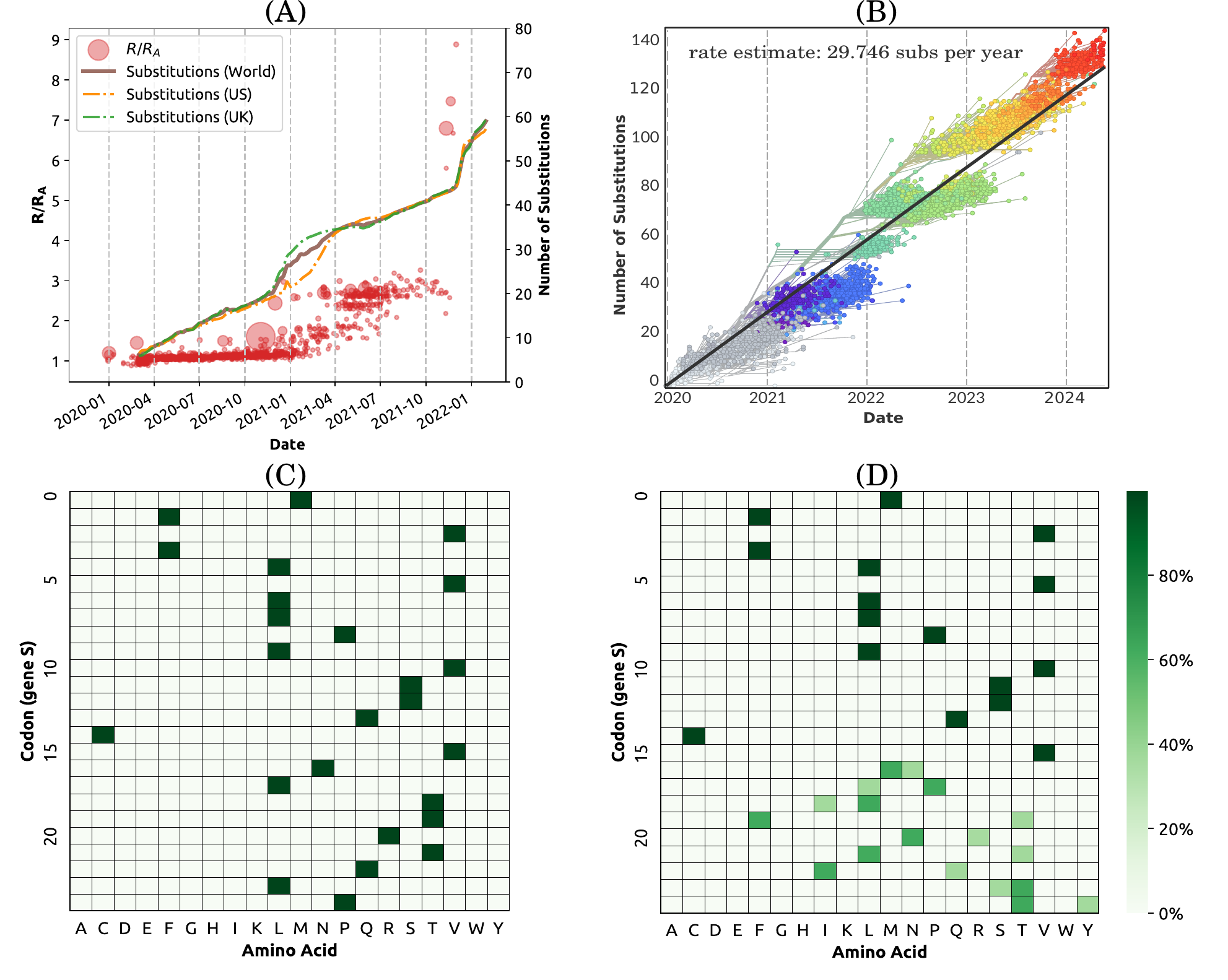}
    \caption{Capability 2. Pathogen fitness with respect to its phylodynamics. (A) Punctuated growth of fitness, measured relative to the basic reproductive {number} of the ancestral strain $R_A$ \cite{Obermeyer_2022} (represented as the $R/R_A$, red circles), overlaid with the average accumulated nucleotide substitutions across the population (represented by the Hamming distance $\widehat{D}$), also measured relative to the ancestral strain.
    (B) A time-scaled phylogeny of representative SARS-CoV-2 sequences mapped onto the number of accumulated substitutions relative to the ancestral genome. We refer to Fig~\ref{fig:obj1} for clade colours. For visualisation purposes, we show around 4,000 genomes. Data retrieved from Nextstrain~\cite{nextstrain}. 
    (C, D) Heatmaps of amino acids over the first 25 codons in gene S of the COVID-19 ancestral strain (C), and 631 randomly selected genomes from GenBank and the Robert Koch Institute between 17 and 31 December 2023 (D) \cite{nextstrain_data}.}
    \label{fig:obj2}
\end{figure}

\begin{figure}
    \centering
    \includegraphics[width=\columnwidth]{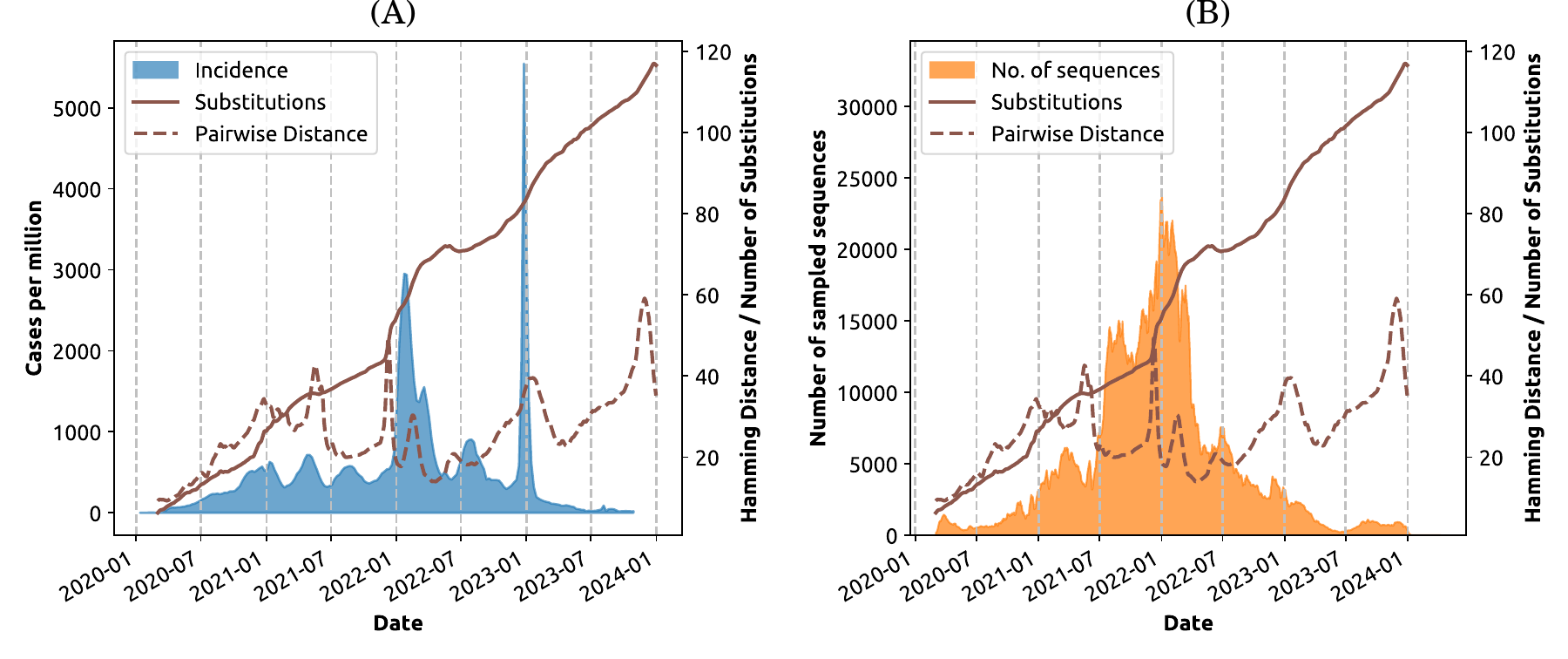}
    \caption{Capability 3 (i). Temporal alignments between phylodynamics and disease dynamics. (A) The global COVID-19 incidence (new weekly cases per million, shaded blue area) overlaid with accumulated mutations $\widehat{D}$ (represented by the average distance between circulating genomes and the ancestral strain, solid brown line) and genomic diversity $\overline{D}$ (represented by the average pairwise distance between two randomly selected genome sequences, dashed brown line). 
    (B) The number of sequences, measured using a 7-day moving average, captured by the open SARS-CoV-2 sequence database processed by Nextstrain \cite{nextstrain_data} (shaded orange area) overlaid with accumulated mutations $\widehat{D}$ across the genome (solid brown line), and genomic diversity $\overline{D}$ (dashed brown line).}
    \label{fig:obj3-i}
\end{figure}

\begin{figure}
    \centering
    \includegraphics[width=\columnwidth,trim={0 0 0 0},clip]{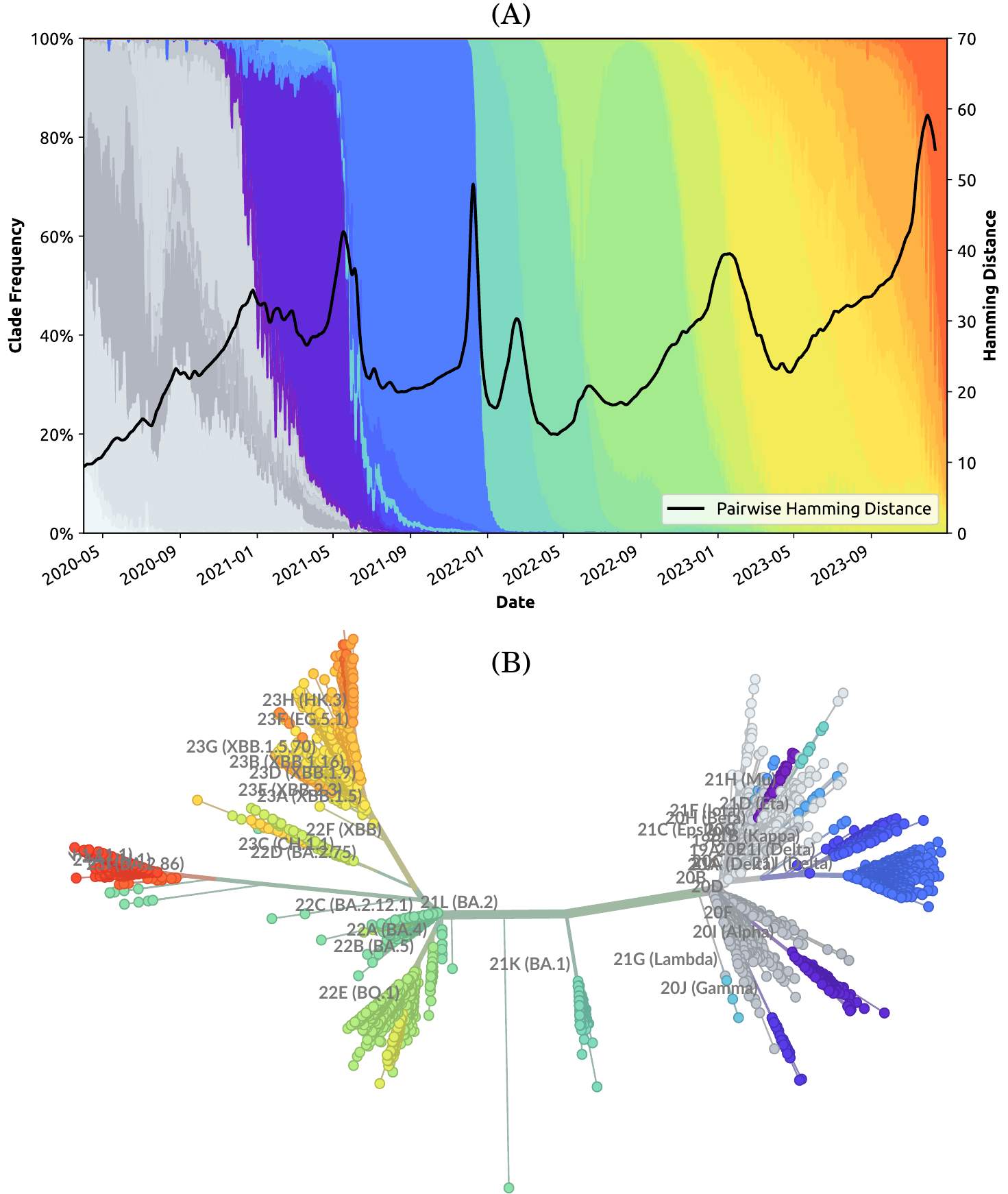}
    \caption{Capability 3 (ii). Emergence and dominance of variants of concern. (A) Genomic diversity {$\overline{D}$ (i.e., the average pairwise distance between two randomly selected genome sequences on a given day,} solid black line) overlaid with the clade frequency between 2020 and 2024. (B) Phylogenetic tree of SARS-CoV-2, built with Nextstrain from worldwide sequences collected between 2020 and 2024~\cite{nextstrain}. The phylogenetic tree is constructed from approximately 4,000 genomes.}
    \label{fig:obj3-ii}
\end{figure}

A robust phylodynamic ABM needs to be validated by comparing its target outcomes against the ground truth dynamics. Once validated, it can be used to explore diverse counterfactual scenarios with respect to phylodynamic, demographic and immuno-epidemiological characteristics. 
We illustrate the ground truth dynamics, matching the three capabilities, in Figs~\ref{fig:obj1} to \ref{fig:obj3-ii}, by using available genomic and disease surveillance data on SARS-CoV-2 and COVID-19 respectively from 2020 to 2024 (we note that the detected incidence is affected by the testing capacity). 

To demonstrate \textbf{Capability 1}, a modelling framework is expected to generate pandemic or epidemic patterns aligned with ground truth (i.e., observed data). For example, Fig~\ref{fig:obj1} shows patterns observed in global data reported during the  COVID-19 pandemic, highlighting prominent peaks and recurrent incidence waves. We point out that each incidence peak is temporally aligned with the emergence of a new variant of concern. Notably, the two most prominent incidence peaks occurred in early 2022 and early 2023, corresponding to the dominance of Omicron BA.1 and Omicron XBB, respectively.  These observations suggest a complex interplay between pandemic patterns and evolutionary phylodynamic features of the viral variants in circulation.

\textbf{Capability 2} demands a dynamic analysis of the pathogen fitness in terms of observable phylodynamic characteristics. For example, Fig~\ref{fig:obj2} relates the growing transmissibility of  SARS-CoV-2 to (i) the mutations accumulated relative to its ancestral strain, and (ii) the associated genomic diversity of the evolved pathogen. It also places the fitness dynamics in the context of the genomics.

Punctuated evolution of the novel coronavirus was observed even during the first year of the pandemic \cite{surya_2023}. Fig~\ref{fig:obj2}A shows a rapid punctuated increase in fitness for the first two years of the pandemic, with two significant surges in the relative transmissibility (measured relative to the basic reproductive number of the ancestral strain $R_A$), and the accumulated mutations, observed during early 2021 and early 2022. The accumulated mutations continued to grow after 2022, reaching 130 substitutions by mid-2024 at the rate of roughly 30 substitutions per year, according to linear regression (Fig~\ref{fig:obj2}B).

Furthermore, notable changes in the amino acid presence were observed across the genome, particularly the spike region, between the start of the pandemic and late 2023, as illustrated for the 5$'$ end of the S gene in Fig~\ref{fig:obj2}C--D. These shifts occurred preferentially at specific positions, deviating significantly from the ancestral genome, and thus, potentially contributing to the increase in fitness.

\textbf{Capability 3} is focused on the emergence and dominance of variants of concern, in the context of phylodynamic and epidemiological dynamics. {Two quantities can be observed on any given day: the average mutations accumulated by the evolved genomes relative to the ancestral strain, denoted $\widehat{D}$; and the average pairwise distance among the genomes of co-circulating pathogens on that day, denoted $\overline{D}$ (see Section \ref{sec:methods})}. For example, Fig~\ref{fig:obj3-i}A traces a continuous increase in the accumulated mutations which is contrasted with the fluctuating genomic diversity. Unlike the mutations $\widehat{D}$ which accumulated relative to the ancestral strain, the genomic diversity, measured as the {daily} pairwise distance $\overline{D}$, shows only a marginal increase during the observed period. We extracted phylodynamic features using the data recorded until early 2024, noting a significant drop in the number of sequenced samples worldwide from 2024 onwards (Fig~\ref{fig:obj3-i}B).  

In general, abrupt changes in genomic diversity are related to the frequency of different variants reported during the pandemic period (Fig~\ref{fig:obj3-ii}A),  and considered in context of the corresponding phylogenetic tree (Fig~\ref{fig:obj3-ii}B). For example, we observed that during the rapid evolution of SARS-CoV-2, sudden decreases in circulating diversity correspond to specific lineages becoming dominant, whereas new variants are more likely to emerge during periods of increasing circulating diversity (i.e., increasing pairwise genomic distance).
Importantly, the observed changes in the pairwise distance are also reflected in the phylogenetic tree produced from representative sub-sampling of global sequences (Fig~\ref{fig:obj3-ii}B). Branches corresponding to new variants of concern tend to descend from more basal lineages than the main lineages circulating immediately prior to their emergence. These evolutionary saltations may explain step changes in transmissibility and virulence.

\section{Results}
\label{results}
We present an agent-based modelling (ABM) framework for computational modelling of pathogen phylodynamics, focusing on communicable diseases within heterogeneous populations. The key component is \phasetrace, \textbf{Ph}ylodynamic \textbf{A}gent-based \textbf{S}imulator of \textbf{E}pidemic \textbf{Tra}nsmission, \textbf{C}ontrol, and \textbf{E}volution, a versatile multi-scale simulator for modelling rapid pathogen evolution. The framework also includes several phylodynamic measures aimed to identify the emergence of novel pathogen variants. 

\phasetrace was developed upon several existing large-scale pandemic simulators,  including the Australian Census-based Epidemic Model (\acemod) of pandemic influenza~\cite{cliff2018nvestigating,zachreson2018urbanization,zachreson2020interfering}, and the Agent-based Model of Transmission and Control of the COVID-19 pandemic in Australia (AMTraC-19)~\cite{chang_modelling_2020,amtract_user_guide}. These ABMs have been successfully validated and used in simulating multiple waves of influenza \cite{cliff2018nvestigating} and COVID-19 \cite{chang_modelling_2020,rockett2020revealing,zachreson_how_2021,chang_simulating_2022,BMC}, mitigated by various interventions, including mass-vaccination roll-outs and non-pharmaceutical interventions.

Similar to these models, \phasetrace simulates the disease transmission in discrete time in an artificially generated population with census-based demographic characteristics and commuting patterns \cite{fair_creating_2019}. 
Going beyond the existing models, \phasetrace is capable of simulating the inter-host transmission of multiple pathogen variants within a heterogeneous population, the within-host evolution of pathogens, and immuno-epidemiological feedback.

In this section, we overview the model's multi-scale approach (subsection \ref{sec:model_overview}). We then apply \phasetrace to a case study of SARS-CoV-2, and evaluate the simulated phylodynamics against the modelling capabilities and objectives (subsection~\ref{sec:case_study_covid19}). Finally, we explore counterfactual scenarios by varying specific assumptions of the case study, e.g., the role of chronic infections and population sizes (S1 Text: Counterfactual modelling). 

\subsection{Model overview}
\label{sec:model_overview}

Distinct from many computational models that focus solely on epidemiological dynamics within host populations or evolutionary dynamics within pathogen populations, \phasetrace simulates dynamic feedback across three scales: (i) micro-scale: within-host evolution and the evolutionary landscape of circulating variants;  (ii) meso-scale: agent-to-agent interactions and inter-host transmission; and (iii) macro-scale: public health interventions (i.e, non-pharmaceutical interventions and vaccination) at the population level. 

\begin{figure}
    \centering
    \includegraphics[width=\columnwidth]{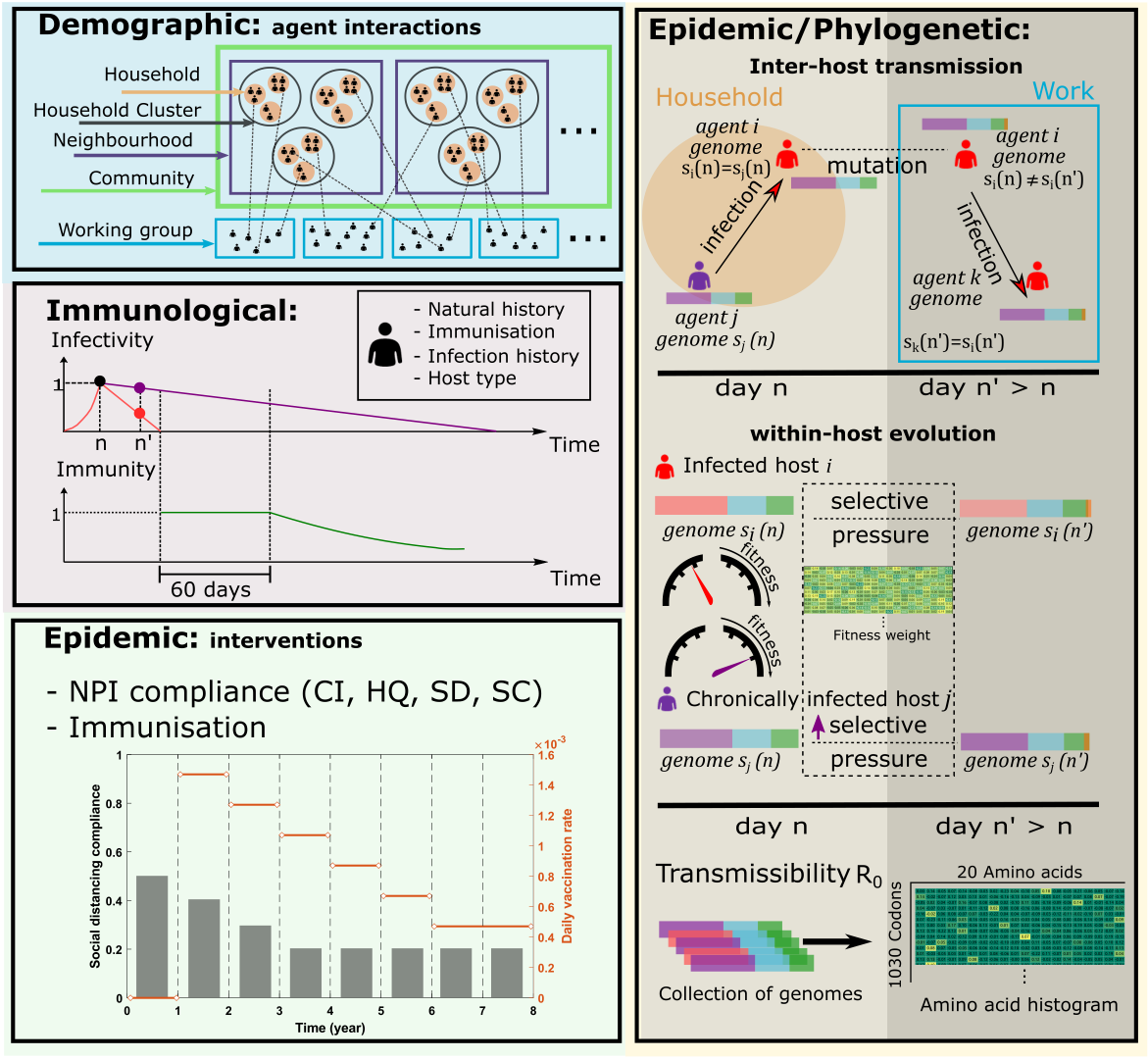}
    \caption{The demographic layer assigns agents to social contexts, including household, household cluster, neighbourhood, community, and working group/school. The immunological layer keeps track of the agents' immunisation records, infection histories, and the host types. {An example of the infectivity profiles is illustrated for a typical infected host (red line) and a chronically infected host (purple line). On day $n$, the infectivity peaks (i.e., reaches 1.0) for both host types (black dot). On day $n'>n$, the infectivity of the chronically infected host (purple dot) reduces more slowly than that of the typical infected host (red dot). Upon recovery, the typical infected host attains perfect immunity of 1.0 for 60 days, after which the immunity starts to decline (green line).} The epidemic layer describes an intervention scenario, in terms of NPIs and their adoption levels: case isolation (CI), home quarantine (HQ), social distancing (SD), and school closure (SC), along with a dynamic immunisation schedule. Both epidemic and phylogenetic layers affect inter-host transmission (e.g., on day $n$, a chronically infected host $j$ infects their family member $i$ and passes genome $s_j(n)$ to agent $i$). The phylogenetic layer determines the genomic mutations under various strengths of selective pressure, and traces changes in transmissibility (reflected in reproductive number). The fitness of each genome is determined by a weight table, quantifying the contributions of each amino acid in a given codon position.
}
    \label{fig:overview}
\end{figure}

This multi-scale feedback is realised by incorporating four concurrent dynamics implemented in distinct processing layers: demographic, epidemic, immunological, and phylogenetic, as illustrated in Fig~\ref{fig:overview}. The demographic layer defines the population structure (i.e., host type) and social groups constraining agent interactions. The transmission is simulated by the epidemic layer, tracing individual interactions occurring within and across different social contexts, including residential (e.g., household) and professional settings (e.g., working group). 
In addition, the epidemic layer sets out an intervention scenario, with varying NPI adoption levels and vaccination strategies, thus reflecting changes in health policy and public opinion \cite{chang_persistence_2023,chang2024impactopiniondynamicsrecurrent}.
The infection transmission depends on the agents' immunity levels derived from their individual dynamic history of exposures and immunisation, determined by the immunological layer. There are two infected host categories: typical infected hosts and chronically infected hosts, with the latter representing hosts with \textit{persistent} SARS-CoV-2 infections arising due to some underlying factors and resulting in longer recovery periods~\cite{gonzalez_reiche_sequential_2023, ghafari_prevalence_2024} (see S1 Text: Infected host categories). The phylogenetic layer determines the within-host evolution of pathogens in terms of mutation and selective pressure. Section \ref{sec:methods} provides a detailed description and implementation of these four layers.

\subsection{Case study: Rapid punctuated evolution of SARS-CoV-2 }
\label{sec:case_study_covid19}
\phasetrace was calibrated to the COVID-19 pandemic and the evolutionary trajectory of SARS-CoV-2 over the period of four years, from 2020 to 2023. The calibration explored ranges of over 90 parameters in four categories corresponding to modelling layers. 
Demographic parameters were census-calibrated \cite{cliff2018nvestigating,
zachreson2018urbanization,
fair_creating_2019, BMC}. Most of the epidemiological and immunological parameters were calibrated in our prior studies~\cite{chang_modelling_2020, chang_simulating_2022, chang_persistence_2023, BMC}. For example, parameter ranges related to non-pharmaceutical interventions were established based on prior studies \cite{chang_modelling_2020, chang_simulating_2022, chang_persistence_2023, BMC} and public health policies during the COVID-19 pandemic. In calibrating the natural history of the disease for this work, we used contemporary evidence, including studies of chronic infections \cite{wilkinson_recurrent_2022,ghafari_prevalence_2024}). In contrast, all phylogenetic parameters were specifically calibrated for this study. The majority of phylogenetic parameters were determined using reported genomic evidence \cite{nextstrain} and recent phylogenetic analyses \cite{markov_evolution_2023,Obermeyer_2022, kistler_rapid_2022, thadani_learning_2023,smith_landscape_2021,sender_total_2021}. Some phylogenetic inputs that are specific to our model, such as the weight table quantifying the fitness contribution of amino acids, were calibrated by comparing the simulation outcomes corresponding to different parameterisations and selecting the most fitting parameterisations. The phylogenetic parameters are summarised in Table \ref{sm:tab:phylo_parameters}.

The results comprise simulations over a six-year period (2020 -- 2025), across three artificial agent populations: 1.7 million agents (small), 8 million agents (medium), and 25.4 million agents (large), as described in S1 Text (Artificial agent-based population). 

In calibrating \phasetrace, we examined COVID-19 pandemic and the size of susceptible, recovered and vaccinated populations (Capability 1); phylodynamic characteristics including transmissibility, fitness, the number of accumulated mutations relative to the ancestral strain, and genomic diversity (Capability 2); and the emergence of variants, analysed using the phylogenetic tree and statistical methods (Capability 3). The robustness of the model is established through sensitivity analysis, by varying key parameters and assessing their impact on simulation outcomes. See Table \ref{sm:tab:phylo_parameters} for the list of phylogenetic parameters used in the SARS-CoV-2 case study and S1 Text (Sensitivity analysis) for more information on sensitivity analysis. 

\begin{figure}[t]
    \centering
    \includegraphics[width=\columnwidth]{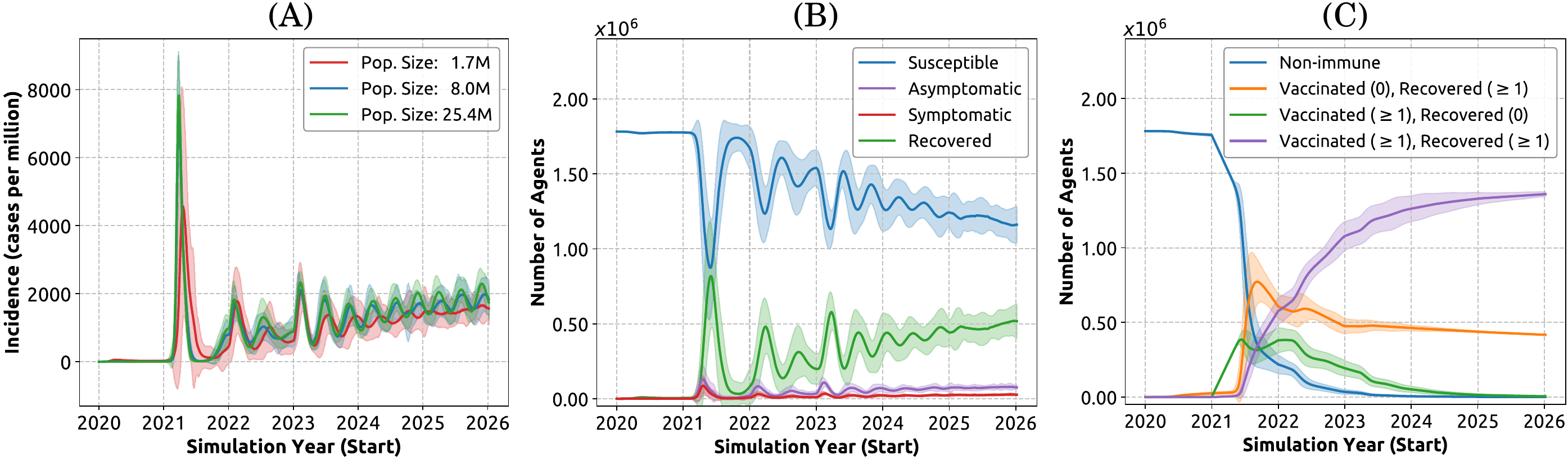}
    \caption{Simulated pandemic patterns (Capability 1) shown as mean (solid line) and standard deviation (shaded area). (A) Detected incidence simulated in population sets of 1.7 million (red), 8 million (blue), and 25.4 million (green). (B) Population in different health states, including susceptible (blue), asymptomatically infected (purple), symptomatically infected (red), and recovered (green). (C) Population with different immunisation and infection history. Numbers in brackets denote the number of vaccination or infection records. Individuals with multiple vaccinations or infections (more than 2) are grouped together for simplicity. (B) and (C) are generated using the population set of 1.7 million. Mean and standard deviation were obtained from approximately 30-50 realisations. }
    \label{fig:sim-obj-1}
\end{figure}
\textbf{Capability 1.} 
Our results show recurrent incidence waves as illustrated in Fig~\ref{fig:sim-obj-1}A and {\ref{sm:fig:inc_log}}, in concordance with empirical observations (Fig~\ref{fig:obj1} and  {\ref{sm:fig:inc_log}}). Although the first simulated incidence peak (detected during 2021) is noticeably higher in larger populations, the incidence converged to around 2,000 cases per million after 2023, indicating endemicity in all simulated populations. 

The main contributing factor to the recurrent incidence is 
the fluctuating number of recovered individuals (Fig~\ref{fig:sim-obj-1}B) that replenish the susceptible population after their immunity wanes. This subsequently leads to an increasing number of re-infection cases. Additionally, due to the diminishing immunity, declining vaccine uptake, and declining adoption of social distancing in the population (simulated to be decreasing from 50\% of hosts in 2020 to 20\% from 2024 onwards), the overall incidence slightly increases over the simulated timeframe after the first peak (Fig~\ref{fig:sim-obj-1}A). Notably, simulation results suggest that chronic infections strongly affect the magnitude of incidence oscillations (\ref{sm:fig:counter_immuno_sim}). This finding may also explain the dampened oscillations observed in smaller populations. Emerging strains tend to dominate more strongly across a smaller population, resulting in: (i) reduced genomic diversity of the pathogen, (ii) a more homogeneous and consistent immune response across the population, and consequently (iii) a smoother incidence profile (see discussion of Capability 3 below).

We also observe an increasing number of individuals with multiple vaccination and infection records (Fig~\ref{fig:sim-obj-1}C). These dynamics slow down the transmission but are unable to completely eliminate the spread, due to the waning immunity and reduced vaccine efficacy against mutated variants. 

Figs~\ref{fig:sim-obj-1}B and \ref{fig:sim-obj-1}C only show dynamics for the 1.7 million population. We note that the pandemic patterns are consistent across different population sizes, as shown in  Fig~\ref{sm:fig:compartments}. 

\begin{figure}[t]
    \centering
    \includegraphics[width=\columnwidth]{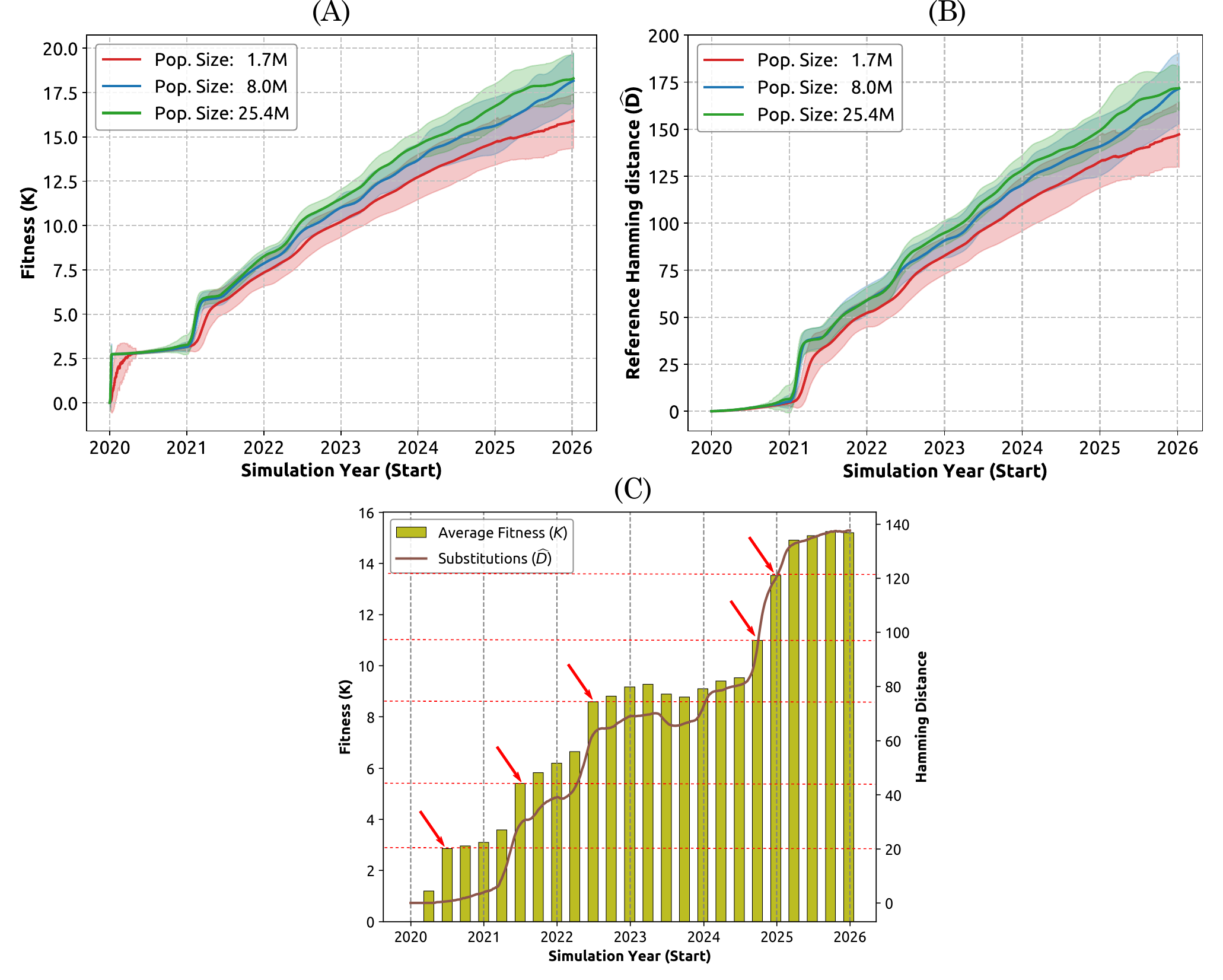}
    \caption{Simulated phylodynamic measures (Capability 2(i)), shown as mean (solid line) and standard deviation (shaded area) for different populations: 1.7 million (red), 8 million (blue), and 25.4 million (green). (A) Growing pathogen transmissibility, interpreted as fitness $K$. (B) Accumulated mutations $\widehat{D}$, measured at each time point by the average reference Hamming distance between the circulating variants and the reference (ancestral) variant. The mean and average were obtained from approximately 30-50 realisations. (C) Single realisation, showing alignment of the average transmissibility, i.e., fitness ($K$, olive bars), and the accumulated mutations ($\widehat{D}$, solid brown line). }
    \label{fig:sim-obj-2i}
\end{figure}

\textbf{Capability 2.} 
Here we investigate how the pathogen fitness changes during the simulation, given the selective pressure on circulating strains which ``compete'' in terms of their transmissibility. We begin by exploring how the phylodynamic characteristics, such as accumulated mutations, develop along the observed increase in simulated fitness (Capability 2 (i)).

Importantly, we observe a concurrent punctuated increase (i.e., `jump') in both transmissibility, i.e., fitness $K$ (Fig~\ref{fig:sim-obj-2i}A), and accumulated mutations $\widehat{D}$ (Fig~\ref{fig:sim-obj-2i}B) at the start of 2021. This is a pattern which is also observed in empirical data illustrated in Fig~\ref{fig:obj2}A.  The timing of the simulated `jump' aligns well with the first incidence peak in 2021, illustrated in Fig~\ref{fig:sim-obj-1}A. This coincidence can be explained by the accumulation of fitness-increasing mutations in chronically infected hosts, due to a higher selective pressure on pathogens evolving in these hosts; consequently generating a highly transmissible strain (i.e., a variant of concern) (see S1 Text: Infected host categories). This is confirmed by a comparison with a counterfactual scenario without chronically infected hosts in which the jump is not observed, as shown in Fig~\ref{sm:fig:counter_immuno_sim} and analysed in S1 Text (Counterfactual modelling). We also traced the fitness contributions of different genomic regions, by computing the fitness per codon (Fig~\ref{sup_fig:genome_structure}). Evidently, the spike and epitope regions, which contribute to both transmissibility and antigenicity, have the highest fitness $K$ per codon (Fig~\ref{sm:fig:fitness_per_codon}) and the highest number of accumulated mutations $\widehat{D}$ per codon (Fig~\ref{sm:fig:fitness_per_codon}). Notably, this high contribution is observed despite the short genome length of this region (45 codons), indicating that the fitness gain is a result of combining the increased transmissibility and the higher immune escape. In comparison, the spike and non-epitope region (55 codons) which contributes only to transmissibility has lower fitness $K$ and accumulated mutations $\widehat{D}$ per codon.

Similar jumps are observed at later pandemic stages in a majority of individual simulation runs (see Figs~\ref{fig:sim-obj-2i}C, and \ref{sm:fig:sim-obj-2i-individual_fitness_Ref_Haming-1M7} to \ref{sm:fig:sim-obj-2i-individual_fitness_Ref_Haming-25M4}), but their timing differs across realisations, and so the average profiles shown in Fig~\ref{fig:sim-obj-2i}A and Fig~\ref{fig:sim-obj-2i}B smooth the saltations. The very first saltation is mostly aligned across the individual realisations, and so it is quite prominent in both $K$ and $\widehat{D}$. Fig~\ref{fig:sim-obj-2i}C shows that the fitness $K$ and accumulated mutations $\widehat{D}$ are strongly temporally aligned (with Pearson correlation $0.99$ over $2,191$ data points), suggesting that saltations in transmissibility are produced by accumulated mutations.

We also observe that the fitness and accumulated mutations are slightly higher with increasing the population size, although these changes are much smaller than the population differences. All three population sets produced saltations in fitness and accumulated mutations. However, in smaller populations (i.e., 1.7 million),  profiles of these two measures are not only slightly lower but also somewhat less abrupt, relative to the profiles produced in larger populations (8 million and 25.4 million).

\begin{figure}
    \centering
    \includegraphics[width=0.7\columnwidth]{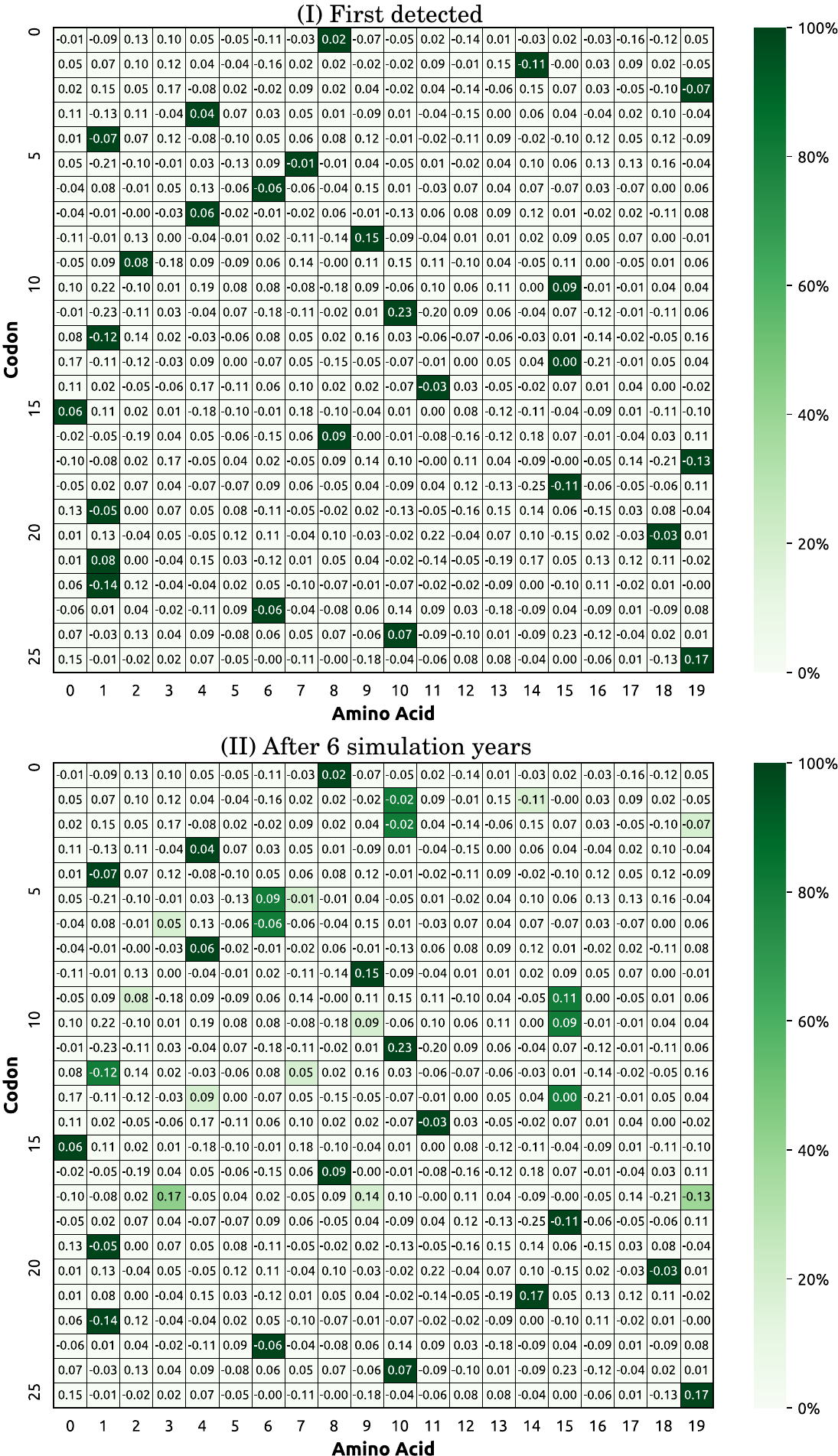}
    \caption{Snapshots of amino acids -- codons histograms at different simulation time steps, obtained from a single realisation using the population of 1.7 million agents (Capability 2(ii)).  Amino acids in the first 26 codons (i.e., the first 78 nucleotides) are shown here for simplicity. Note that the numerical fitness contributions are generated following distributions specified in S1 Text (Contribution to fitness), prior to each simulation run. (I) The histogram produced for the first day when cases were detected, and (II) the evolved histogram computed from 500 randomly selected genomes after 6 simulation years. The colour intensity represents the frequency of a given amino acid--codon combination in the sampled genome(s). The value shown in each histogram cell represents the pre-defined amino acid contribution to fitness. }
    \label{fig:sim-obj-2iii}
\end{figure}

To explain the punctuated increase in fitness, we traced changes in the distribution of fitness contributions across the simulated genome (Capability 2(ii)) by comparing the relative frequencies of amino acids in simulated ancestral genome (Fig~\ref{fig:sim-obj-2iii} (I)) and the evolved distributions in circulating genomes after six simulation years (Fig~\ref{fig:sim-obj-2iii} (II)).
In simulated dynamics, we observed a clear shift towards amino acids with positive fitness contributions, in accordance with empirical observations (Figs~\ref{fig:obj2}C and \ref{fig:obj2}D). This observation indicates that the viral mutations are subject to selective pressure transitioning to a higher point in the viral fitness landscape. Section \ref{sec:methods} and S1 Text section \ref{sm:sec:fitness_model} provide a detailed description of the fitness contribution method.

\begin{figure}
    \centering
    \includegraphics[width=0.65\columnwidth]{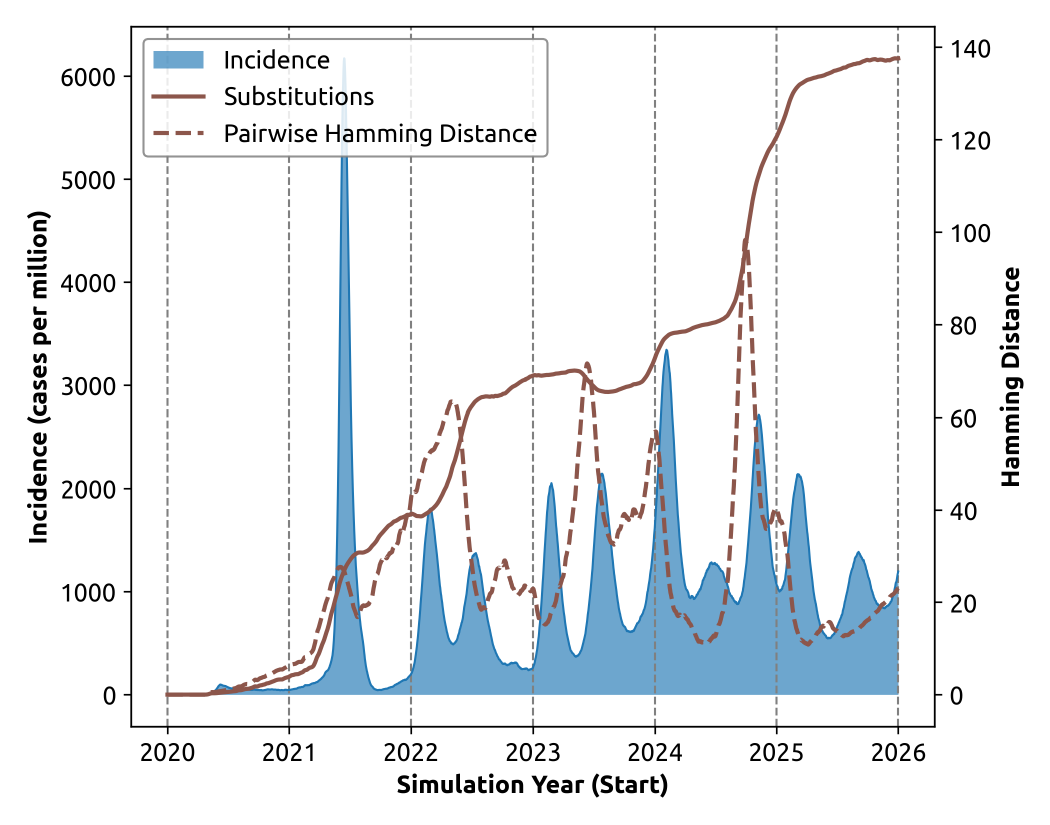}
    \caption{Temporal alignment of phylodynamics and epidemiological dynamics (Capability 3i), simulated for a 1.7 million population over a 6-year period.  Alignment of epidemic incidence (shaded blue area), accumulated mutations ($\widehat{D}$, solid brown line), and genomic diversity ($\overline{D}$, dashed brown line).   Profiles are shown for a single realisation.}
    \label{fig:sim-obj-3i}
    \end{figure}

\textbf{Capability 3.}
In pursuing our final objective, we explore the emergence and dominance of variants of concern in the simulated dynamics. We begin by examining whether the simulated phylodynamic characteristics are temporally aligned with the disease incidence. 
We note that saltations in the accumulated mutations $\widehat{D}$ can be matched by peaks in  the genomic diversity $\overline{D}$, as shown by
Fig~\ref{fig:sim-obj-3i}. These diversity peaks indicate a rise of a new variant (i.e., increase in diversity), followed by its dominance (i.e., decrease in diversity) until yet another variant emerges. In turn, these abrupt changes correspond to incidence peaks, that is, there is a notable synchrony between dynamics of the incidence and the genomic diversity $\overline{D}$. 
At the same time, not all incidence peaks can be explained by the changes in genomic diversity or the saltations in fitness. 

These observations are well aligned with the empirical data for the first four COVID-19 pandemic years. In particular, Fig~\ref{fig:obj3-i}A suggests that, while the accumulated mutations monotonically increase, the genomic diversity fluctuates along a volatile pattern with periods of drift (i.e., steady increase) followed by rapid increases and sudden collapses in the pairwise Hamming distance. Our simulation produced a similar alignment between the growing accumulated mutations $\widehat{D}$ and the fluctuating genomic diversity $\overline{D}$, as illustrated by Fig~\ref{fig:sim-obj-3i}. 
We note that Fig~\ref{fig:sim-obj-3i} illustrates the genomic diversity dynamics based on one simulation realisation (for 1.7 million agents).  Figs \ref{sm:fig:sim-obj-2i-pairwise_1M7} to \ref{sm:fig:sim-obj-2i-pairwise_25M4} show simulation results across multiple realisations and different population sizes.

\begin{figure}
    \centering
    \includegraphics[width=\columnwidth]{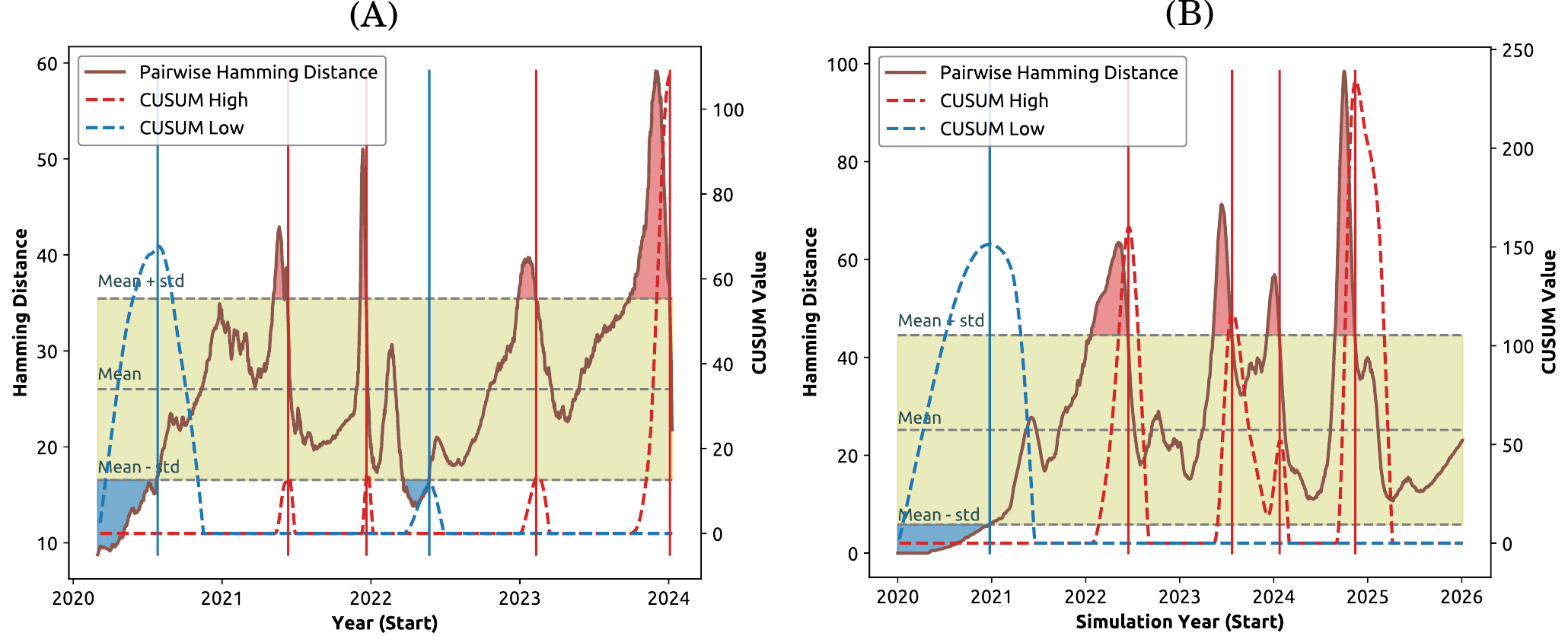}
    \caption{Detection of emerging variants (Capability 3 ii) by applying CUSUM to the genomic diversity quantified by pairwise Hamming distance ($\overline{D}$). (A) The empirical genomic diversity based on Nextstrain datasets~\cite{nextstrain_data}. (B) Simulated genomic diversity in a population of 1.7 million. Vertical lines represent variants detected using CUSUM high (red) and CUSUM low (blue). The yellow shaded area shows the control range around the mean $\overline{D}$ within one standard deviation. Simulated profile in (B) is shown for a single realisation.
    }
    \label{fig:sim-obj-3ii}
\end{figure}

Aiming to demonstrate Capability 3 (ii), we quantified the emergence and dominance of variants by analysing the genomic diversity dynamics with a statistical technique based on deviations from Cumulative Sum (CUSUM), and visualising the phylogenetic tree (see Section \ref{sec:methods}). The analysis was applied to both empirical and simulated data. For empirical data, we detected six notable deviations  (Fig~\ref{fig:sim-obj-3ii}A), identifying six emergent and dominant variants between 2020 and 2024. For the simulated dynamics, five deviations were detected between 2020 and 2026 (Fig~\ref{fig:sim-obj-3ii}B).

\begin{figure}
    \centering
    \includegraphics[width=\columnwidth]{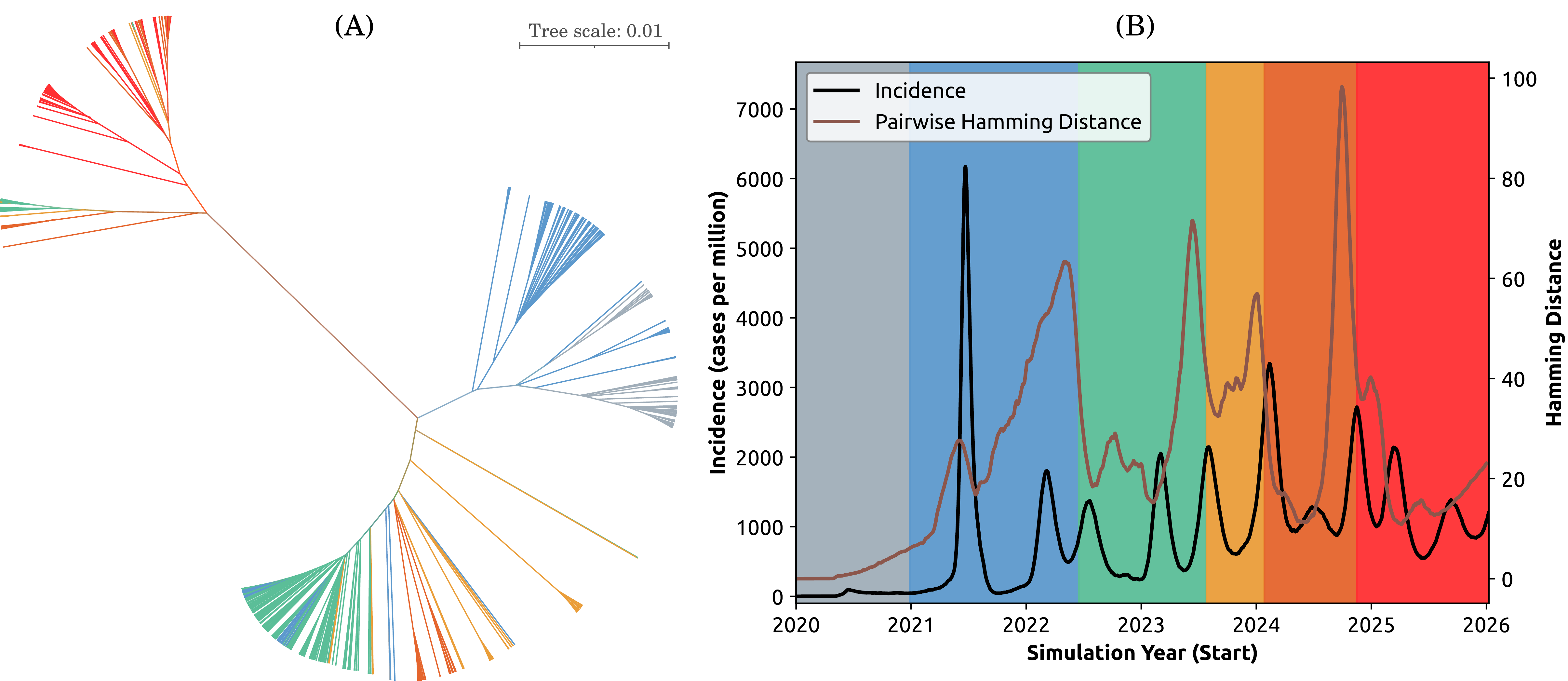}
    \caption{Evaluation of the emergence and dominance of variants (Capability 3 ii). (A) Phylogenetic tree constructed using BEAST \cite{bouckaert_beast_2019,oude_munnink_rapid_2020} and iTol \cite{letunic_interactive_2024}, 
    depicting the genomes of the most transmissible strains sampled every half-day. A total number of 4,394 genomes are plotted. (B) Alignment of simulated incidence (solid black line), the genomic diversity ($\overline{D}$, solid brown line), and the detected variants identified in Fig~\ref{fig:sim-obj-3ii} (B). Both (A) and (B) are colour-coded according to the day ranges determined by the CUSUM peak detection in Fig~\ref{fig:sim-obj-3ii}.}
    \label{fig:sim-obj-3ii_2}
\end{figure}

By aligning the genomic diversity with the incidence dynamics (Fig~\ref{fig:sim-obj-3ii_2}B), we observe that {distinct} transitions between the variants which emerged during simulation {coincide with sharp changes in the corresponding genomic diversity}. We also note stationarity of the genomic diversity in smaller populations, although the time series become less stationary in larger populations (see S1 Text: Stationarity of genomic diversity). It is well known that the spatiotemporal synchrony of disease spread between communities is correlated to the population size of the communities \cite{cliff2018nvestigating}. In our study, the smaller population (1.7 million) represents the state of South Australia, with its capital (Adelaide) comprising a significant fraction of this population (see S1 Text: Artificial agent-based population). Consequently, the population size of local government areas is relatively homogeneous and the population is more well-mixed. Our conjecture is, therefore, that a newly emerged and more transmissible strain tends to dominate more easily, reducing the average genomic diversity. This yields a more homogeneous immune response within a smaller population, dampening the incidence waves (see discussion of Capability 1 above). 

Finally, we constructed a phylogenetic tree using simulation results. The tree is shown in Fig~\ref{fig:sim-obj-3ii_2}A, with the branches colour-coded by the variants identified using CUSUM, corresponding to Fig~\ref{fig:sim-obj-3ii}B. Notably, the phylogenetic tree  reveals that (1) the new variants emerge at a distance from the ancestral strain, while branching away from more basal lineages, and (2) 
the variants detected in later years are closer to each other. This branching structure could be explained by the time it takes a new variant to accumulate novel antigenic features and increase its fitness due to these antigenic advantages. These observations are consistent with the branching of the phylogenetic tree constructed using empirical data (Fig~\ref{fig:obj3-ii}B).

Overall, these simulation results, aligned with the empirical observations, imply that a rapid punctuated evolution with prominent saltations in transmissibility is driven by accumulation of fitness-increasing mutations relative to the ancestral strain, aided by persistent, chronic infections. The simulated phylodynamics produces a fluctuating genomic diversity, associated with the emergence and dominance of variants of concern, at least during a mid-term pathogen evolution. 

\section{Discussion} \label{sec:discussion}
In this study, we identified key requirements for a new class of multi-scale pandemic models. The three considered dynamics included: (1) pathogen evolution, punctuated by the emergence of new pathogen variants, (2) human interactions
within heterogeneous populations, and (3) public health interventions aimed at controlling disease transmission. We described specific modelling capabilities, and 
developed a computational framework, implemented in a comprehensive simulator --- \phasetrace ~\ --- which meets these objectives. \phasetrace is capable of modelling the spread of infectious diseases while accounting for the evolutionary trajectory of pathogens across diverse demographics. We validated the framework with a COVID-19 case study, by calibrating \phasetrace to the phylodynamic and epidemiological characteristics of SARS-CoV-2, and simulating the corresponding immuno-epidemiological and phylodynamic patterns across heterogeneous population settings, scaling the demographics to different sizes.

We then applied different phylodynamic measures to analyse the simulated dynamics to detect emerging and dominating pathogen variants, aligning the outcomes with empirical observations. In general, while the epidemiological layer can be calibrated and validated relatively quickly, the phylogenetic and immunological layers require a greater effort to calibrate using information collected over a longer time period. However, once this process is complete, this model can be used not only for retrospective phylodynamic analysis, but also for an investigation of future longer-term evolutionary and epidemiological trends and risks. \phasetrace can also be calibrated to pandemic scenarios related to other respiratory diseases with known phylodynamic and epidemiological characteristics. In cases where phylodynamic and epidemiological evidence is lacking, \phasetrace can be used to generate a suitable model and investigate various ``what-if'' scenarios under different phylodynamic hypotheses. However, scenarios related to non-respiratory diseases (e.g., foodborne epidemics, vector-borne diseases) warrant further research.

In particular, \phasetrace reproduced recurrent incidence waves with salient initial peaks and a transition to endemicity (Capability 1). These observations showed that, given the waning immunity and reduced vaccine efficacy against emerging variants, the adopted NPIs and vaccination roll-outs would not eliminate the spread completely. Importantly, the simulated phylodynamics produced a rapid punctuated evolution, and this was explained in terms of the accumulation of fitness-increasing mutations within chronically infected hosts. This increase was further confirmed by a clear shift in the simulated distribution of amino acids towards fitness-increasing mutations, appearing due to selective pressure (Capability 2). Finally, we related the emergence and dominance of variants of concern to prominent changes in genomic diversity. This analysis was supported by the visualisation of the phylogenetic tree and relevant stationarity tests (Capability 3). 

\phasetrace is designed to support the testing of different hypotheses by simulating counterfactual modelling scenarios. For example, chronic infections have been hypothesised to contribute to the rapid punctuated evolution of SARS-CoV-2, being often associated with accelerated substitution rates, and a higher genetic diversity and selective  pressure~\cite{markov_evolution_2023}. We applied \phasetrace to investigate the potential role of chronic infections on the pathogen evolution by comparing resultant phylodynamics with and without chronically infected hosts (see S1 Text: Chronic infections).

Computationally, \phasetrace involves a nested stochastic simulation, where the (micro-scale) within-host pathogen evolution is simulated in every artificial agent, while the disease transmission is simulated based on the (meso-scale) agent interactions within a heterogeneous population. Finally, the (macro-scale) public health interventions, such as NPIs and vaccination, are simulated at the population level, constraining the actions of individual agents.
The addition of immunological and phylogenetic layers significantly extended the capabilities of \phasetrace relative to state-of-the-art pandemic ABMs, although at the cost of higher computational complexity and the need to integrate fragmented data inputs. Nevertheless, the approach retains computational tractability and scalability (see S1 Text: Computational complexity and implementation). 

We acknowledge several limitations of the current study.
We assumed that at any given time, each infected host can only carry a single strain, without exploring the possibility of co-infection (i.e., a simultaneous infection by multiple strains).
In our case study, we simplified the {genome representation}, partitioning it into spike and non-spike regions, with an overlapping region representing epitopes containing genetic information relevant to triggering immune responses (i.e., transmissibility and antigenicity). We assigned fitness based on a linear combination of contributions from individual amino acids, without modelling transcription. This precludes nonlinear contributions, which may be more relevant as a large number of mutations accumulate.
Our design also does not accommodate biologically important fitness effects due to RNA secondary structure, codon usage bias and GC content\cite{Wang_2020}, gene regulation, and ribosomal frameshifts.
A potential extension of the model could investigate a more comprehensive definition of fitness.

Our mutation model assumes that only point mutations occur, with no structural changes such as insertions, deletions and recombination. We also adopt a simple Jukes-Cantor substitution model with equal rates between nucleotides. An extended model that incorporated more nuanced fitness and transmissibility effects would need to revisit these simplifying assumptions.

We implemented within-host selective pressure by generating numerous mutant strains and selecting a strain among the most transmissible of these candidates. This allowed us to investigate whether a bias towards higher transmissibility in a subset of hosts was sufficient to generate the observed lineage dynamics. In reality, the host environment selects for factors other than transmissibility, particularly in chronically infected hosts (who might be immunocompromised and receiving antiviral or antibody therapy). In our model, each infected host only has a single viral genome associated with them at any given simulation time step, and the only ``functions'' derivable from our synthetic genome representation are transmissibility and antigenic distance. A possible extension might model virulence and directly simulate competition between a population of viral lineages in each host. This would require considerable work, calibration, and computational resources.

We made a simplifying assumption that the imported infections carry a strain with the highest transmissibility among the strains circulating during the preceding simulation month. This component could be made stochastic by considering multiple imported strains, chosen in proportion to their incidence.

For computational efficiency, we modelled phylodynamics of a pandemic pathogen in a population of up to 25.4 million, whereas the 
relevant population size is much larger, approaching the size of the world population. However, our results demonstrated that the key simulation outcomes (such as substitution rate, fitness level, accumulated mutations, and genomic diversity) scaled sublinearly, indicating convergence as the population size grows (Fig~\ref{fig:sim-obj-2i}). Secondly, we reduced the genome length to a tenth of the SARS-CoV-2 genome length. This may partially explain why the genomic diversity becomes less stationary in larger populations, compared to stationarity of the genomic diversity computed using empirical data (see S1 Text: Stationarity of genomic diversity). 

Furthermore, we did not exhaustively explore the impact of different weight tables, defining the fitness contribution of amino acids, on the genomic diversity over time and its resultant (non-)stationarity. In addition, the employed stationarity test (ADF) is known to be sensitive to the number of included lags in the time series. 
These limitations may be overcome in future studies, further enhancing the modelling scope and range of applicability of \phasetrace.

\section{Methods}
\label{sec:methods}
\subsection{Multi-layer architecture of \phasetrace}
\label{sec:methods_architecture}
\begin{figure}[th]
    \centering
    \includegraphics[width=0.9\textwidth]{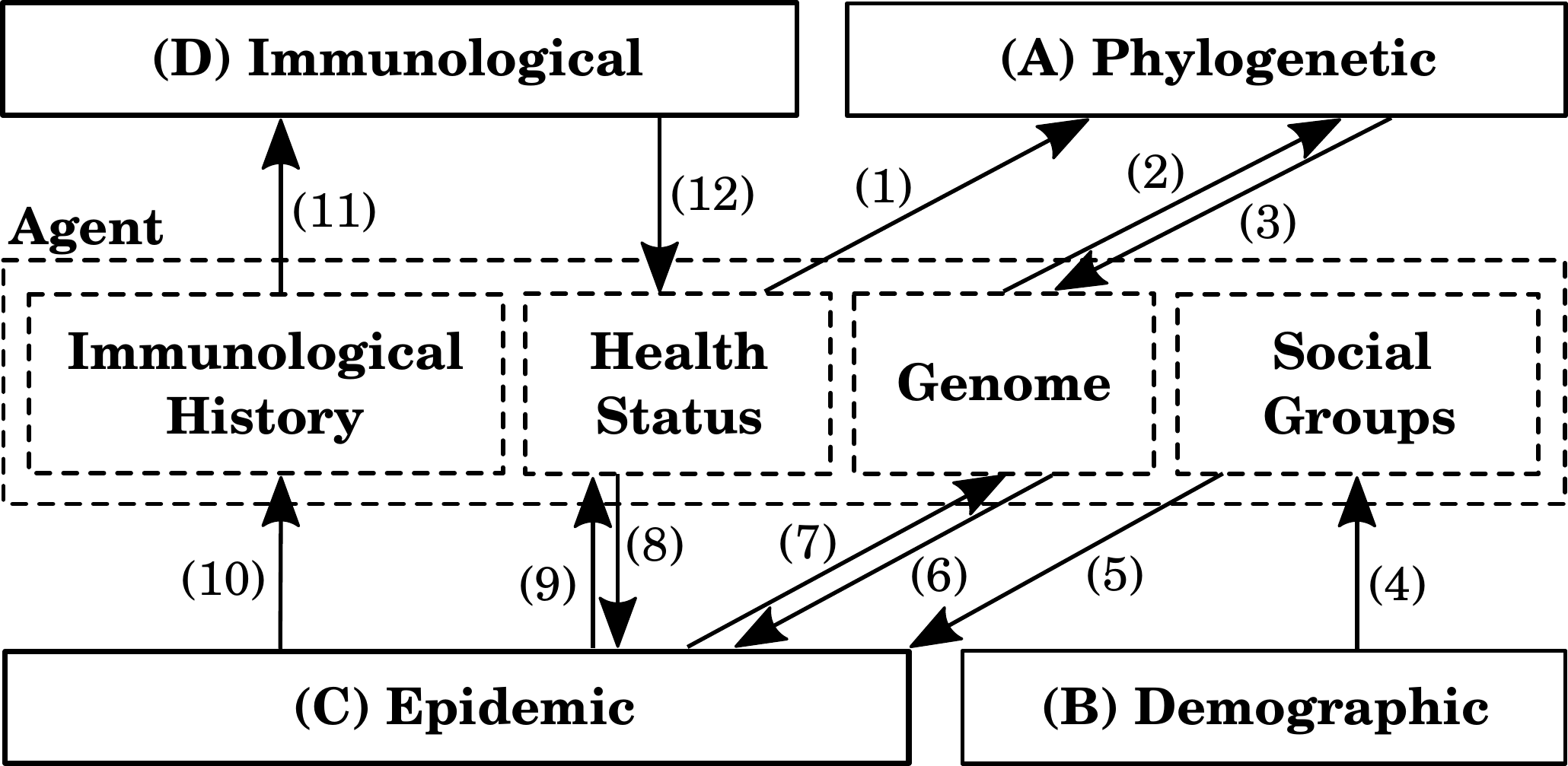}
    \caption{Architecture of \phasetrace. Four processing layers (A–D) and twelve data flows update four core attributes of the \textit{Agent} objects.}
    \label{fig:4-layer-model}
\end{figure}

\phasetrace is a large-scale stochastic simulator developed to model the mid- to long-term phylodynamics of pathogens within a heterogeneous population of agents. Computationally, each agent in \phasetrace is represented as an object with multiple attributes which can be modified by processing layers: (A) Phylogenetic, (B) Demographic, (C) Epidemic, and (D) Immunological, as illustrated in Fig~\ref{fig:4-layer-model}. 

Each agent has four core attributes: (i) immunological history which records past infections and vaccinations, as well as the associated time stamps of these events; (ii) health status, which tracks the agent's current health state (Susceptible, Asymptomatically Infectious, Symptomatically Infectious, or Recovered), infected host category (typical infected or chronically infected), and the current immunity against circulating variants; (iii) genome profile of the variant carried by the agent (if infected); and (iv) the social groups, which indicate the social contexts where interactions occur.

The flows between processing layers and agent attributes are directional, as illustrated in Fig~\ref{fig:4-layer-model}: the flows from processing layers to agent attributes modify the attributes, while the flows from agent attributes to processing layers influence the processing layer, as described below.

\textbf{(A) Phylogenetic}: this layer simulates mutations and selective pressure on the genome carried by an infected agent, as detailed in S1 Text (Phylogenetic model). Once the agent becomes infectious, the Phylogenetic layer receives input on the agent's health status, including the host type category via (1), and the genome profile derived from the infection source via (2). The mutated genome is then saved in the agent's genome attribute via (3).

\textbf{(B) Demographic}: prior to the simulation, this layer generates a heterogeneous artificial population of a specified size, based on 2021 Australian Census data  \cite{ABS_census}. The generated demographics are used in constructing the social groups via (4), determining each agent's social mixing contexts which constrain agent interactions. These contexts include (a) residential contexts, such as households, household clusters, and statistical areas (SAs) at various resolution levels; and (b) studying/workplace environments, such as schools or working groups, depending on the agent's age group. Details on the Demographic layer and artificial population generation are provided in S1 Text (Artificial agent-based population). 

\textbf{(C) Epidemic}: this layer models disease transmission and control, detailed in S1 Text (Multi-strain transmission model). The infection transmission is modelled stochastically, being affected by three agent attributes: social groups which constrain interactions between susceptible and infectious agents, via (5); health status which comprises the agent's health state and immunity level, given previous immunological events, via (8); and genome profile which provides a representation of the pathogen infecting the agent, via (6). Having simulated a transmission of infection between two agents, the epidemic layer updates three agent attributes: health status, modifying the agent's health state, via (9); genome profile, by using the representation of the pathogen variant carried by the source of infection, via (7); and immunological history once the infected agent recovers, via (10).
 
\textbf{(D) Immunological}: this layer reads from an agent's immunological history of past infections and vaccinations, via (11); and updates the immunity level as part of the agent's health status, via (12). S1 Text (Vaccination) provides a detailed description of the vaccination component. The immunity levels are also affected by a non-linear accumulation over multiple immunity-boosting events (i.e., compound immunity) and the associated waning effects, detailed in S1 Text (Immunological layer) and S1 Text (Compound immunity and waning effects).

An efficient implementation of this multi-layer architecture requires a resolution of several computational challenges,  given the demanding simulation timeframe (over 6 years) and memory-intensive tasks associated with storing evolving agent attributes, particularly in large populations. To address these challenges, we employed multi-threading processing that computes multiple attributes in parallel in each \textit{Agent} object, with the attributes configured independently from each other to achieve concurrency. In doing so, we significantly increased the computational efficiency by reducing both simulation time and associated computational costs. S1 Text (Computational complexity and implementation) details an analysis of the performance, scalability, and computational resources of \phasetrace.

\subsection{Multi-scale phylodynamic simulation}
This section describes processing layers (A), (C) and (D), highlighting their key dynamic relationships. Since the demographic characteristics of agents, including their social groups, are not updated during simulation, there is no dynamic modelling involved in layer (B), and this layer is described in S1 Text (Artificial agent-based population).

\subsubsection{Phylogenetic layer (A)}
The phylogenetic layer models the within-host evolution (mutation and selection), using pathogen genomes associated with each infected agent, and computes the fitness of circulating variants. An artificial genome contains 3,090 nucleotides (or 1,030 codons) which are associated with 20 known types of amino acids. The genome structure is partitioned into the spike region (100 codons) and the non-spike region (930 codons), with an overlapping region accounting for epitopes (75 codons, with 45 of these located in the spike region and 30 in the non-spike region). S1 Text (Genome structure) provides a detailed description of the genome structure, followed by S1 Text (Mutations) describing point mutations.  

Pathogen fitness is defined in terms of its transmissibility, {proportional to the corresponding basic or effective reproductive number ($R_0$ or $R_{\text{eff}}$)}.
To quantify the individual fitness contributions of amino acids at each codon position, 
we employ a weight table, specifying weights
$a_{i,j}$ of $N_A = 20$ types of amino acids across $N = 1,030$ codon positions, 
as detailed in S1 Text (Contribution to fitness). 
The overall fitness $K$ of a strain with genome $s$ is determined as the sum of these individual contributions:
\begin{equation}
    K(s) = \sum_{i=0}^{N-1}  a_{i,s(i)}
    \label{eq:linear_combo_for_fitness}
\end{equation}
where $a_{i,s(i)}$ is the fitness contribution 
of amino acid $j = s(i)$ located at codon position $i$, with $j \in [0, N_A-1]$.\\ 

To re-iterate, the phylogenetic model distinguishes between two infected host categories: typical infected hosts and chronically infected hosts (S1 Text: Infected host categories). Consequently, there are different selective pressures driving the within-host evolution of pathogens in these categories (S1 Text: Intra-host (within-host) selective pressure). At each mutation step, we generate multiple candidate sequences, rank them by transmissibility, and focus on the group of top candidates. The size of this group indicates the strength of selective pressure. For example, selecting from a smaller group of top candidates implies a stronger selective pressure, while selecting from a larger group of top candidates implies a weaker selective pressure. When simulating selective pressure for chronically infected hosts, the size of the top group is assumed to be smaller, indicating a stronger selective pressure, compared to the selective pressure for typically infected hosts. We elaborate on this distinction and the limitation of these assumptions in Section \ref{sec:discussion}.

We distinguish between mutation and substitution rates. 
While the mutation rate refers to the frequency of new mutations arising in a genome per unit of time, the substitution rate is the rate at which these new mutations are retained over time within the population~\cite{bromham_why_2009, nature_molecular_clock, markov_evolution_2023}. 
In our case study, the mutation rate is a key input parameter, whereas the average substitution rate emerges as a simulation outcome. We use the attained substitution rate to calibrate the model, by matching the regression coefficient observed in empirical observations, illustrated in Fig~\ref{fig:obj2}B. 

\subsubsection{Epidemic layer (C)} 
At the start of simulation, pathogens with the ancestral genome are ``seeded'' by infecting agents residing around international airports. Every month, the simulation updates the genomes seeded around airports, selecting the variant with the highest transmissibility detected during the preceding month.

As described in S1 Text (Multi-strain transmission model), disease is transmitted among agents that interact across different social contexts. The transmission process is simulated in discrete half-day time steps: ``daytime'' cycles during which agents interact in workplaces or educational settings (e.g., class, grade, school), and ``nighttime'' cycles during which agents interact in residential settings (e.g., household, household cluster, neighbourhood, and community). Each agent progresses through several health states: Susceptible, Infectious (asymptomatic or symptomatic), and Recovered, following the natural history of the disease.

At simulation cycle $n$, the infection probability $p_i(n)$ for a susceptible agent $i$ is determined across all its social contexts $g \in G_i$ (see S1 Text: Susceptible-infectious transition):
\begin{eqnarray}
\begin{aligned}
    p_i(n) &= 1 - \prod_{g \in G_i(n)} \; \prod_{j \in A_g\backslash\{i\}} \left( 1 - p_{j \rightarrow i}(n,g) \right)
\end{aligned}
\label{eq:general_infection_prob}
\end{eqnarray}
where $G_i(n)$ denotes the set of all social contexts $g$ that agent $i$ interacts with during the time cycle $n$, $A_g\backslash\{i\}$ denotes the set of agents in $g$ (excluding agent $i$), and $p_{j \rightarrow i}(n,g)$ denotes the probability of infection transmission from infectious agent $j$ to susceptible agent $i$ within their social context $g \in G_i(n)$.  The probability $p_{j \rightarrow i}(n,g)$ is defined as follows:
\begin{eqnarray}
\label{eq:individual_transmission_prob}
    p_{j \rightarrow i}(n,g) = K(s_j) \ f_j(n-n_j) \ q_{j \rightarrow i}(g)
\end{eqnarray}
where $q_{j \rightarrow i}(g)$ is the age-dependent interaction probability within $g$ (see S1 Text: Susceptible-infectious transition);
$n_j$ denotes the infection onset time for agent $j$; the agent-specific function $f_j(n-n_j)$ is the natural history of the disease, reflecting the infectivity of agent $j$ as its infection progresses (see S1 Text: Infectious-recovered and recovered-susceptible transitions); and $K(s_j)$ represents the transmissibility of pathogen variant $s_j$, proportional to the corresponding basic reproductive number $R_0$ {or effective reproductive number $R_{\text{eff}}$}, i.e., $K(s_j)$ is the fitness of genome $s_j$ carried by agent $j$, as defined by Eq \ref{eq:linear_combo_for_fitness}. 

Once susceptible agent $i$ becomes infected, it is possible to assign (i.e., identify) a specific infectious agent as the source of infection. This is simulated by weighted random sampling of an infection source from all potential infectious agents $j$ across all social contexts $g \in G_i(n)$ in which agent $i$ interacted during this cycle. Then the pathogen genome profile $s_j$ (carried by the
identified infection source agent $j$) is copied to agent $i$ (see S1 Text: Susceptible-infectious transition). 

Various interventions may change the infection probabilities across social contexts. Modelling non-pharmaceutical interventions (NPIs) is described in Section Non-pharmaceutical interventions.

\subsubsection{Immunological layer (D)} 
\label{methods:immuno}
The immunological layer simulates a vaccination rollout, including vaccination coverage, schedule and rates, given vaccine efficacy. Furthermore, the immuno-epidemiological model quantifies the agent immunity resulting from multiple immunity-boosting events (i.e., vaccinations and infections), as described in S1 Text (Vaccination). 
In this work, we broaden the concept of ``hybrid immunity'' to ``compound immunity'', in order to capture non-linear immunity accumulation over various combinations of prior vaccinations and infections. Compound immunity may result from one or multiple vaccinations, one or multiple past infections, or a combination of both vaccination(s) and past infection(s)~\cite{chang2024impactopiniondynamicsrecurrent}.

We decompose the compound immunity via three separate sub-components, quantifying reductions of different risks: susceptibility, symptomatic infection, and forward transmission. For example, the compound immunity against symptomatic infection, denoted $M_{i}^\textrm{c}$, for susceptible agent $i$ interacting with infectious agent $x$, is defined as follows:
\begin{equation}
    M_{i}^\textrm{c}(n,H_i, s_x) = \text{min}\left(\sqrt{\sum_{r \in H_i} \left [m_{i}^\textrm{c}(n,r,s_x) \right ]^2}, 1\right)
    \label{eq:immunity_symptomatic_accumulation}
\end{equation}
where $m_{i}^\textrm{c}(n,r,s_x)$ is the immunity against symptomatic infection induced by past immunological event $r$ (vaccination or infection); $H_i$ is the immunological history formed by past records $r$ up to cycle $n$; and $s_x$ is the genome carried by infectious agent $x$. In addition, the compound immunity wanes over time and depends on the genetic distance (see Eq 12 in S1 Text: Compound immunity against symptomatic infection). 
S1 Text (Compound immunity and waning effects) provides more details accounting for different vaccine efficacy components contributing to the compound immunity, and its effects on infection probabilities.

\subsection{Phylodynamic measures}
\subsubsection{Hamming distance}
In this study, we used Hamming distance as the primary measure to count nucleotide differences between two genomes \cite{PINHEIRO2012129}. The average reference Hamming distance between evolved genomes and the reference genome (i.e., ancestral genome), denoted $\widehat{D}$, is used to account for the mutations accumulated during the simulation timeframe, while the average pairwise Hamming distance among evolved genomes, denoted $\overline{D}$, is used to quantify the genomic diversity. The computation of these two measures is described below. 

To quantify the accumulated mutations $\widehat{D}$, for each simulated day $n$:

\begin{itemize}
    \item Select all genome profiles obtained within a one-week forward window starting on day $n$. 
    \item For each of the profiles, record the number of differences against the reference genome (i.e., ancestral genome, NCBI GenBank accession number MN908947 \cite{Wu2020,nextstrain}), producing the reference Hamming distance.
    \item Compute the average reference Hamming distance across all circulating variants. 
\end{itemize} \vskip5pt

To quantify the genomic diversity $\overline{D}$, for each simulated day $n$:
\begin{itemize}
    \item Randomly select 10,000 pairs of profiles 
    obtained within a one-week forward window starting on day $n$.
    \item For each pair of genomes, record the number of differences between them as the pairwise Hamming distance.
    \item Compute the average pairwise Hamming distance of all pairs. 
\end{itemize}\vskip5pt

Figs~\ref{fig:obj2} and \ref{fig:obj3-i} trace the reference and pairwise Hamming distances for actual SARS-CoV-2 sequence data \cite{nextstrain_data}. 

In order to trace Hamming distances for the simulated dynamics
we followed slightly altered workflows, without applying the weekly windows for genome selection. When dealing with actual sequence data, these windows were needed to filter out sampling inconsistencies. Simulated data include pathogen genomes from all detected hosts, sampled on each simulation day, thus reducing sampling inconsistencies. The reference Hamming distance $\widehat{D}$ was computed against the simulated ancestral strain, constructed for each realisation (as described in Section~\ref{sec:methods}).
Figs~\ref{fig:sim-obj-2i}, \ref{fig:sim-obj-3i} -- \ref{fig:sim-obj-3ii_2} trace the reference and pairwise Hamming distances for the simulated phylodynamics. 

\subsubsection{Statistical stationarity tests}
To examine stationarity of the pairwise Hamming distance, we performed statistical stationarity tests, specifically Augmented Dickey-Fuller (ADF) test and one-sided Cumulative Sum (CUSUM) analysis. This allowed us to identify saltations as punctuated changes in the pairwise Hamming distance, which have been found to be closely related to the emergence and dominance of pathogen variants \cite{Nielsen}.

\textbf{ADF test.} An ADF test detects {non-stationarity in a time series} \cite{dickey_distribution_1979}. We use the following null and alternative hypotheses to determine stationarity: 
\vskip5pt
\begin{itemize}
    \item[] $H_0$: Pairwise Hamming distance is non-stationary.
    \item[] $H_1$: Pairwise Hamming distance is stationary.
\end{itemize}\vskip5pt

We computed p-value from the ADF test and compared it against a chosen significance level (i.e., 0.05), with p-value smaller than the significance level rejecting $H_0$ and confirming stationarity. Results of ADF are shown in S1 Text (Stationarity of genomic diversity) and Fig~\ref{sm:fig:ADF}. 

\textbf{CUSUM.} A one-sided Cumulative Sum (CUSUM) control chart ~\cite{page_continuous_1954, ross_chapter_2014} can identify \textit{anomalies} in the observed time series. In this study, we applied CUSUM on pairwise Hamming distance as shown in Fig~\ref{fig:sim-obj-3ii}.

Let ${W(n)}$ be the pairwise Hamming distance on day $n$, with the mean $\mu$ and standard deviation $\sigma$. We converted ${W(n)}$ into \textit{high} CUSUM ($S^X$) and \textit{low} CUSUM ($S^Y$), as follows:
\begin{align}
    S^X(n+1) = \textrm{max}\left(0, S^X(n) + W(n+1) - \mu - \sigma\right) \label{eq:high_CUSUM}\\
    S^Y(n+1) = \textrm{max}\left(0, S^Y(n) - W(n+1) + \mu - \sigma\right) \label{eq:low_CUSUM}
\end{align}
where $S^X(0)=S^Y(0)=0$.\\

The high and low CUSUM values, $S^X$ and $S^Y$, are traced in Fig~\ref{fig:sim-obj-3ii}. We then applied a positive peak detection in $S^X$ and $S^Y$ (i.e., considering CUSUM value greater than 0), to detect the anomalies. Specifically, the peaks identified in $S^X$ indicate the emergence of a more transmissible variant, whereas the peaks identified in $S^Y$ indicate the dominance of a variant within the population. The number of dominant variants is equivalent to the total number of peaks detected in both $S^X$ and $S^Y$. 

We note that CUSUM is a simplified approach and can only identify one variant of concern during a defined period. In other words, it cannot trace the frequency of multiple co-circulating variants and the associated transitions.

\section*{Declarations}
The authors declare no competing interests.
\section*{Data availability}
Simulation and post-processing data are available at Zenodo:
10.5281/zenodo.14279413.
\section*{Code availability}
The source code of \phasetrace is available at Zenodo: 10.5281/zenodo.14279368.
\section*{Acknowledgement}
This work was supported by the Australian Research Council grant DP220101688 (MP, QDN, SLC, VS, TS). The simulations involved in this work were carried out on the high-performance computing cluster (Artemis) provided by the Sydney Informatics Hub at the University of Sydney. The authors would like to thank Christina M. Jamerlan, Tim Germann, Sara Del Valle, Michael Lachmann, Stuart Ridge, and Stuart Kauffman for many insightful discussions and comments. The authors are also grateful to Oliver Cliff, Cameron Zachreson and Nathan Harding for contributing to the development of AMTraC-19, a predecessor of \phasetrace.

\appendix
\newpage
\paragraph*{S1. Supplementary Text.}
\section*{Overview of Supplementary Materials}

The phylodynamic agent-based model implemented in \phasetrace synthesises two parts: the within-host evolution using a phylogenetic model, and the inter-host transmission using a high-resolution agent-based model. This simulator comprises four processing layers: (A) Phylogenetic, (B) Demographic, (C) Epidemic, and (D) Immunological, each with various data flows updating core attributes of  \textit{Agent} objects. These agent attributes include:  (i) health status, (ii) genome representation of the pathogen if infected, (iii) the social groups with which an agent may interact, and (iv) immunological history of past infections and vaccinations. The \phasetrace architecture is described in Methods (subsection \ref{sec:methods_architecture}). 

The processing layers are detailed in Supplementary Materials as follows:
\begin{itemize}
    \item  (A) Phylogenetic Layer: Section \ref{sm:sec:phylogenetic} (Phylogenetic model) describes the model of the within-host evolution.
    \item  (B) Demographic Layer: Section \ref{sm:sec:pop} (Artificial population) details the generation and structure of the artificial heterogeneous population.
    \item  (C) Epidemic Layer: Section \ref{sm:sec:trac} (Multi-strain transmission and control) describes the disease transmission model in terms of agent interactions, in presence of multiple co-circulating pathogens characterised by different genomes (subsection \ref{sm:sec:multi}), as well as the control model of multiple non-pharmaceutical interventions and their impact on disease transmission (subsection \ref{sm:sec:NPI}).
    \item (D) Immunological Layer: Section \ref{sm:sec:imm-epi} (Immuno-epidemiological model) details modelling of vaccination roll-outs and vaccine-induced immunity (subsection \ref{sm:sec:vac}), followed by a description of how the immunity changes over time during pathogen transmission and evolution (subsection \ref{sm:sec:hybrid_and_waning}). This section also describes how the within-host pathogen evolution affects inter-agent infection transmission, under the effect of compound and waning immunity generated by the immunological history.\\
\end{itemize}

Supplementary text also includes some additional results, sensitivity analysis and details on the computational complexity and implementation of \phasetrace.

\section{Phylogenetic model}
\label{sm:sec:phylogenetic}
A key part of \phasetrace is a phylogenetic model developed to simulate pathogen evolution under diverse selective pressures.
The phylogenetic model is incorporated within an agent-based model (ABM) of multi-strain epidemic transmission and control (detailed in Section \ref{sm:sec:multi}), and consists of five key components:
\begin{itemize}
    \item Genome structure represented by a nucleotide sequence with positions assigned to different functions. 
    \item Point mutations occurring at a rate that is averaged across the genome. 
    \item Pathogen fitness determined based on amino acids at specific regions of the genome. 
    \item Infected host categories, differentiating between typical infected hosts and chronically infected hosts. 
    \item Within-host (or intra-host) selective pressure generating advantageous pathogen mutations. 
\end{itemize}

These components capture complex evolutionary dynamics relating selective pressures, mutations, and varying levels of immune response emerging in the heterogeneous host population facing various disease scenarios.\\ 

In this study, we calibrated the model to match viral characteristics and epidemic dynamics of SARS-CoV-2, using the COVID-19 pandemic as a case study. This model can be adapted to explore the evolutionary dynamics of other pathogens causing communicable diseases.

\subsection{Genome structure}
\label{sm:sec:genome}

Each infected agent (i.e., host) is assigned an artificially constructed genome. Importantly, the simulated genome is designed to model evolutionary dynamics but is \textit{not} mapped to real features of the viral genome such as known coding sequences, epitopes, or viral functions. Furthermore, for reasons of computational efficiency, we simulate a sequence of only 3,090 nucleotide positions, approximately 10\% of the true size of the viral genome~\cite{nextstrain}. For simplicity, we model only one viral genome per host at any given time step, i.e., we assume that a single viral population dominates in the host, and there is a narrow transmission bottleneck~\cite{markov_evolution_2023}.

\begin{figure}[t]
    \centering
    \includegraphics[width=\columnwidth]{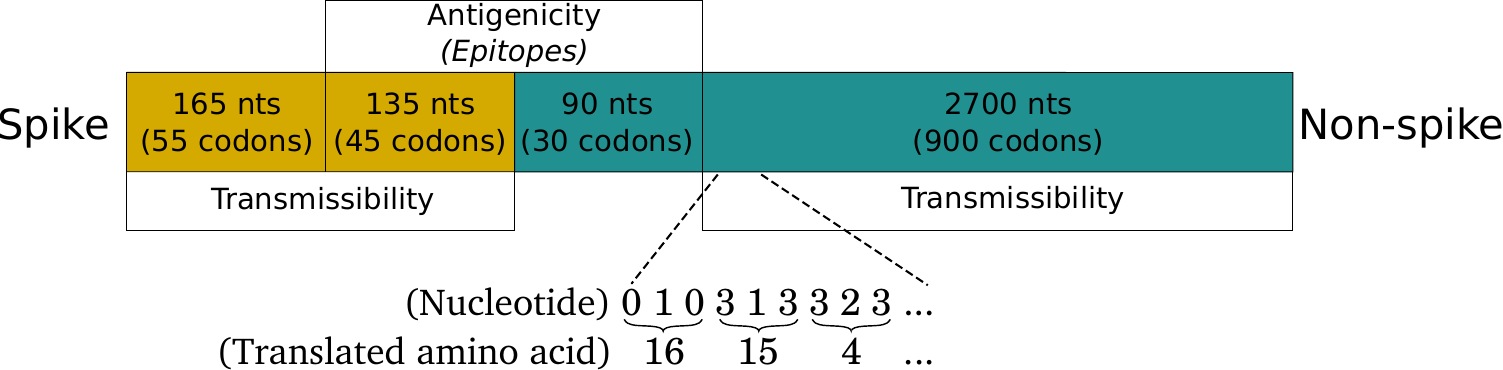}
    \caption{Simulated genome structure consisting of 3,090 nucleotides (nts), equivalent to 1,030 codons. The genome is partitioned into spike (orange) and non-spike (blue) regions. Additionally, the nucleotides can be grouped according to functions: (i) regions consisting of 1,000 codons (100 from spike region and 900 from non-spike region), contributing to the pathogen fitness and the resultant transmissibility; and (ii) regions consisting of 75 codons (45 from spike region and 30 from non-spike region) contributing to antigenicity (i.e., epitopes). The inset displays examples of grouping nucleotides into codons followed by translation to amino acids. For example, nucleotides 0 1 0 form a codon, which is translated as amino acid 16.}
    \label{sup_fig:genome_structure}
\end{figure}

Each nucleotide position in the simulated genome is represented by a number between 0 and 3. Every three nucleotides is interpreted as a codon (3,090 nucleotides correspond to 1,030 codons, see Fig \ref{sup_fig:genome_structure}), encoding a sequence of amino acids following the standard translation rules \cite{Smith_nucleic}. We annotate the 20 known amino acids by numerical labels from 0 to 19 instead of the conventional letters \cite{Smith_nucleic}. Although we model translation to an amino acid sequence, we do not look for coding regions or otherwise interpret or annotate the simulated sequence in the context of the real virus.

The simulated genome is partitioned into two distinct regions (Fig~\ref{sup_fig:genome_structure}): the ``spike'' region (consisting of 100 codons) and the ``non-spike'' region (consisting of 930 codons). This partitioning affects only two aspects of the simulation: (1) vaccine-related selective pressure is restricted to the spike region, and (2) mutations in the spike region have a greater 
chance of enhancing transmissibility
(Section \ref{sm:sec:fitness_model}). 75 codons (45 codons in the spike region and 30 codons in the non-spike region) are interpreted as epitopes relevant for modelling of immunity~\cite{harvey_sars-cov-2_2021,smith_landscape_2021}. 1,000 codons (including 45 that are also in epitopes) are interpreted as relevant to transmissibility. Section \ref{sm:sec:immunity_infection} provides more details.

\subsection{Mutations}
\label{sm:sec:mutation_substitution}
In this study, we model mutations as random substitutions at single nucleotide positions. We do not consider insertions, deletions, recombination, or other structural changes to the genome. Fitness and antigenicity are computed from the amino acid sequence, meaning that synonymous nucleotide substitutions will not result in a functionally different simulated genome. We also assume that the transmissibility and antigenicity of viral variants can be computed based on independent contributions from each mutation with respect to the ancestral strain.

In each half-day simulation cycle, the total number of mutations ($L$) in each of the viral genomes within infected hosts is sampled from a Poisson distribution. The mean of the distribution is set based on an average mutation rate of 0.001 per nt per year across the genome, i.e., $3090 \times 0.001/(365 \times 2) \approx 0.00423$. The average mutation rate was chosen within the reported range for SARS-CoV-2 of 0.000219 to 0.0012 per nt per year~\cite{sender_total_2021,markov_evolution_2023}.
The relationship between mutations and their contribution to fitness is described in Section \ref{sm:sec:fitness_model}.

\subsection{Contribution to fitness}
\label{sm:sec:fitness_model}
Individual mutations in the genome affect the overall viral fitness, resulting in growing transmissibility and higher reproductive number \cite{Obermeyer_2022, kistler_rapid_2022, thadani_learning_2023}. 
Obermeyer et al. \cite{Obermeyer_2022} assessed the relative fitness of SARS-CoV-2 lineages (measured as the fold increase in $R$ over the ancestral strain) by linearly combining fitness contributions of individual amino acid substitutions, some of which were found to be fitness-increasing and recognised as spike mutations and non-spike mutations within the nucleocapsid and nonstructural proteins. Additionally, \cite{thadani_learning_2023} identified a general distribution of amino acid selection at each position in SARS-CoV-2 sequences and highlighted positions that are prone to viral escape and mutations.

Informed by these studies, we employ a weight table to quantify the individual contributions to fitness of amino acids at each codon position. The overall fitness of a strain, determined by its genome, is modelled as the sum of these individual contributions. This approach allows us to determine fitness of any simulated genome with arbitrary mutations. Specifically, the table assigns a weight $a_{i,j}$ for each of the $N = 1,030$ codon positions and $N_A = 20$ types of amino acids. For the COVID-19 case study, we assigned the weights as follows:
\begin{itemize}
    \item sample $\mathcal{N}(0, 0.085)$ for $i \in [0, 99]$, the spike region, 
    \item 0 for $i \in [100, 129]$, the non-spike epitopes, and
    \item sample $\mathcal{N}(0, 0.07)$ for $i \in [130, 1029]$, the non-spike region.
\end{itemize}

Here, we chose a higher-variance distribution for the weights in the spike region so that those codons have a higher probability of contributing to the viral fitness relative to the non-spike region. This is based on the observation of a higher substitution rate in the spike region compared to other positions
\cite{ghafari_prevalence_2024, amicone_mutation_2022}.

\begin{figure}
    \centering
    \includegraphics[width=\columnwidth]{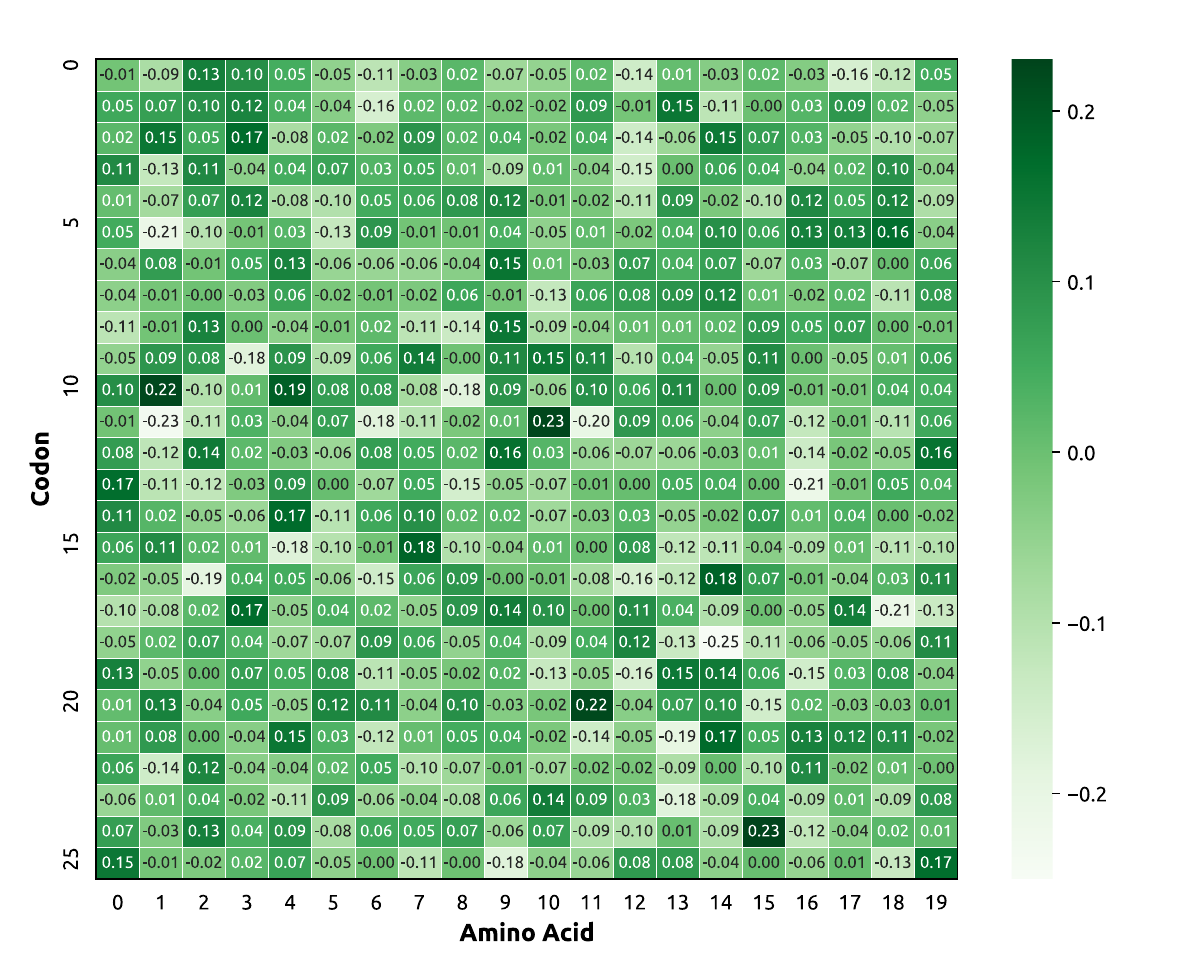}
    \caption{A section of the weight table that quantifies the fitness contributions of 20 amino acids across the genome. Only the first 26 codon positions (vertical axis) are shown for illustrative purposes. Each cell value indicates the potential contribution from each amino acid if present (horizontal axis) at the corresponding codon position (vertical axis). The cell values are sampled from a normal distribution with a mean of 0 (i.e., most random mutations are neutral). Cell colour indicates the magnitude of fitness, with darker colour representing fitness increase and lighter colour representing fitness decrease.}
    \label{sup_fig:sim-weight-table}
\end{figure}

Fig \ref{sup_fig:sim-weight-table} shows a section of the weight table (the first 26 codon positions).
The weight table generated for a simulation regulates the pathogen's evolution. In the absence of other pressures, one would expect substitutions that increase fitness to be preferentially fixed as a simulation progresses, leading to a gradual increase in transmissibility.
Using the weight table, we calculate the overall fitness $K$ of a genome $s$ by combining the individual contributions of the present amino acids:
\begin{equation}
    K(s) = \sum_{i=0}^{N-1} \ a_{i,s(i)}
    \label{sup_eq:linear_combo_for_fitness}
\end{equation}
where $a_{i,s(i)}$ is the fitness contribution 
of amino acid $j = s(i)$ located at codon position $i$, with $j \in [0, N_A-1]$.
We note that $a_{i,s(i)}$ can be zero, positive, or negative, representing mutations of neutral, advantageous, or disadvantageous contribution to pathogen fitness. We note that Eq \ref{sup_eq:linear_combo_for_fitness} and Eq \ref{eq:linear_combo_for_fitness} in the main manuscript are equivalent.

Prior to simulating the COVID-19 case study, we generated a genome for the ancestral SARS-CoV-2 strain sampled such that its viral fitness, calculated using Eq \ref{sup_eq:linear_combo_for_fitness} and the weight table, fell between 2.65 and 2.85, matching the transmissibility range of the ancestral SARS-CoV-2 strain modelled in our previous work~\cite{chang_modelling_2020}. We used this ancestral genome as a starting point from which other variants originate, thus assuming a shared ancestry for all  variants~\cite{attwood_phylogenetic_2022,yang_species_2023}. To optimise memory usage, given the high number of simulated infections, each strain is represented by a list of mutations against the ancestral genome instead of a full-length genome.

\subsection{Infected host categories}
\label{sm:sec:chronic_inf}
Host factors, including individual variation in immune response, are known to strongly affect epidemic progression~\cite{bull_contribution_2012,voloch_intra-host_2021}. In particular, chronic carriage of SARS-CoV-2 by immunocompromised hosts is thought to have had a key role in the emergence of new viral variants during the pandemic~\cite{markov_evolution_2023}.
In a host with compromised or suppressed immune function, the virus is subjected to different selective pressures and within-host dynamics.
A host immune system that is not able to effectively clear the virus is likely to result in a longer infection duration, a higher viral load within the host, and a greater diversity of viral sub-populations~\cite{markov_evolution_2023}.
These conditions may facilitate more substitutions and larger increases in viral fitness (saltational evolution) compared to the evolution of the virus over the same interval of time when transmitted by a chain of typical hosts and subjected to transmission bottlenecks and competent immune systems.
In the case of SARS-CoV-2, chronically infected hosts can have a prolonged recovery period of several months, compared to one to two weeks for typical infected hosts~\cite{gonzalez_reiche_sequential_2023, ghafari_prevalence_2024}. 

\begin{figure}[t]
    \centering
    \includegraphics[width=0.7\columnwidth]{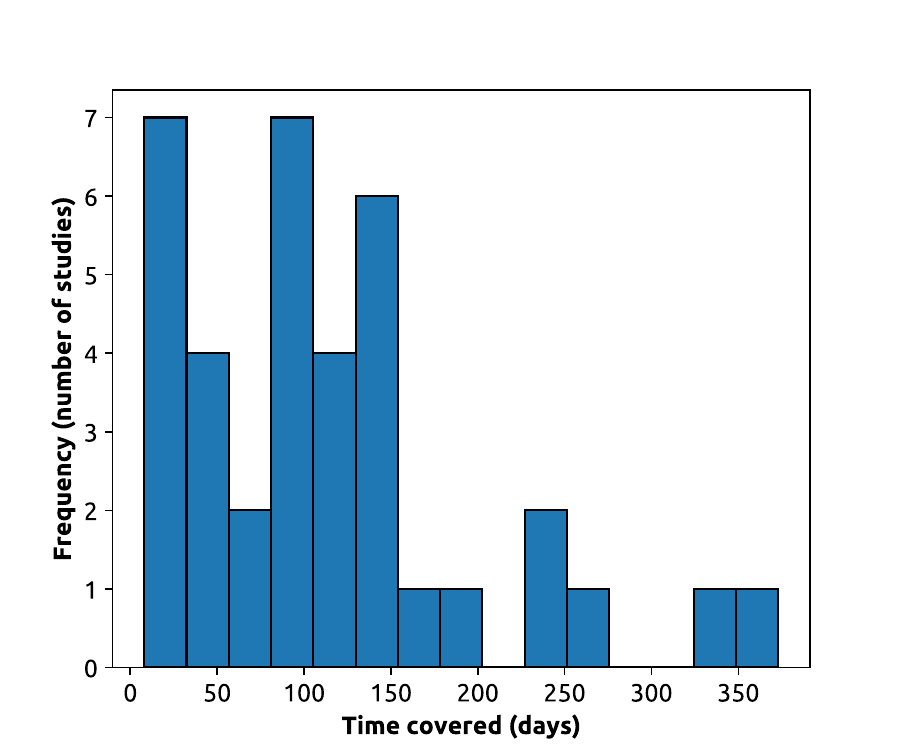}
    \caption{Histogram of infection period of the COVID-19 cases reported by various studies. Datasets were obtained from \cite{wilkinson_recurrent_2022,wilkinson_recurrent_2022_dataset}, updated as of December 25, 2022.}
    \label{sup_fig:prolonged_infection_time_covered}
\end{figure}

To examine the possible role of chronic infections on the pathogen evolution, we differentiate between two infected host categories: chronically infected hosts and the typical infected hosts. The main distinctions include the following aspects: (i) chronically infected hosts are assigned a different natural history model with a significantly longer recovery period, randomly sampled from a uniform distribution ranging from 60 to 370 days reported in many studies (Fig \ref{sup_fig:prolonged_infection_time_covered})~\cite{ghafari_prevalence_2024,wilkinson_recurrent_2022,wilkinson_recurrent_2022_dataset}, and (ii) chronically infected hosts may have a higher intra-host selective pressure (detailed in Section \ref{sm:sec:sensitivity} and
Fig~\ref{sm:fig:sim-Topxx}).

Prior study \cite{ghafari_prevalence_2024} reported that chronically infected hosts make up a small fraction of the infected hosts (between 0.1 and 0.5\% of COVID-19 infections), and only 32\% of these hosts may experience strong positive intra-host selection. Considering the high attack rate of the COVID-19, we assume that 0.1\% of the entire population is susceptible to chronic infection with COVID-19. We also assume a stronger within-host selective pressure in chronically infected agents, compared to typical infected hosts.

To evaluate the model robustness and quantify the impact of the chronically infected hosts on the overall phylodynamics, we performed sensitivity analysis varying the percentage of chronically infected hosts between 0\% to 5\% (see Section \ref{sm:sec:sensitivity}, and Fig~\ref{sm:fig:sim-fraction_chronic}).

\subsection{Intra-host (within-host) selective pressure}
\label{sm:sec:selection}
The emergence of new variants requires mutations with a fitness advantage capable of breaking through the transmission bottleneck~\cite{markov_evolution_2023}. Mutant strains require time to replicate and out-compete other viral sub-populations within the host in order to be transmitted onwards to other hosts. The selection dynamics within the host are influenced by host characteristics and the length of infection \cite{wilkinson_recurrent_2022,markov_evolution_2023,ghafari_prevalence_2024}.

A virus that has less opportunity to develop within-host diversity will be less effective at exploring the fitness landscape.
We emulate this by biasing the fitness of mutant sequences generated from a parent sequence.
At each time step where we need to generate mutations in the viral sequence associated with an infected host, we first independently generate $M$ candidate sequences and rank them by their fitness.
We then sample one sequence from the top $X \le M$ candidates to be the new sequence.
The difference $M - X + 1$ specifies the selective pressure strength.
For example, if $M = X$, this is equivalent to generating only a single candidate sequence, indicating a weak selective pressure.
On the other hand, if $M = 100$ and $X = 1$, then the selected sequence will be the fittest of 100 randomly generated candidate sequences, indicating a strong selective pressure and bias towards higher fitness. In other words, small values of $X$ imply more efficient exploration of the fitness landscape as might be expected during a chronic infection.
This mechanism emulates within-host dynamics that are not explicitly modelled, since each infected agent is associated with only a single viral genome at each simulation cycle. We discussed this distinction in Section \ref{sec:discussion}.

Each cycle, for every infected host agent carrying genome $s$ and acquiring $L$ independent mutations (governed by the mutation rate as described in Section \ref{sm:sec:mutation_substitution}), we simulate the within-host selective pressure through a three-step procedure: 
\begin{enumerate}
\item  Randomly generate $M - X + 1$ mutated genome sequences by independently selecting and changing $L$ positions in sequence $s$ for each, by substituting the corresponding nucleotide number by a uniformly sampled number between $0$ and $3$. These generated mutated genome sequences become candidates for the upcoming selection process.  
\item Compute the fitness of each of the $M - X + 1$ generated mutated genome sequences using the fitness weight table (see Section \ref{sm:sec:fitness_model}). Rank these generated mutated genome sequences by fitness in descending order.
\item Select the genome sequence with the highest rank (fittest) among the generated mutated genome sequences. Replace $s$ by this selected sequence.\\
\end{enumerate}

At the start of their infection, both typical infected hosts and chronically infected hosts are processed using the described selective pressure procedure with the same selective strength $X$ where $X \lessapprox M$ (representing low selective pressure). However, beginning from 60 days post-infection, we differentiate chronically infected hosts by assigning them a higher selective strength (i.e., $X \ll M$). In comparison, by this point typical infected hosts have already recovered. This setup aligns with the observation that a 60-day post-infection interval is observed before the intra-host pressure begins to impact the SARS-CoV-2 transmission bottleneck significantly \cite{markov_evolution_2023, ghafari_prevalence_2024}. We note that chronically infected hosts have a lengthened recovery period between 60 and 370 days (see Section \ref{sm:sec:chronic_inf}), and the described selection process continues until the host recovers.

\section{Artificial agent-based population}
\label{sm:sec:pop}
Prior to the simulation, we stochastically generated three distinct artificial populations of anonymous agents (corresponding to layer (B) Demographic shown in Fig \ref{fig:4-layer-model}), used to examine evolutionary and epidemiological dynamics:
\begin{itemize}
    \item (large) approx. 25.4 million, comparable to the Australian population;
    \item (medium) approx. 8 million, comparable to the population of New South Wales (a relatively population-dense state in Australia); 
    \item (small) approx. 1.7 million, comparable to the population of South Australia (a relatively population-sparse state in Australia).
\end{itemize}

We selected these population sizes for two reasons: (i) to investigate the relationship between the population size and phylodynamics, and (ii) to ensure computational feasibility. These considerations are based on the observation that the SARS-CoV-2 variants of concern may have emerged in countries of different population sizes (e.g., South Africa, with a population of approximately 59 million; UK, with a population of around 56 million; and Botswana, with a population over 2.4 million). In addition, our preliminary test simulations revealed that the computational cost increases significantly with the population size (see Section \ref{sm:sec:comp}). In this study, we focus on revealing the representative evolutionary characteristics of SARS-CoV-2, while maintaining computational efficiency by using the three populations listed above. However, given sufficient computational resources, much larger populations (e.g., over one billion population representing countries such as India) may also be generated 
in order to quantify the impact of population size on phylodynamic and disease dynamics.

The constructed artificial population captures essential demographic characteristics and commuting patterns represented in the latest Australian census and other datasets, as well as agent-to-agent interactions in various social mixing contexts, thus representing the population heterogeneity. Specifically, we generated the artificial population using demographic and travel data sourced from the Australian Bureau of Statistics (ABS) 2021 Census \cite{ABS_census}, international air traffic reports (detailing incoming passenger flows at Australian airports) \cite{BITRE_airport_data}, and educational registration records (including data on schools and students) \cite{ACARA_Data}. Each agent in the artificial population was assigned multiple demographic attributes, such as age, gender, and residency location, alongside social mixing contexts across various settings: residential (e.g., household, household cluster, neighbourhood and statistical area which maps to a local government area), educational (e.g., classroom/school for agents aged 18 years or younger), and workplace (for agents aged over 18 years). Fig~\ref{fig:overview} provides a visual representation of the considered social mixing contexts. For each agent, the residential contexts are determined based on residential demographics, while the workplace and educational contexts are assigned based on commuting patterns (commute to work and class/school, depending on the agent's age group).

The population generation used by \phasetrace can also cover a specific geographical area. In other words, in addition to an artificial population matching the demographics of the entire Australian population, it is possible to generate smaller populations corresponding to individual states and territories, while maintaining their demographic attributes and travel patterns. The population generation algorithm resolves several known discrepancies (introduced by privacy-protection algorithms employed by government agencies), and maintains integrity across different public datasets \cite{fair_creating_2019, BMC}. A detailed description of the population generation methodology and the population data structure can be found in the Supplementary Materials of a prior study~\cite{BMC} and the user guide of our open-source software \cite{amtract_user_guide_v90}.

\section{Multi-strain transmission and control}
\label{sm:sec:trac}

\subsection{Multi-strain transmission model}
\label{sm:sec:multi}
The multi-strain transmission model of \phasetrace stochastically simulates pathogen transmission between hosts (Epidemic Layer shown in Fig~\ref{fig:4-layer-model}). This model introduced several new features to our previous ABM, which was implemented in AMTraC-19~\cite{chang_modelling_2020, chang_simulating_2022, chang_persistence_2023, BMC}. In this section, we describe how transmission is affected by the phylogenetic model described above.

At the start of the simulation, all agents in the population are susceptible. Initial infections carrying the generated ancestral genome are seeded in metropolitan statistical areas (around international airports) as imported cases. These seeding events occur periodically throughout the simulation. The viral genome of the imported infections is taken to be the variant with the highest transmissibility from the preceding month of simulation, with additional mutations randomly added (using a weak selective pressure, see Table \ref{sm:tab:phylo_parameters}) to represent the evolution of the virus outside of the simulated population.

From the initial infections, the pathogen propagates through the population as susceptible and infected agents interact. These interactions occur in various social contexts in discrete half-day time cycles. During ``daytime'' cycles, agents interact in workplaces or educational settings (e.g., class, grade, school). ``Nighttime'' cycles instead involve interactions in residential settings (e.g., household, household cluster, neighbourhood, and community). Weekdays consist of a daytime cycle and a nighttime cycle, whereas weekend days consist of two nighttime cycles (i.e., no interactions in workplace/educational contexts).

A transmission event changes the health state of the newly-infected agent from Susceptible to Infectious (asymptomatic or symptomatic). Once the infection is cleared, the agent transitions to the Recovered state. Recovered agents are again susceptible to re-infection, but any subsequent transition back to Infectious is moderated by the immunity level (see following sections for more detail).

\subsubsection{Susceptible-infectious transition}
\label{sm:sec:S_I_trans}
Transmission from agent $j$ to agent $i$ depends on the probability of their interaction, their ages, and the context $g$ in which the agents interact. These daily context- and age-dependent interaction probabilities $q_{j \rightarrow i}(g)$ have been defined and calibrated in previous studies \cite{chang_modelling_2020,chang_persistence_2023}.
At cycle $n$, the transmission probability, $p_{j \rightarrow i}(n,g)$, is determined as follows:
\begin{equation}
p_{j \rightarrow i}(n,g) = K(s_j) \ f_j(n-n_j) \ q_{j \rightarrow i}(g)
\label{sm:eq:inf_j_i}
\end{equation}
where $K(s_j)$ represents the transmissibility of pathogen variant $s_j$, proportional to the corresponding \textcolor{black}{basic of effective reproductive number}, i.e., $K(s_j)$ is the fitness of genome $s_j$ carried by agent $j$, as defined by Eq \ref{sup_eq:linear_combo_for_fitness}; $n_j$ is the time cycle when agent $j$ started experiencing the onset of infection; and the function $f_j(n-n_j)$ determines agent $j$'s infectivity over time according to the natural history model. At the time cycle $n$, agent $j$'s infectivity is $0 < f_j(n-n_j) \leq 1$ with the infectivity peak equal to 1. For uninfected agents, $f(\cdot)=0$. For simplicity, we assume that all variants share the same progression of disease, but differ in their transmissibility (determined by fitness $K$), and that only one variant can be transmitted during the infection process. We note that Eq \ref{sm:eq:inf_j_i} and Eq \ref{eq:individual_transmission_prob} in the main manuscript are equivalent.

We then extend the scenario to consider the probability of agent $i$ getting infected by interacting with all other agents sharing the same social mixing contexts across all levels at time cycle $n$, defined as follows: 
\begin{equation}
p_i(n) = 1 - \prod_{g \in G_i(n)} \; \prod_{j \in A_g\backslash\{i\}} \left( 1 - p_{j \rightarrow i}(n,g) \right)\\
\label{sm:eq:simple_case_infection_prob}
\end{equation}
where $G_i(n)$ is the set of social mixing contexts where interactions between agent $i$ and other agents may occur (depending on factors such as weekday or weekend, daytime or nighttime), and $A_g\backslash\{i\}$ is the list of all agents in the context $g \in G_i(n)$ except agent $i$. We shall expand Eq \ref{sm:eq:simple_case_infection_prob} in later sections to incorporate the effects of non-pharmaceutical interventions (Eq \ref{sm:eq:infection_prob_compliant_agent}) and effects of immunity (Eq \ref{sm:eq:p_I_NPI_Immunity}). We note that Eq \ref{sm:eq:simple_case_infection_prob} and Eq \ref{eq:general_infection_prob} in the main manuscript are equivalent.

Eq \ref{sm:eq:simple_case_infection_prob} defines the infection probability for agent $i$ interacting with other infected agents across all of its social contexts. However, this probability is an aggregate of the interactions and does not identify the specific source of infection, which must be known in order to copy its viral sequence to the newly infected host. We assign the source of infection by constructing a discrete distribution to randomly sample one agent from all potential infected agents $j$ across all social contexts $g \in G_i(n)$ to which agent $i$ belongs. The probability of selecting agent $j$ is given by $$\frac{p_{j \rightarrow i}(.)}{\sum_j{p_{j \rightarrow i}(.)}}.$$

An infectious agent is either symptomatic or asymptomatic. The probability of being symptomatic, $z_i(n)$, depends on the infection probability $p_i(n)$ and an age-dependent scaling factor $\sigma_i$ characterising the proportion of symptomatic cases among all infections in the corresponding age group:
\begin{equation}
    z_i(n) = \sigma_i \ p_i(n)
    \label{sm:eq:infection_prob_symptomatic}
\end{equation}
where $\sigma_i = \sigma^a$ for adults ($\text{age} > 18$)  and  $\sigma_i = \sigma^c$ for children ($\text{age} \leq 18$). The asymptomatic infectious agents  have lower infectivity (i.e., lower $f_i(n-n_i)$, compared to the symptomatic agents, which also affects the strength of transmission to other susceptible agents.

Following our prior studies \cite{chang_modelling_2020, chang_persistence_2023, BMC}, we assumed that only a fraction of total infections, especially asymptomatic infections, are detected daily. In our model, we set a lower detection probability for asymptomatic cases ($\pi'$) than the symptomatic cases ($\pi$), such that $\pi \gg \pi'$. In addition, we assumed that chronically infected cases have the same detection probability as asymptomatic cases. In the COVID-19 case study, we used detection rates of $\pi = 0.13$ and $\pi' = 0.01$, calibrated to the Omicron variant of SARS-CoV-2~\cite{chang2024impactopiniondynamicsrecurrent}.

\subsubsection{Infectious-recovered and recovered-susceptible transitions}
\label{sm:sec:natural-hist}

\begin{figure}[t]
    \centering
    \includegraphics[width=0.8\textwidth]{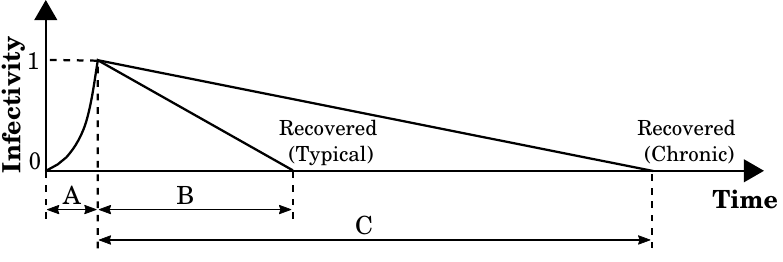}
    \caption{A schematic representation of the natural history describing the SARS-CoV-2 infection progression in an agent. The infectivity initially increases exponentially from the onset of infection, reaching the peak, and subsequently declining linearly to zero until recovery. For each infected host, the duration from the infection onset to the infectivity peak is sampled from a lognormal distribution (A) with parameters $\mu=1.013$ and $\sigma=0.413$. The recovery period is sampled from a uniform distribution, ranging from 7 to 11 days (B) for typical infections, or from 60 to 370 days (C) for chronic infections.}
    \label{sm:fig:natural_history_of_disease}
\end{figure}

Following the natural history model, an infectious agent recovers after a certain recovery period dependent on the host category (see Section \ref{sm:sec:sensitivity}). We assume that both typical and chronically infected hosts develop infectivity fairly quickly from the onset of infection (following a lognormal distribution). However, chronically infected hosts experience a significantly longer recovery than the typical infected hosts. The natural history model describing the disease progression for both host categories is illustrated in Fig~ \ref{sm:fig:natural_history_of_disease}.

For a period of 60 days following recovery, agents are immune to re-infection, consistent with empirical definitions that attempted to distinguish between reinfection and serial testing of an initial infection for SARS-CoV-2~\cite{CDC_inref, NSW_reinf}. After 60 days, the immunity induced by infection starts to wane, and the recovered agent becomes susceptible again. The re-infection probability for a recovered agent on subsequent exposure depends on their infection and vaccination history both in terms of the genomic similarity of viral strains and the time elapsed (see Section \ref{sm:sec:vac} for more details).

\subsection{Non-Pharmaceutical Interventions}
\label{sm:sec:NPI}
The model includes several non-pharmaceutical interventions (NPIs), such as case isolation (CI), home quarantine (HQ), school closures (SC), and social distancing (SD). The implementation of each NPI is governed by: (i) a macro-distancing parameter determining the population fraction adhering to the intervention (Table \ref{sup_mat:tab:macro-npi}), and (ii) a set of micro-distancing parameters that quantify the altered (typically decreased) interaction strengths between NPI-compliant individuals in a given social context~(Table \ref{sup_mat:tab:micro-npi}).

Following Eq. \ref{sm:eq:simple_case_infection_prob}, the NPI-affected infection probability for agent $i$ without prior infections or vaccinations is given by
\begin{eqnarray}
    p_i(n) =  1 - \prod_{g \in G_i(n)} \left [1 - F_g(i) \left ( 1 - \prod_{j \in A_g\backslash\{i\}} (1 - F_g(j) \ p_{j \rightarrow i}(n,g)) \right ) \right ]
    \label{sm:eq:infection_prob_compliant_agent}
\end{eqnarray}
where $F_g(j)$ denotes the NPI-affected strength of interaction between agent $j$ and other agents in mixing context $g$. For an agent $j$ adopting NPIs, $F_g(j) \neq 1$, denoting a modified infection probability from agent $j$. For an agent $j$ not adopting any NPIs, $F_g(j) = 1$.

The assignment of NPI-compliant agents is determined based on a Bernoulli process with the probability specified by the macro-distancing parameters. While an agent may comply with multiple NPIs, their interaction strengths, $F_g(j)$, can only be adjusted to one NPI. We therefore use the parameters for only the first NPI to which an agent complies in the following ordered list: CI, HQ, SD, and SC. The micro and macro-distancing parameters of CI vary depending on the host type, where chronically infected hosts are fully compliant (i.e., macro-distancing level is set to $1.0$), with significantly reduced interaction strengths across all social contexts (Tables~\ref{sup_mat:tab:micro-npi} and \ref{sup_mat:tab:macro-npi}). This parametrisation agrees with the public health recommendations for chronically infected hosts to take extra precautions to prevent severe illness~\cite{CDC_chronic}.  

\begin{table}[t]
    \centering
    \begin{tabular*}{.8\linewidth}{@{\extracolsep{\fill}} lS[table-format=1.2]S[table-format=1.2]S[table-format=1.2]l}
        \toprule
        & \multicolumn{4}{l}{Micro-distancing (interaction strengths)}\\
        {Intervention} &    {Household}    & {Community} & {Workplace/School} & {Duration $t$}   \\
        \midrule
        CI (typical)      & 1.0 & 0.25 & 0.25 & $D(i)$ \\
        CI (chronic)      & 0.01& 0.0  & 0.0  & $D(i)$ \\
        HQ                & 2.0 & 0.25 & 0.25 & 7 -- 30   \\
        SC                & 1.0 & 0.5  & 0    & dynamic\\
        SD                & 1.0 & 0.25 & 0.1  & dynamic\\
        \bottomrule
    \end{tabular*}
    \caption{The micro-distancing parameters (interaction strengths) for the considered NPIs. The micro-duration of CI is limited by the disease progression in the affected agent $i$,  $D(i)$. Interaction strengths for CI are set to be significantly lower for chronically infected hosts. CI: Case Isolation; HQ: Home Quarantine; SC: School Closure; and SD: Social Distancing.}
    \label{sup_mat:tab:micro-npi}
\end{table}

We define SD as a broad behavioural-driven NPI that reduces interaction strengths among individuals, due to stay-at-home orders and other measures (including physical distancing, mask wearing, etc.) adopted during (partial) lockdowns. In other words, the SD compliance level can be interpreted as the fraction of the population that follow the restrictions imposed during (partial) lockdowns. We designed an intervention scenario with a dynamically adjustable SD compliance profile representative of NPIs implemented in many countries during the COVID-19 pandemic~\cite{chang_persistence_2023}. Specifically, we set the SD compliance level to gradually decline annually over the 6-year simulation period, from 50\% at the start of the pandemic in 2020 to 20\% in the endemic stage from 2023 onwards. In addition, the activation of SD was set to be triggered by a sufficiently high prevalence (e.g., exceeding 200 cases), with SD deactivated once the disease prevalence falls below a certain threshold (e.g., 100 cases).  In our simulations, the deactivation of SD has not been triggered. Fig~\ref{fig:overview} describes a detailed setup and parameterisation.

\begin{table}[t]
    \centering
    \begin{tabular*}{.8\linewidth}{@{\extracolsep{\fill}} lS[table-format=1.2]S[table-format=1.1]S[table-format=1.1]S[table-format=1.1]}
        \toprule
         &\multicolumn{4}{l}{Simulation year}\\
        Intervention      & {2020} & {2021} & {2022} & {2023-onwards} \\
        \midrule
        CI (typical)           & 0.7  & 0.7  & 0.7  & 0.7\\
        CI (chronic)           & 1.0  & 1.0  & 1.0  & 1.0\\
        HQ                     & 0.5  & 0.5  & 0.5  & 0.5\\
        SC (students/teachers) & 1.0  & 0.0  & 0.0  & 0.0\\
        SC (parents)           & 0.25 & 0.0  & 0.0  & 0.0\\
        SD                     & 0.5  & 0.4  & 0.3  & 0.2\\
        \bottomrule
    \end{tabular*}

    \caption{The macro-distancing parametrisation (population fractions) for the considered NPIs over the 6-year simulation period for SARS-CoV-2. The CI-compliant population fraction is lower for typically infected hosts than for chronically infected hosts. Students/teachers are assumed to fully comply with SC, while parents of school-aged children have a reduced compliance level. CI: Case Isolation; HQ: Home Quarantine; SC: School Closure; and SD: Social Distancing}
    \label{sup_mat:tab:macro-npi}
\end{table}

\section{Immuno-epidemiological model}
\label{sm:sec:imm-epi}

\subsection{Vaccination}
\label{sm:sec:vac}
Following our prior work \cite{zachreson_how_2021, chang_persistence_2023, BMC}, the model includes a vaccination scheme under which agents are immunised according to an average daily vaccination rate. The vaccination rate is demographically stratified by age, with agents aged 18 to 35 years old vaccinated at a 10-fold higher rate than young or elderly agents, and by immunisation history, with 60\% of vaccines going to agents with some history of vaccination~\cite{lin_vaccinated_2023}. Agents are eligible for vaccination only if they are free from infection for at least three months. We did not model vaccination schedules consisting of multiple doses within a season, instead assuming that a single vaccination event confers the cumulative immunity of the entire recommended dosing regimen in a year.

Vaccinated agents have diminishing levels of protection due to two factors: (i) vaccine escape as circulating variants diverge genetically from the vaccine strain at specific epitopes, and (ii) immunity from vaccination wanes over time. The vaccine used each year in the simulation is chosen to target the dominant variant circulating in the preceding simulated year (starting from 2021). We model only one vaccine type administered at any one time. We assumed that peak immunity occurs immediately after vaccination~\cite{cao_rapid_2022}.

\begin{figure}[t]
    \centering
    \includegraphics[width=0.7\linewidth]{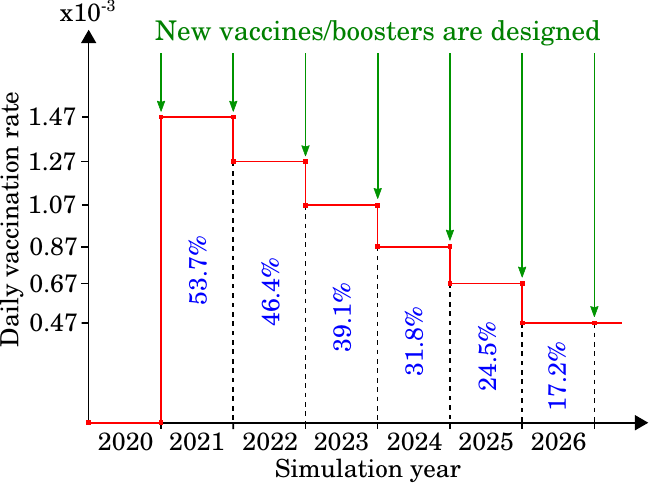}
    \caption{A simulated mass vaccination roll-out with varying daily vaccination rates (y-axis) and multiple boosting events targeted for COVID-19 (green arrows). The percentage (in blue) within each bar represents the approximate percentage of the vaccinated population at the end of the year.}
    \label{sm:fig:vaccine_schedule}
\end{figure}

We simulated a mass vaccination roll-out with a variable daily vaccination rate updated annually, starting from zero (i.e., no available vaccine) in 2020, rising to 0.147\% of the entire population per day in 2021 (when vaccines first became available), and then linearly declining to 0.047\% of the population per day in 2026 (Fig~\ref{sm:fig:vaccine_schedule}).

\subsection{Compound immunity and waning effects}
\label{sm:sec:hybrid_and_waning}
Individuals who have been vaccinated and/or experienced prior infections develop immunity against the disease, subsequently reducing their immediate susceptibility to infection \cite{szanyi2022log}. Immunity resulting from both vaccination and natural infection in the same individual, known as hybrid immunity, offers more robust protection compared to immunity derived solely from vaccination or natural infection. Hybrid immunity against SARS-CoV-2, for example, is found to have stronger and longer-lasting immune responses than the immunity granted from vaccination or infection alone~\cite{larkin_hybrid_2023, WHO_hybrid_immunity}. As described in Methods (subsection~\ref{methods:immuno}), we use the term ``compound immunity'' to encompass immunity due to previous exposures, vaccination, and hybrid immunity~\cite{chang2024impactopiniondynamicsrecurrent}.

For each agent $i$ with a history $H_i$ containing all vaccination and past infection records $r \in H_i$, 
we characterise their immunity by two distinct components: (i) compound immunity against symptomatic infection ($M^\textrm{c}_i$, described in section \ref{sm:sec:immunity_symptomatic}), and (ii) compound immunity against forward transmission ($M^\iota_i$, described in section \ref{sm:sec:immunity_forward}). We then integrate the compound immunity into the infection probability in Section \ref{sm:sec:immunity_infection}. These immunity components range between 0 and 1 with zero indicating no immunity and one indicating full immunity.

\subsubsection{Compound immunity against symptomatic infection}
\label{sm:sec:immunity_symptomatic}

To determine the immunity against symptomatic infection of a susceptible agent $i$, exposed to a potential infection from an infectious agent $x$, we need to consider the genome sequence of the variant $s_x$ carried by $x$. 
At time cycle $n$, given a single vaccination or infection record $r$ for agent $i$, we determine their immunity against symptomatic infection as follows:
\begin{equation}
    m_i^\textrm{c}(n,r,s_x) = m^\textrm{c}(r) \Bigl[1 - \text{min} \Bigl(1,  \epsilon^c (n-n_r)  \Bigr) \Bigr] \Bigl[1 - \text{min} \Bigl(1,\tau \Delta_r(s_x,s_T)  \Bigr) \Bigr]
    \label{sm:eq:waning_immunity_symptomatic}
\end{equation} 
\noindent where 
\begin{itemize}
    \item $m^\textrm{c}(r)$ denotes the peak immunity against symptomatic infection developed from either vaccination or past infection recorded in $r$ against a target variant with genome $s_T$ (see Table \ref{sm:tab:immunity_parameters}),
    \item $\epsilon^c$ is the waning rate of immunity (proportion of immunity lost) against symptomatic infection per simulation cycle,
    \item $(n-n_r)$ is the number of cycles between the current cycle $n$ and the record cycle $n_r$, 
    \item $\Delta_r(s_x,s_T)$ is the genetic distance measured by number of amino acids that differ between genomes $s_T$ and $s_x$ in the epitope regions, either in spike regions only~\cite{du_spike_2009, cankat_search_2024} (if $r$ is a vaccination event), or in both spike and non-spike regions (if $r$ is a previous infection)~\cite{kojima_protective_2022},
    \item factor $\tau$ is the immunity reduction per amino acid difference between these two genomes. 
\end{itemize}

\begin{figure}[t]
    \centering
    \includegraphics[width=0.85\linewidth]{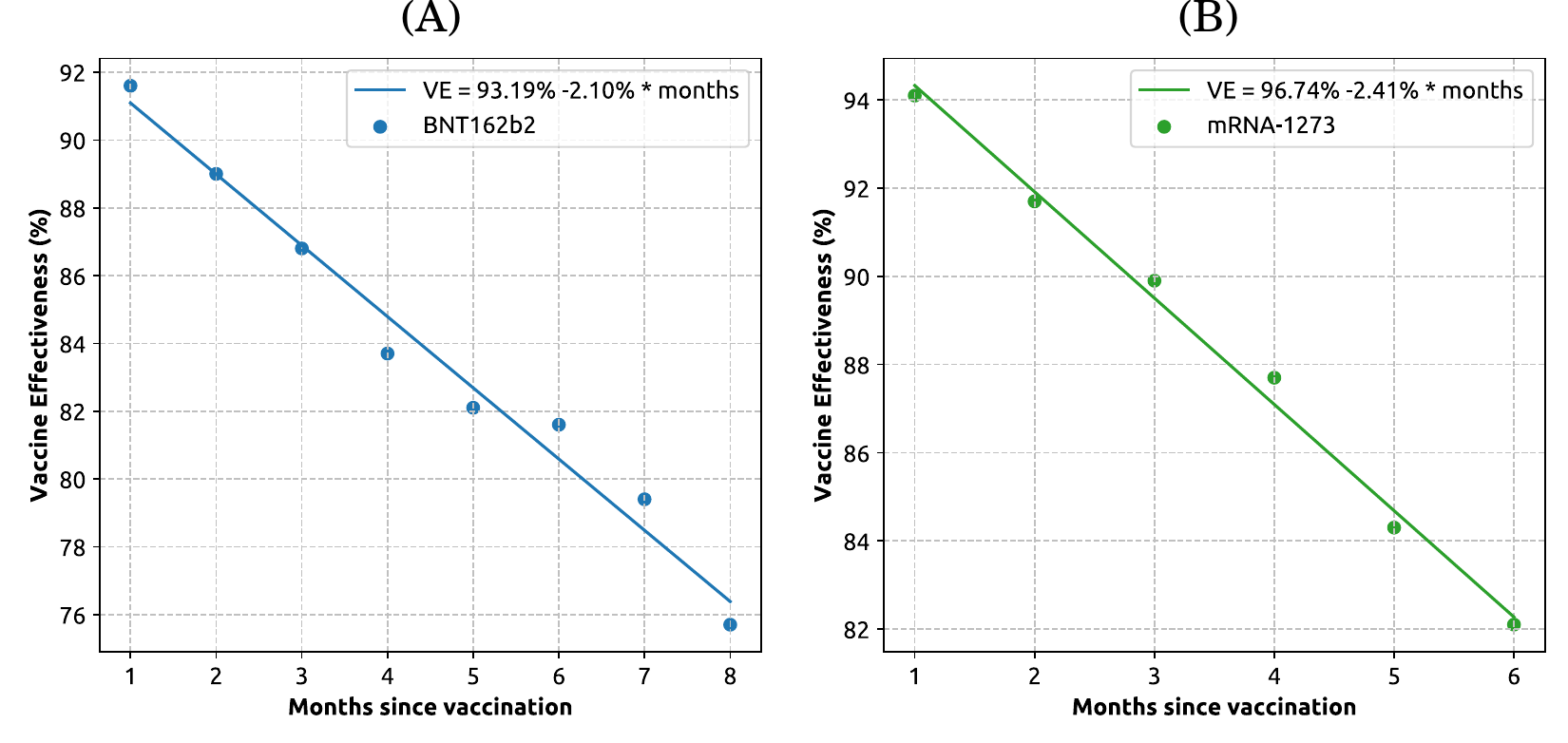}
    \caption{Vaccine effectiveness against infection reduction over time. Two widely used mRNA vaccines are studied: (A) Pfizer (BNT162b2), and (B) Moderna (mRNA-1273), plotted and fitted using data from \cite{menni_covid-19_2022}. The equation shown in each plot represents the fitted linear regression for each vaccine, estimating a linear reduction in effectiveness from 2.1\%  to 2.4\% per month.}
    \label{sm:fig:waning_effect}
\end{figure}

Eq \ref{sm:eq:waning_immunity_symptomatic} shows that $m_i^\textrm{c}$ reduces linearly over time, governed by two constant rates: immunity waning rate $\epsilon^c$ over time, and immunity reduction factor $\tau$ over genetic distance. This assumption matches observations reported in various studies suggesting a linear reduction between (i) COVID-19 vaccine effectiveness and time~\cite{kodera_estimation_2022,cromer_predicting_2023, menni_covid-19_2022}, and (ii) genetic distance and immune evasion~\cite{cao_rapid_2022}.  Fig~\ref{sm:fig:waning_effect} shows a linear regression fitting of vaccine effectiveness over time for different types of vaccines, constructed using supplementary data from \cite{menni_covid-19_2022}. We note that the waning immunity term can be easily adjusted to follow a non-linear reduction, such as exponential decay \cite{khoury_neutralizing_2021, menegale_evaluation_2023} or based on a gamma distribution \cite{feng_modelling_2022}.

We then compute the compound immunity for agent $i$ induced by multiple vaccination and/or infection events. The compound immunity against symptomatic infection, $M_{i}^\textrm{c}(n,H_i,s_x)$, accumulates non-linearly with an upper bound of $1$ (i.e., perfect immunity):
\begin{equation}
M_{i}^\textrm{c}(n,H_i, s_x) = \text{min}\left(\sqrt{\sum_{r \in H_i} \left [m_{i}^\textrm{c}(n,r,s_x) \right ]^2}, 1\right)
\label{sm:eq:immunity_symptomatic_accumulation}
\end{equation}
We note that this equation is equivalent to Eq \ref{eq:immunity_symptomatic_accumulation} in the main manuscript.

In our model, the level of protection provided by $M^c_i$ is further decomposed into two components: the immunity against susceptibility ($M^\theta_i$), and the immunity against symptomatic infection given the infection ($M^\zeta_i$). These two components may also be referred to as the susceptibility-reducing immunity and disease-preventing immunity, respectively. Following prior studies~\cite{zachreson_how_2021, chang_simulating_2022, chang_persistence_2023}, we calculate $M^\theta_i$ and $M^\zeta_i$ as:
\begin{equation}
    M^\zeta_i(n,H_i,s_x) = M^\theta_i(n,H_i,s_x) = 1 - \sqrt{1-M^c_i(n,H_i,s_x)}
    \label{sm:eq:waning_immunity_symptomatic_s_d}
\end{equation}
which satisfies a general relationship:
\begin{equation}
    M^c_i(n,H_i,s_x) = M^\zeta_i(n,H_i,s_x) + M^\theta_i(n,H_i,s_x) - M^\zeta_i(n,H_i,s_x) M^\theta_i(n,H_i,s_x)
    \label{sm:eq:waning_immunity_general}
\end{equation}
We note that Eq \ref{sm:eq:waning_immunity_general} reduces to Eq \ref{sm:eq:waning_immunity_symptomatic_s_d} when  $M^\zeta_i(n,H_i,s_x) = M^\theta_i(n,H_i,s_x)$. 

\subsubsection{Compound immunity against forward transmission}
\label{sm:sec:immunity_forward}
Following \cite{harris_effect_2021, zachreson_how_2021, nguyen_general_2022}, we also consider the immunity component against forward transmission. At time cycle $n$, for an infected agent $j$, the immunity against forward infection from one vaccination/infection event, represented by record $r$, is calculated as follows:
\begin{equation}
    m_j^\iota(n,r)= m^\iota(r) \Bigl[1 - \text{min} \Bigl(1, \epsilon^\iota (n-n_r) \Bigr)  \Bigr]
    \label{eq:sm:eq:waning_immunity_forward}
\end{equation}
where $m^\iota(r)$ denotes the peak immunity against forward transmission developed from either vaccination or past infection recorded in $r$; the rate $\epsilon^\iota$ is the forward-transmission immunity waning rate per simulation cycle; and $n_r$ is the record cycle of $r$. We assume that $m_j^\iota(n,r)$ is dependent only on the immunological history of the infection source agent $j$, and is independent of the immunological history of the potentially infected agent.

We then consider the compound immunity against forward transmission, accounting for all records in $H_i$, which follows a non-linear combination of all individual components $m_j^\iota(n,r)$. Similar to the calculation of the compound immunity against symptomatic infection, we calculate the compound immunity against forward transmission where the upper bound equals $1$ (i.e., perfect immunity):
\begin{equation}
M_{j}^\iota(n,H_j) = \text{min}\left(\sqrt{\sum_{r \in H_j} \left [m_{j}^\iota(n,r) \right ]^2}, 1\right)
\label{sm:eq:immunity_forward_accumulation}
\end{equation}

\subsubsection{Incorporating compound immunity in infection probability}
\label{sm:sec:immunity_infection}
Here, we incorporate the compound immunity, against infection ($M^\theta$) and forward transmission ($M^\iota$), within the infection probability, extending the NPI-affected probability of infection (Eq \ref{sm:eq:infection_prob_compliant_agent} in Section \ref{sm:sec:NPI}).

The impact of compound immunity on infection probability is modelled by distinguishing two factors that affect the infection of a susceptible agent $i$: 

\begin{description}
    \item[$(\cal{P}^{I})$] the impact of forward infection from all infected agents $j$ with whom this agent has contact, affected by $M^\iota_j$, and
    \item [$(\cal{P}^{II})$] the agent $i$'s immunity against infection from these sources, affected by $M^\theta_i$.
\end{description}

We modelled the impact of factors (${\cal{P}}^{I}$) and (${\cal{P}}^{II}$) by decomposing the transition from Susceptible state to Infectious state into two consecutive steps: 
\begin{enumerate}[label=(\arabic*)]
\item considering factor (${\cal{P}}^{I}$) alone, the model identifies whether agent $i$ is \textit{potentially} infected, and if so, it determines the most likely source of this potential infection; and 
\item considering (${\cal{P}}^{II}$), the model ascertains whether infection is \textit{actually} transmitted from this source to agent $i$.
\end{enumerate} 

\textbf{Step (I).} We calculate the probability of agent $i$ becoming \textit{potentially} infected at time cycle $n$, $p^E_i(n)$, while accounting for NPI-compliance and the immunity against forward infection of other agents, i.e., quantifying factor (${\cal{P}}^{I}$), as:
\begin{equation}
    p^\textit{I}_i(n) = 1 - \prod_{g \in G_i(n)} \left[
        1 - F_g(i) \left(
            1 - \prod_{j \in A_g\backslash\{i\}} \bigg(
                1 - \big( 1 - M_j^\iota(n, H_j) \big) \ F_g(j) \ p_{j \rightarrow i}(n,g)
            \bigg)
        \right)
    \right]
    \label{sm:eq:p_E}
\end{equation}
where $F_g(j)$ denotes the strength of interaction between agent $j$ and other agents in mixing context $g$, and $M_j^\iota(n, H_j)$ represents the immunity against forward transmission based on the history $H_j$ of past infections and vaccinations of agent $j$. We note that $p^I_i(n)$ accounts for the immunity against forward transmission developed by any agent $j$ sharing a social context with agent $i$, thus accounting for factor (${\cal{P}}^I$).

If agent $i$ is determined to be \textit{potentially} infected according to the Bernoulli trial with $p^I_i(n)$, the source potentially infecting agent $i$ is identified by sampling from a discrete distribution that includes all infectious agents $j$ sharing a social context $g$ with agent $i$. The probability of selecting an agent $j$ as a potential source of infection from this distribution is given by 
\[\frac{ \left(1 - M_j^t(n, H_j) \right) \ F_g(j) \ p_{j \rightarrow i}(n,g)}{\sum_j{ \left(1 - M_j^\iota(n, H_j) \right) \ F_g(j) \ p_{j \rightarrow i}(n,g)}} \ . \]

\textbf{Step (II).} Given the identified source agent $e$  potentially transmitting infection to agent $i$, the model identifies the corresponding genome $s_e$, yielding the probability of infection transmission from source $e$ to agent $i$:
\begin{equation}
    p^{\textit{II}}_i(n, H_i, s_e) = 1 - M^\theta_i(n, H_i, s_e)
    \label{sm:eq:p_I}
\end{equation}
where $M^\theta_i(n, H_i, s_e)$ is the susceptibility-reducing immunity of agent $i$ with immunological history $H_i$, determined with respect to genome $s_e$ specifically.

Combining two steps (I) and (II)
determines the infection probability for susceptible agent $i$ (i.e., the probability of its transition from Susceptible to Infectious state), in the context of relevant NPI compliance and  immunological histories within the population:
\begin{equation}
    p_i(n, H_i, s_e) = p^{\textit{II}}_i(n, H_i, s_e)\, p^{\textit{I}}_i(n)
    \label{sm:eq:p_I_NPI_Immunity}
\end{equation}

The probability that susceptible agent $i$ becomes ill (i.e., symptomatic), originally defined in Eq \ref{sm:eq:infection_prob_symptomatic}, is then updated as:
\begin{equation}
    z_i(n, H_i, s_e) = \left(1 - M^\zeta_i(n, H_i, s_e) \right)  \sigma_i \ p_i(n, H_i, s_e)
    \label{sm:eq:infection_prob_symptomatic_w_immunity}
\end{equation}
where $M^\zeta_i$ is the disease-preventing immunity of agent $i$, developed as a result of past infections and vaccinations (represented in $H_i$), determined against the specific source of infection (i.e., the variant characterised by $s_e$).

Table \ref{sm:tab:immunity_parameters} lists key immunity-related parameters used in the  COVID-19 case study. Using a range of reports~\cite{shrestha_necessity_2022, shrestha_coronavirus_2022, franchi_natural_2023}, we assumed that the peak level of vaccine-induced immunity against SARS-CoV-2 infection is comparable to that of natural immunity. 

\begin{table}[t]
    \begin{tabular}{llp{12cm}}
       \toprule
        Parameter   & Value   & Reference/Note \\
        \midrule
        $m^c(r)$       & 0.7     & Peak immunity against symptomatic infection, given prior infection or vaccination record $r$\\
        $\epsilon^c$& 0.00067 & Immunity waning rate per day for $m^c$ \\
        $\tau$      & 0.052   & Immunity reduction per amino acid difference in epitope regions \\
        $m^\iota(r)$       & 0.4     & Peak immunity against infection transmission, given prior infection or vaccination record $r$\\
        $\epsilon^\iota$& 0.00067 & Immunity waning rate per day for $m^\iota$\\
        \bottomrule
    \end{tabular}
    \caption{Simulation parameters for compound immunity.}
    \label{sm:tab:immunity_parameters}
\end{table}

\section{Parametrisation and additional results}
\label{sm:sec:additional}
The phylogenetic parameters used in the SARS-CoV-2 case study are summarised in Supplementary Table \ref{sm:tab:phylo_parameters}. For an in-depth description of the dynamics simulated with 
these parameters, refer to Section \ref{sec:case_study_covid19}. Additionally, see Section \ref{sm:sec:phylogenetic} for parameter justification and calibration.

\begin{table}[t]
     \centering
    \begin{tabular}{ll}
       \toprule
        Parameter   & Value   \\
        \midrule
        Genome length, spike region  (nucleotide positions)      & 300     \\
        Genome length, non-spike region (nucleotide positions)  & 2790 \\
        Genome length, spike and epitope region (nucleotide positions)     & 135   \\
        Genome length, non-spike and epitope region (nucleotide positions)       & 90     \\
        Mutation rate, spike region(mutations per nucleotide per year) & 0.001 \\
        Mutation rate, non-spike region (mutations per nucleotide per year) & 0.001 \\
        Within-host selective pressure, typical infected hosts & $X=99, M=100$ \\
        Within-host selective pressure, chronically infected hosts & $X=30, M=100$ \\
        Within-host selective pressure, imported infections & $X=99, M=100$ \\
        Fraction of chronic infections in the entire agent population & 0.001 \\
        Fitness of ancestral strain, minimum & 2.65 \\
        Fitness of ancestral strain, maximum & 2.85 \\
        \bottomrule
    \end{tabular}
    \caption{Phylogenetic parameters used in the SARS-CoV-2 case study. }
    \label{sm:tab:phylo_parameters}
\end{table}

\begin{figure}[ht]
    \centering
    \includegraphics[width=\linewidth,trim={0 5cm 0 5cm},clip]{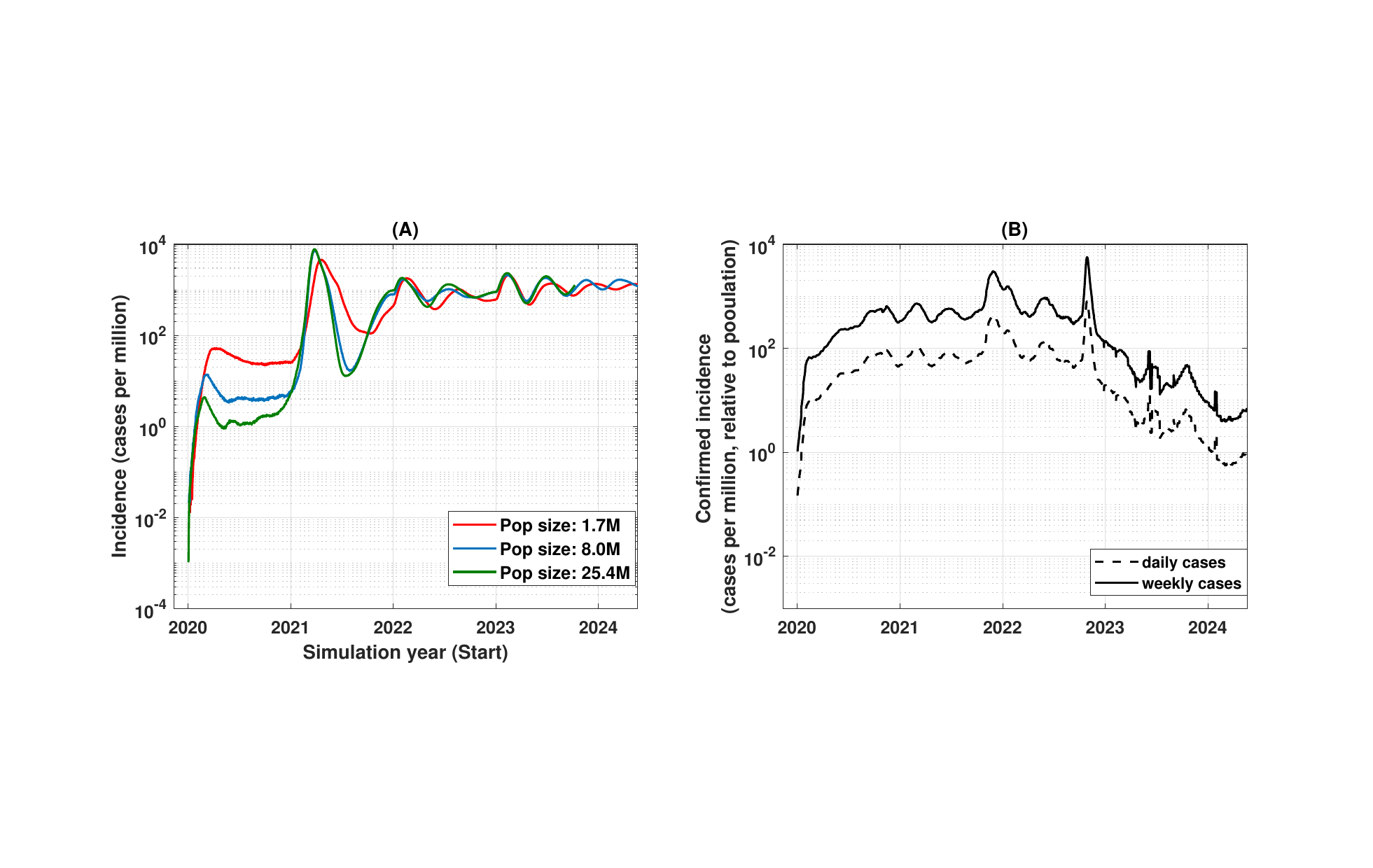}
    \caption{Detected incidence: log scale. (A) Simulated incidence on a log scale in population sets of 1.7 million (red), 8 million (blue), and 25.4 million (green). This figure corresponds to Fig~6A on a linear scale. (B) Worldwide detected incidence \cite{ourworldindata}, measured as new weekly cases per million (solid black line) and new daily cases per million (dashed black line).}
    \label{sm:fig:inc_log}
\end{figure}

\begin{figure}[ht]
    \centering
    \includegraphics[width=0.9\linewidth]{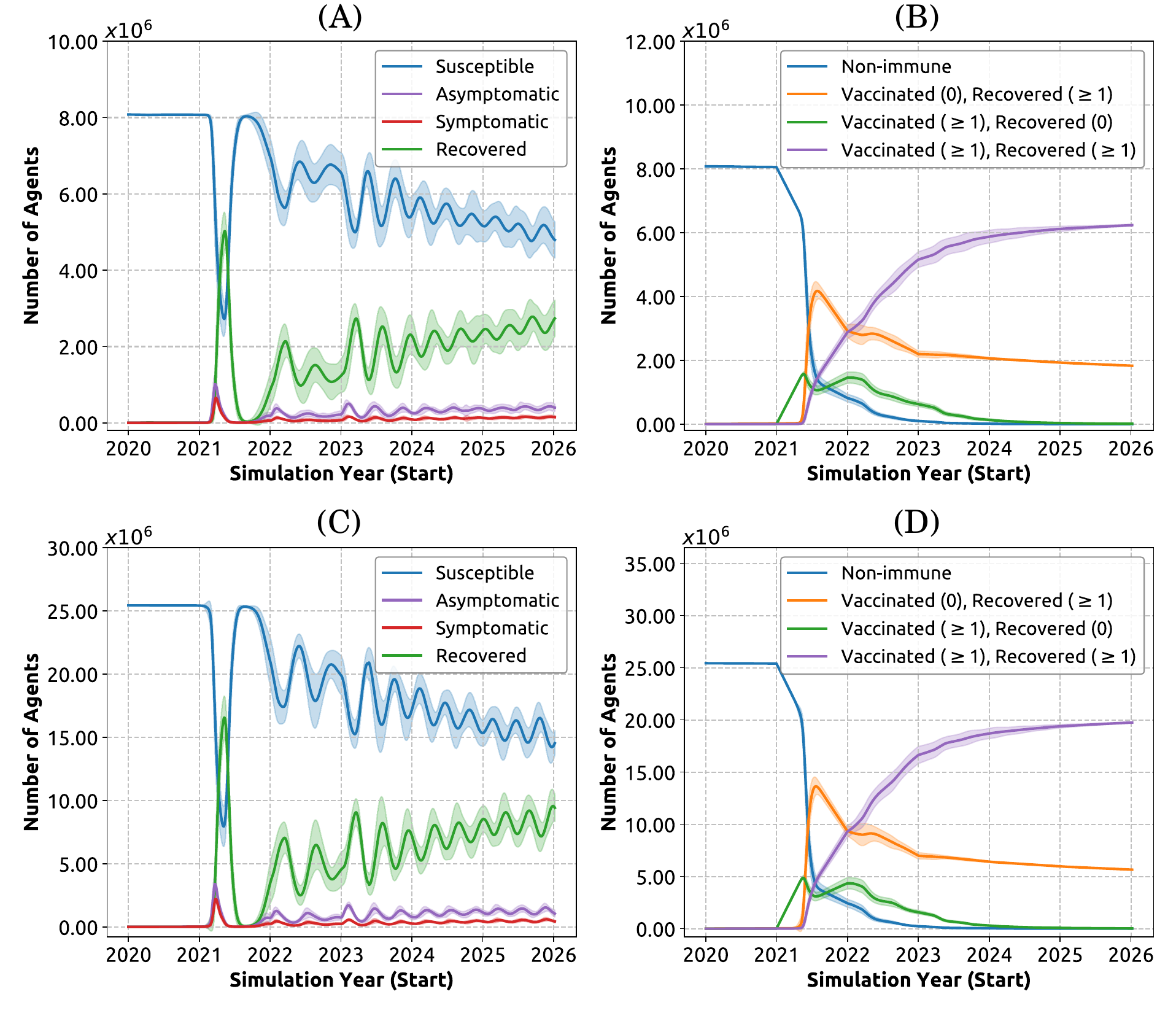}
    \caption{Simulated epidemic patterns (Capability 1) shown as mean (solid line) and standard deviation (shaded area). (A) and (C) Population across different health states, including susceptible (blue), asymptomatically infectious (purple), symptomatically infectious (red), and recovered (green) for the populations of 8 million and 25.4 million, respectively. (B) and (D) Population with different immunisation and infection history for the populations of 8 million and 25.4 million, respectively. Numbers in brackets denote the number of immunological (vaccination or infection) records. Individuals with multiple vaccinations or infections (more than 2) are grouped together for simplicity. The mean and average were obtained from approximately 30 realisations.}
    \label{sm:fig:compartments}
\end{figure}

\begin{figure}[ht]
    \centering
    \includegraphics[width=\columnwidth]{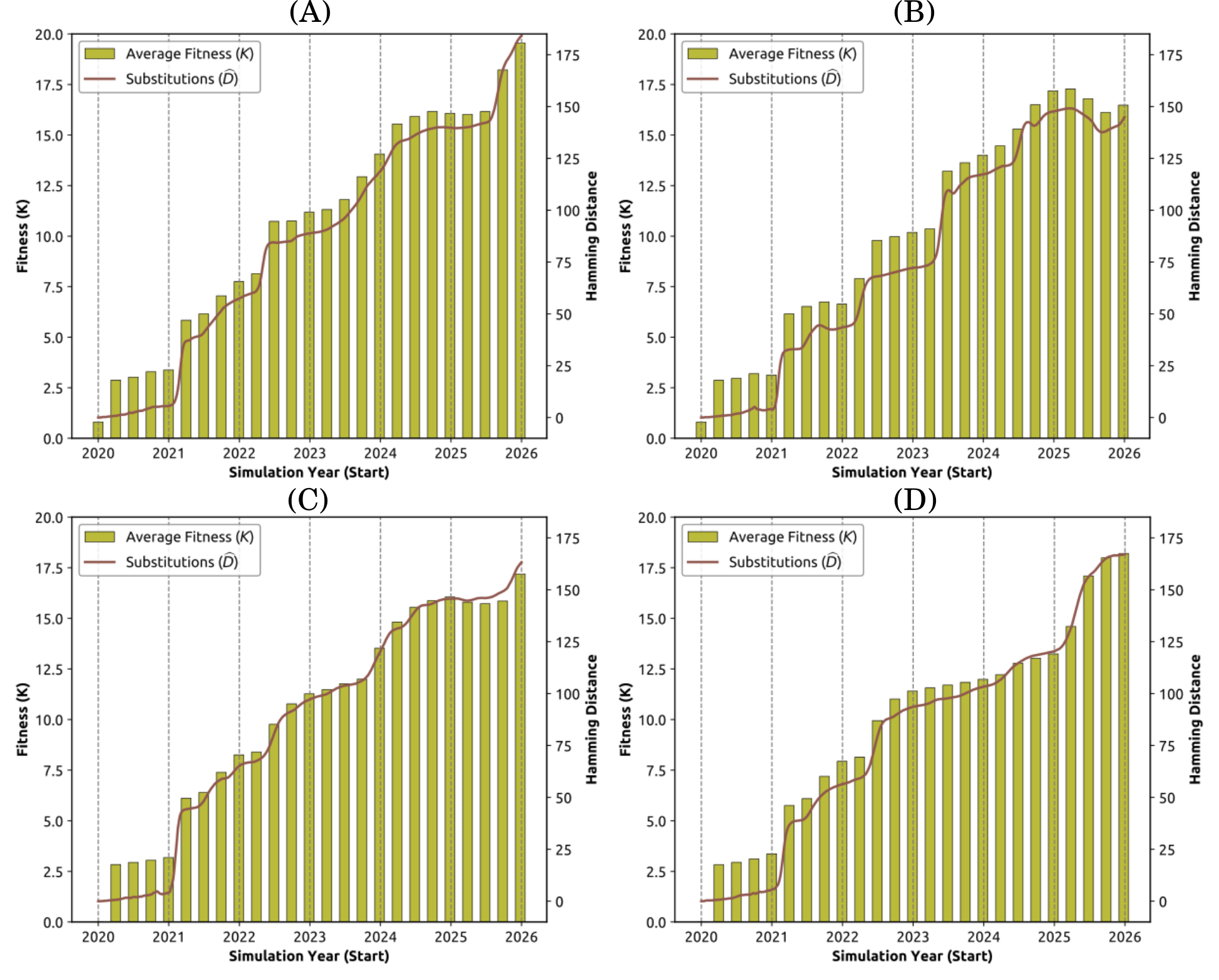}
    \caption{Simulated dynamics of the average transmissibility for a 1.7 million population. Fitness ($K$, olive bars, y-axis on the left), and the accumulated mutations ($\widehat{D}$, solid brown line, y-axis on the right) are plotted from 2020 to 2026. Panels (A)-(D) are profiles plotted using four different realisations.}
    \label{sm:fig:sim-obj-2i-individual_fitness_Ref_Haming-1M7}
\end{figure}

\begin{figure}[ht]
    \centering
    \includegraphics[width=\columnwidth]{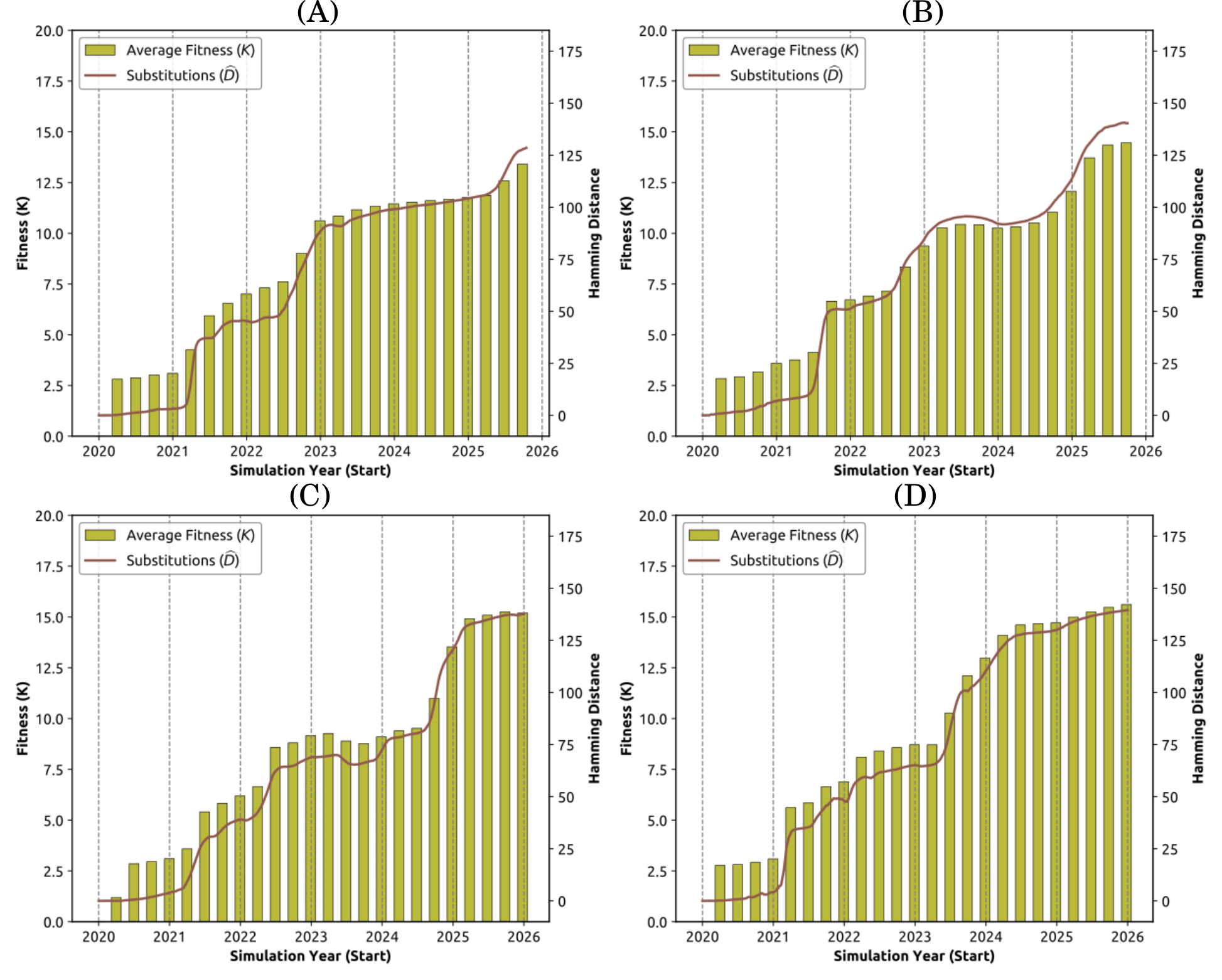}
    \caption{Simulated dynamics of the average transmissibility for an 8.0 million population. Fitness ($K$, olive bars, y-axis on the left), and the accumulated mutations ($\widehat{D}$, solid brown line, y-axis on the right) are plotted from 2020 to 2026. Panels (A)-(D) are profiles plotted using four different realisations.}
    \label{sm:fig:sim-obj-2i-individual_fitness_Ref_Haming-8M}
\end{figure}

\begin{figure}[ht]
    \centering
    \includegraphics[width=\columnwidth]{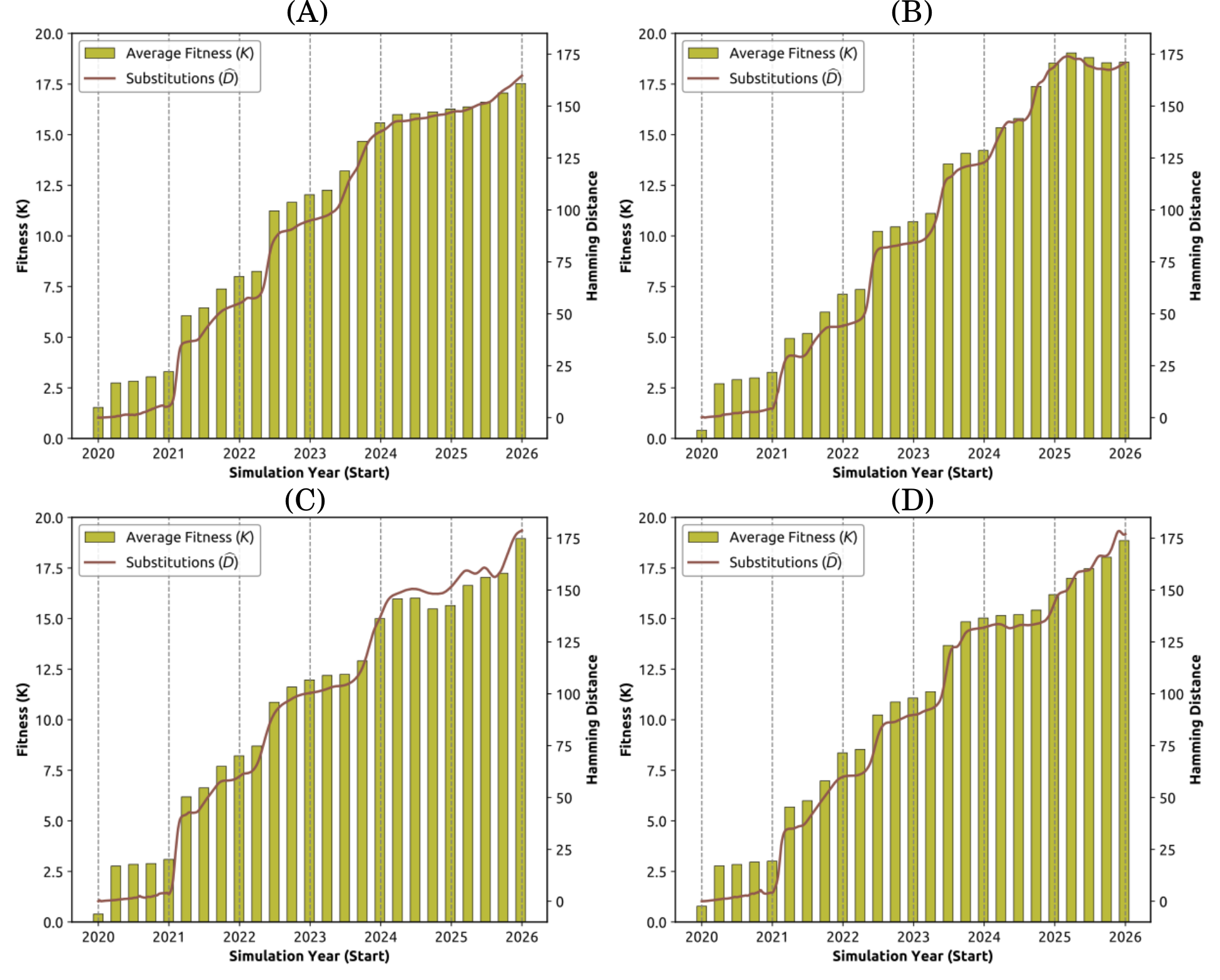}
    \caption{Simulated dynamics of the average transmissibility for a 25.4 million population. Fitness ($K$, olive bars, y-axis on the left), and the accumulated mutations ($\widehat{D}$, solid brown line, y-axis on the right) are plotted from 2020 to 2026. Panels (A)-(D) are profiles plotted using four different realisations.}
    \label{sm:fig:sim-obj-2i-individual_fitness_Ref_Haming-25M4}
\end{figure}

Figs \ref{sm:fig:compartments} to \ref{sm:fig:sim-obj-2i-individual_fitness_Ref_Haming-25M4} illustrate simulated epidemic dynamics, as well as the dynamics of transmissibility and mutation accumulation across populations of varying sizes: small (1.7 million), medium (8.0 million), and large (25.4 million). These figures provide insights into population health states, infection and immunisation histories, and viral transmissibility trends, aligned with phylogenetic dynamics observed in different realisations across diverse population scenarios.

\begin{figure}[ht]
    \centering
    \includegraphics[width=0.9\columnwidth]{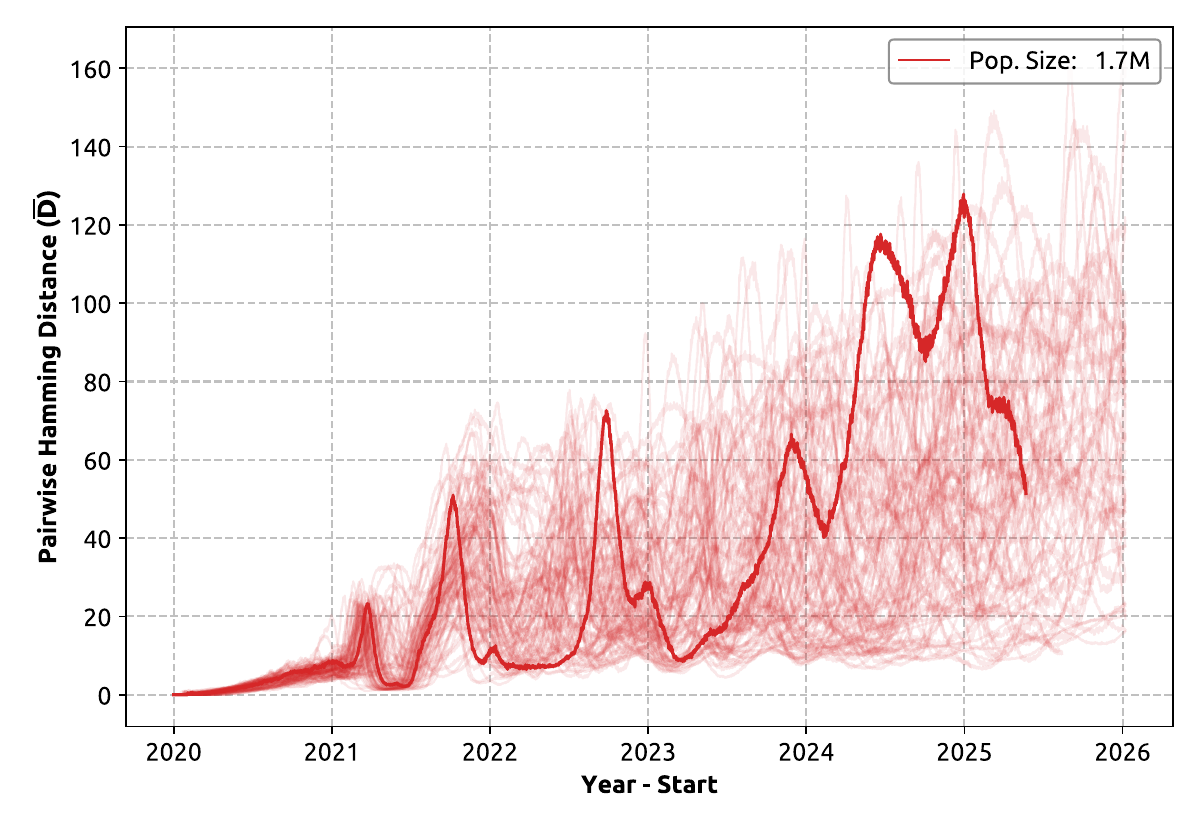}
    \caption{Simulated dynamics of the average pairwise Hamming distance ($\overline{D}$) between two randomly selected genomes from infected hosts in a population of 1.7 million. Approximately 10,000 pairs of genomes are randomly sampled at each simulation time point from 2020 to 2026 (Capability 2(i)). Opaque red lines represent the ensemble of all realisations and the solid red line shows the dynamics of one realisation only.
    }
    \label{sm:fig:sim-obj-2i-pairwise_1M7}
\end{figure}

\begin{figure}[ht]
    \centering
    \includegraphics[width=0.9\columnwidth]{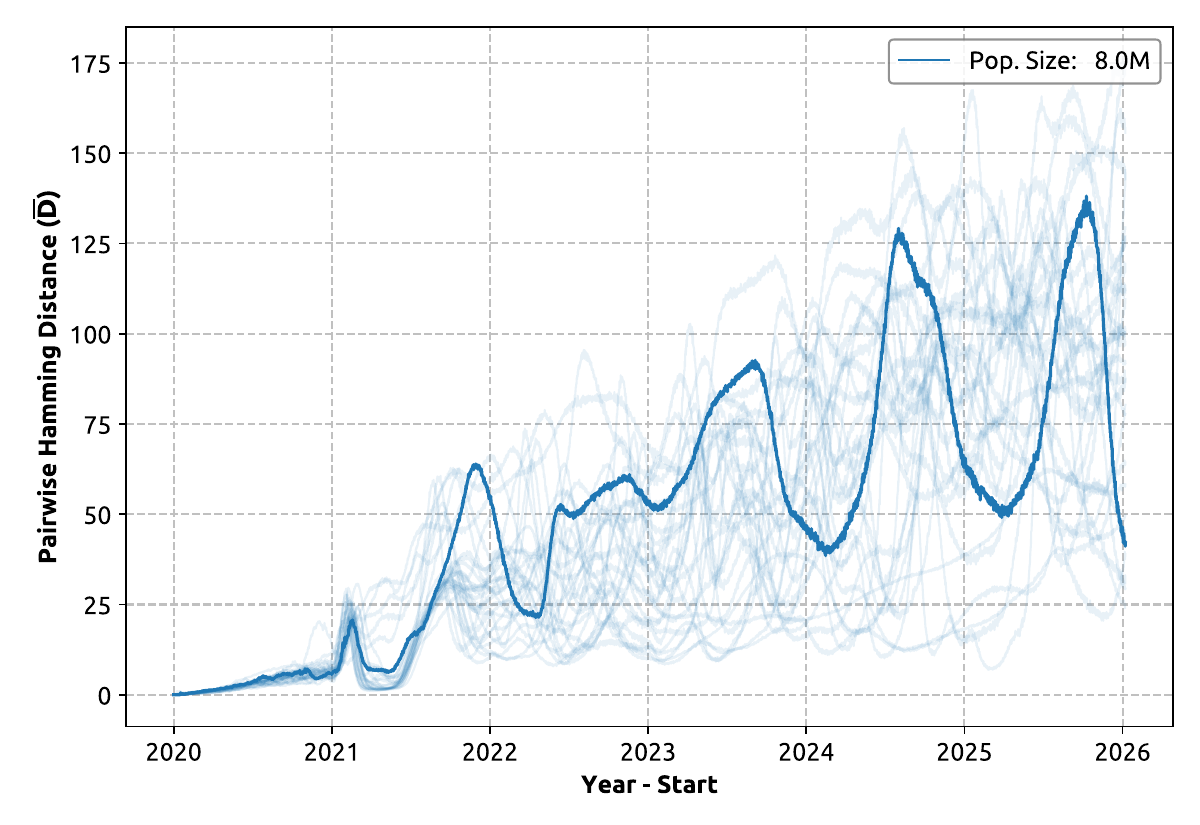}
    \caption{Simulated dynamics of the average pairwise Hamming distance between two randomly selected genomes from infected hosts in a population of 8.0 million. Approximately 10,000 pairs of genomes are randomly sampled at each simulation time point from 2020 to 2026 (Capability 2(i)). Opaque blue lines represent the ensemble of all realisations and the solid blue line shows the dynamics of one realisation only.
    }
    \label{sm:fig:sim-obj-2i-pairwise_8M}
\end{figure}

\begin{figure}[ht]
    \centering
    \includegraphics[width=0.9\columnwidth]{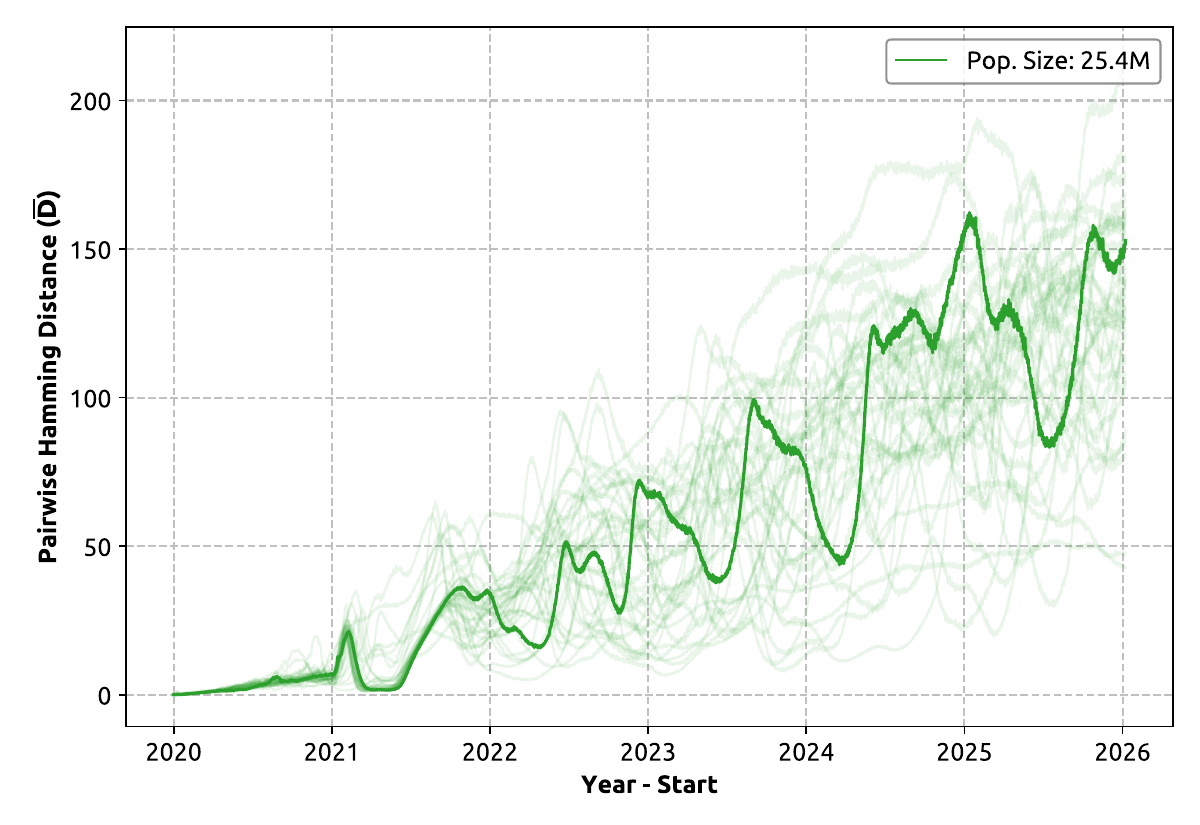}
    \caption{Simulated dynamics of the average pairwise Hamming distance between two randomly selected genomes from infected hosts in a population of 25.4 million. Approximately 10,000 pairs of genomes are randomly sampled at each simulation time point from 2020 to 2026 (Capability 2(i)). Opaque blue lines represent the ensemble of all realisations and the solid blue line shows the dynamics of one realisation only.
    }
    \label{sm:fig:sim-obj-2i-pairwise_25M4}
\end{figure}

Figs \ref{sm:fig:sim-obj-2i-pairwise_1M7} to \ref{sm:fig:sim-obj-2i-pairwise_25M4} present pairwise Hamming distance dynamics generated using identical phylogenetic parameters across three different population sets.
Fig \ref{sm:fig:sim-obj-2i-pairwise_1M7} shows alternative evolution dynamics relative to those shown in Fig. \ref{fig:sim-obj-3i} (A, brown dashed line). The high variability of genomic diversity dynamics across different realisations results from the stochastic emergence of variants; hence, we refrain from averaging genomic diversity dynamics across realisations. Nonetheless, the pattern of genomic diversity dynamics aligns well with the trends discussed in Section \ref{sec:case_study_covid19}, which includes periods of drift, rapid rise, and abrupt collapse. To assess the stationarity of genomic diversity dynamics, we performed the Augmented Dickey–Fuller test. Fig \ref{sm:fig:ADF}
shows that the p-value for stationarity associated with genomic diversity dynamics varies significantly with population size, highlighting the increased difficulty for variants to spread and become dominant in larger populations. 

\begin{figure}[ht]
    \centering
    \includegraphics[width=\columnwidth]{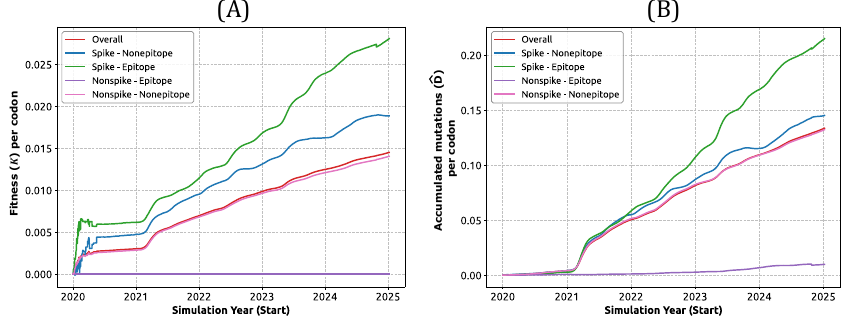}
    \caption{Simulated dynamics of the average transmissibility across different genome regions for a 1.7 million population. (A) Fitness $K$ per codon, and  (B) accumulated mutations $\widehat{D}$ per codon. Five genome regions are plotted: overall genome (red), spike and non-epitope region (blue), spike and epitope (green), non-spike and epitope (purple), and non-spike and non-epitope (pink).}
    \label{sm:fig:fitness_per_codon}
\end{figure}

\newpage
\section{Counterfactual modelling}
\label{sm:sec:results_counter}
In this section, we describe two counterfactual modelling scenarios of SARS-CoV-2: the impact of chronic infections on the SARS-CoV-2 evolution (Section \ref{sm:sec:results_chronic}), and the impact of population size on stationarity of the genomic diversity $\overline{D}$ (Section \ref{sm:sec:results_stationarity}).

\subsection{Chronic infections}
\label{sm:sec:results_chronic}

\begin{figure}[ht]
   \centering
\includegraphics[width=0.9\columnwidth]{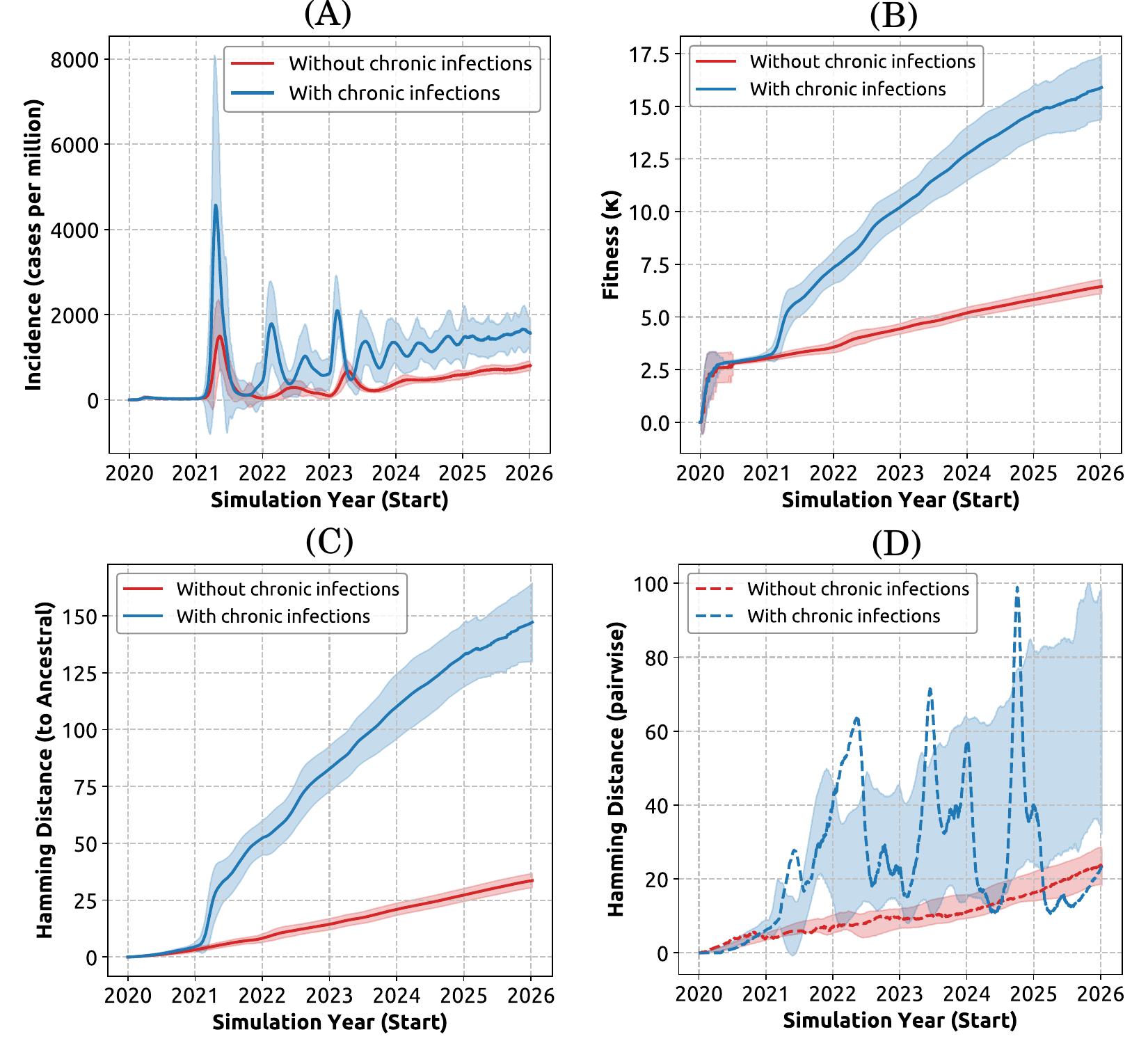}
   \caption{Counterfactual scenario: chronic infections. Simulated dynamics in scenarios with the fraction of agents susceptible to chronic infections being 0.1\% of the entire population (blue curves), and without chronic infections (red curves). The profiles trace detected (A) incidence, (B) average pathogen transmissibility, interpreted as fitness $K$, (C) accumulated mutations ($\widehat{D}$), and (D) genomic diversity ($\overline{D}$) in the population of 1.7 million, over six simulation years. The shaded area represents the range of one standard deviation from the mean values. In (A) to (C), the solid lines denote the mean value across 30-50 realisations. In (D), the dashed line denotes the pairwise Hamming distance from one realisation only, given the high variability between realisations.}
   \label{sm:fig:counter_immuno_sim}
\end{figure}

Fig \ref{sm:fig:counter_immuno_sim} differentiates between the COVID-19 pandemic simulation scenarios with and without chronic infections.  Notably, the absence of chronic infections results in a reduced number of recurrent incidence waves (Fig \ref{sm:fig:counter_immuno_sim} (A)) and significantly reduced fluctuations in the genomic diversity $\overline{D}$,  producing lower variability (i.e., a smaller standard deviation) across realisations (Fig~ \ref{sm:fig:counter_immuno_sim} (D)). Furthermore, the growth of both transmissibility (i.e., fitness $K$) and the accumulated mutations $\widehat{D}$ is significantly slower (Fig~ \ref{sm:fig:counter_immuno_sim} (B) and (C)), impeding the emergence of high-fitness variants. 

We also explored whether the pathogen phylodynamics are correlated with (1) fractions of chronic infection, and (2)  strength of the within-host selective pressure in chronically infected hosts. Fig~\ref{sm:fig:sim-fraction_chronic} shows that a hundred-fold increase in the chronic infection fraction (e.g., from 0.05\% to 5\% of the population) yields only negligible impact on the incidence curves and phylodynamics. This indicates a ceiling effect of the chronic infection fraction on the incidence, fitness, and accumulated mutations. In contrast, Fig~\ref{sm:fig:sim-Topxx} presents a clear positive correlation between the within-host selective pressure and the phylodynamic characteristics. That is, a higher selective pressure (i.e., selecting top 10\% of mutated genome candidates out of their ranked list, $X = 10$, $M = 100$) leads to higher incidence peaks and a more rapid increase in the pathogen fitness and accumulated mutations.

\subsection{Stationarity of genomic diversity}
\label{sm:sec:results_stationarity}

The genomic diversity traced in empirical data from 2020 to 2024  shows fluctuating yet relatively stationary dynamics, without notable trends of increasing or decreasing diversity, as illustrated in Fig~\ref{fig:obj2} (C)). The Augmented Dicky-Fuller (ADF) test (detailed in Section \ref{sec:methods}) applied for the empirical genomic diversity produced a p-value of 0.024, indicating stationarity at the significance level of 0.05. 

\begin{figure}[ht]
    \centering
    \includegraphics[width=0.8\columnwidth]{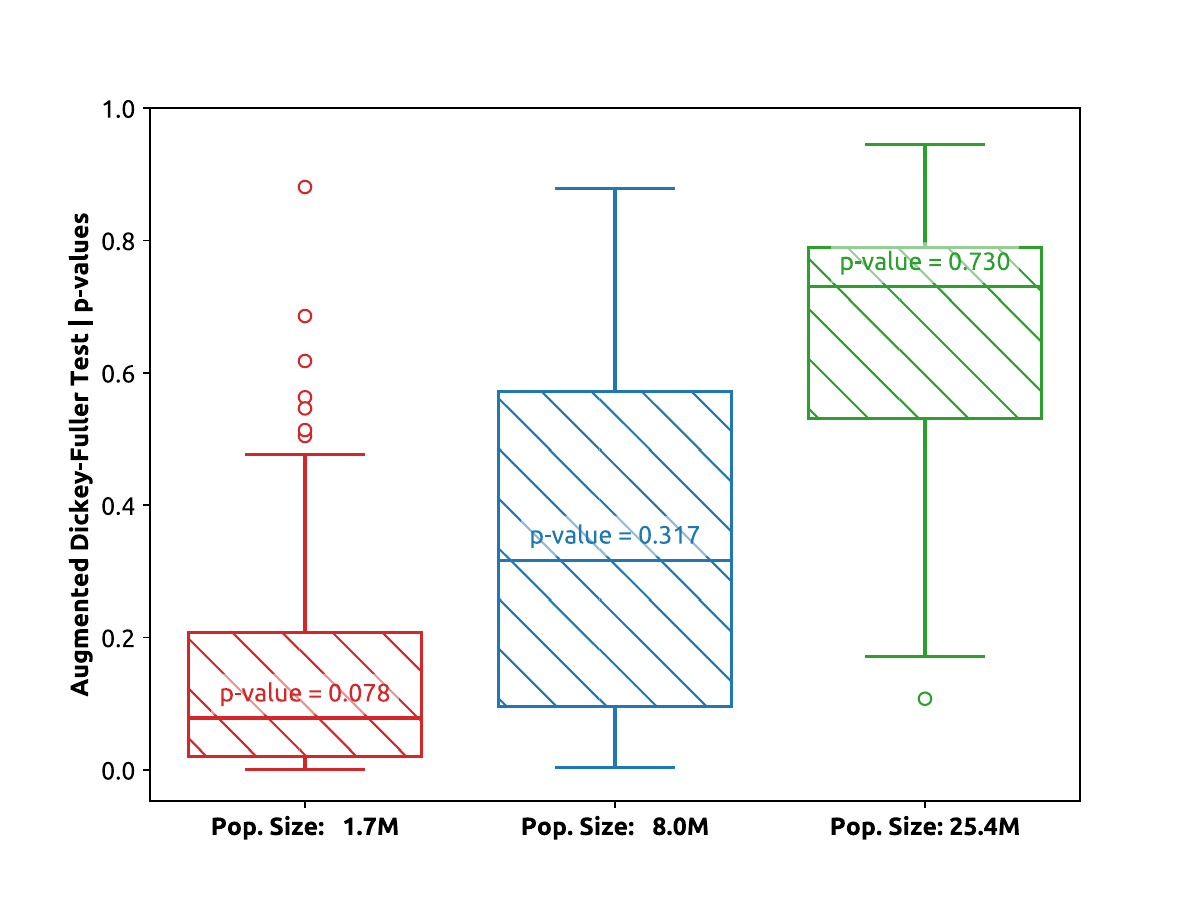}
    \caption{Statistical estimation of stationarity. The stationarity of the time evolution of the average pairwise Hamming distance is computed between two randomly selected genomes. Approximately 10,000 pairs of genome sequences are randomly sampled at each simulation time point from 2020 to 2026 for three simulation scenarios with different population sizes of 1.7 million (A), 8 million (B), and 25.4 million (C) (Capability 3(ii)).
    }
    \label{sm:fig:ADF}
\end{figure}

\begin{figure}[ht]
    \centering
    \includegraphics[width=\columnwidth]{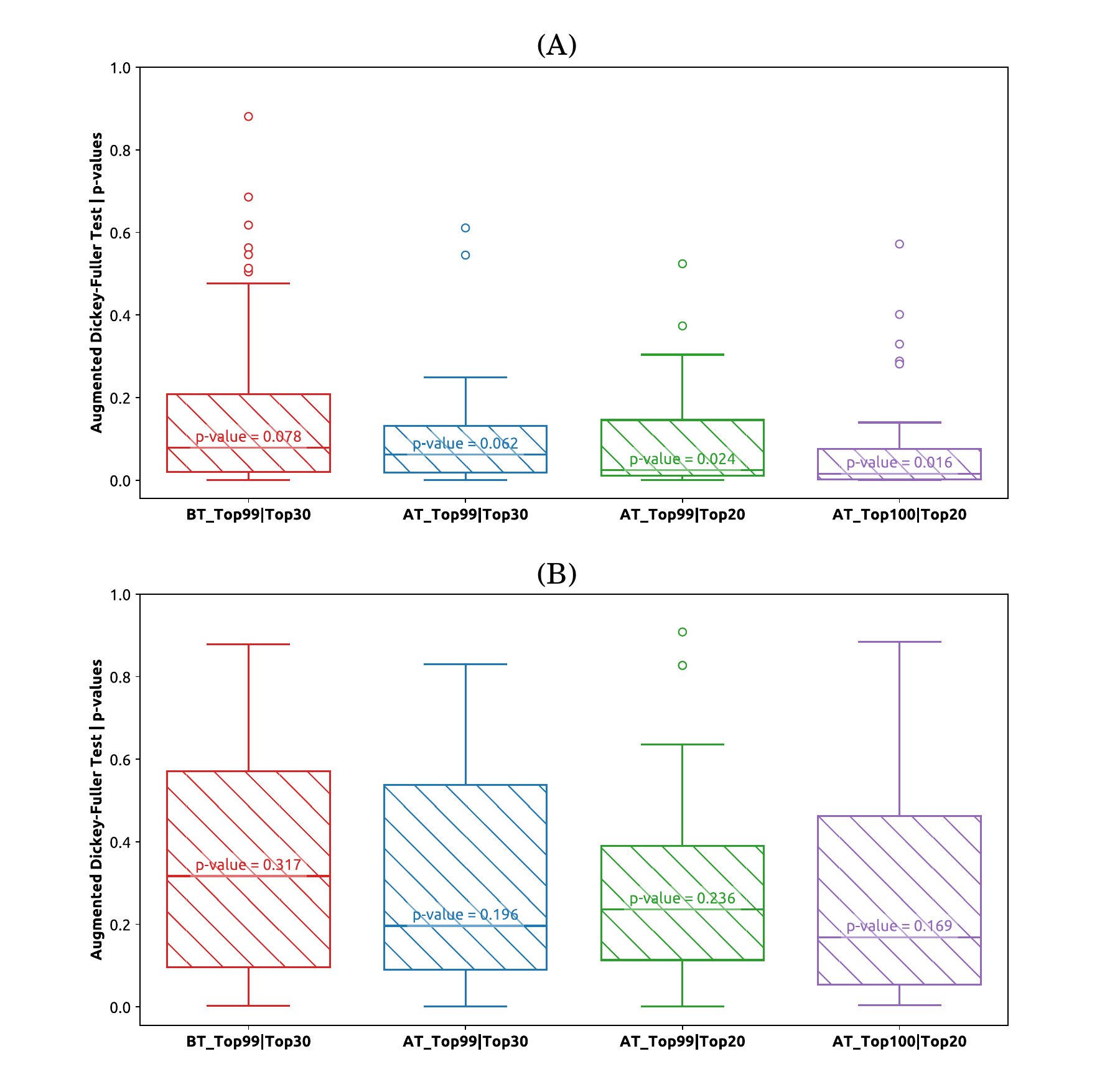}
    \caption{Statistical estimation of the stationarity of the average pairwise Hamming distance between two randomly selected genomes simulated with with different population sizes. (A) populations of 1.7 million and (B) 8.0 million. Approximately 10,000 pairs of genome sequences are randomly sampled at each simulation time point from 2020 to 2026 for four simulation scenarios by varying amino acid weight table and/or within-host selective pressure (see bar legends). We tested two different fitness weight tables following different normal distributions: (i) \textit{BT}: the baseline setting, where each amino acid contribution in spike and non-spike codon positions is sampled from the normal distributions \textit{N(0, 0.085)} and \textit{N(0, 0.07)}, respectively, and (ii) \textit{AT}: the alternative setting, where each amino acid contribution in spike and non-spike codon positions is sampled from the normal distributions \textit{N(0, 0.0489)} and \textit{N(0, 0.0454)}, respectively. Within-host selective pressure is varied for both typical infected hosts and chronically infected hosts, using notation separated by a vertical bar (e.g., $Top99|Top30$ represents a setting where selective pressure at top 99 for typically infected hosts (\textit{X=99, M=100}) and at top 30 for chronically infected hosts (\textit{X=30, M=100}).
    }
    \label{sm:fig:weight_table_and_selection}
\end{figure}

However, stationarity of the simulated genomic diversity is harder to establish for larger population sizes, as illustrated in Fig~\ref{sm:fig:ADF}. For a small population of 1.7 million, the genomic diversity dynamics are closer to stationary (p-value of 0.078).
This outcome is robust to variations of the fitness contributions weight table, as well as changes in the within-host selective pressure, as shown in Fig~\ref{sm:fig:weight_table_and_selection}.  For larger population sizes, the ADF tests produced larger p-values of 0.317 (8 million) and 0.730 (25.4 million), indicating a progressive loss of stationarity.

\section{Sensitivity analysis}
\label{sm:sec:sensitivity}
\begin{figure}[ht]
    \centering
    \includegraphics[width=\columnwidth]{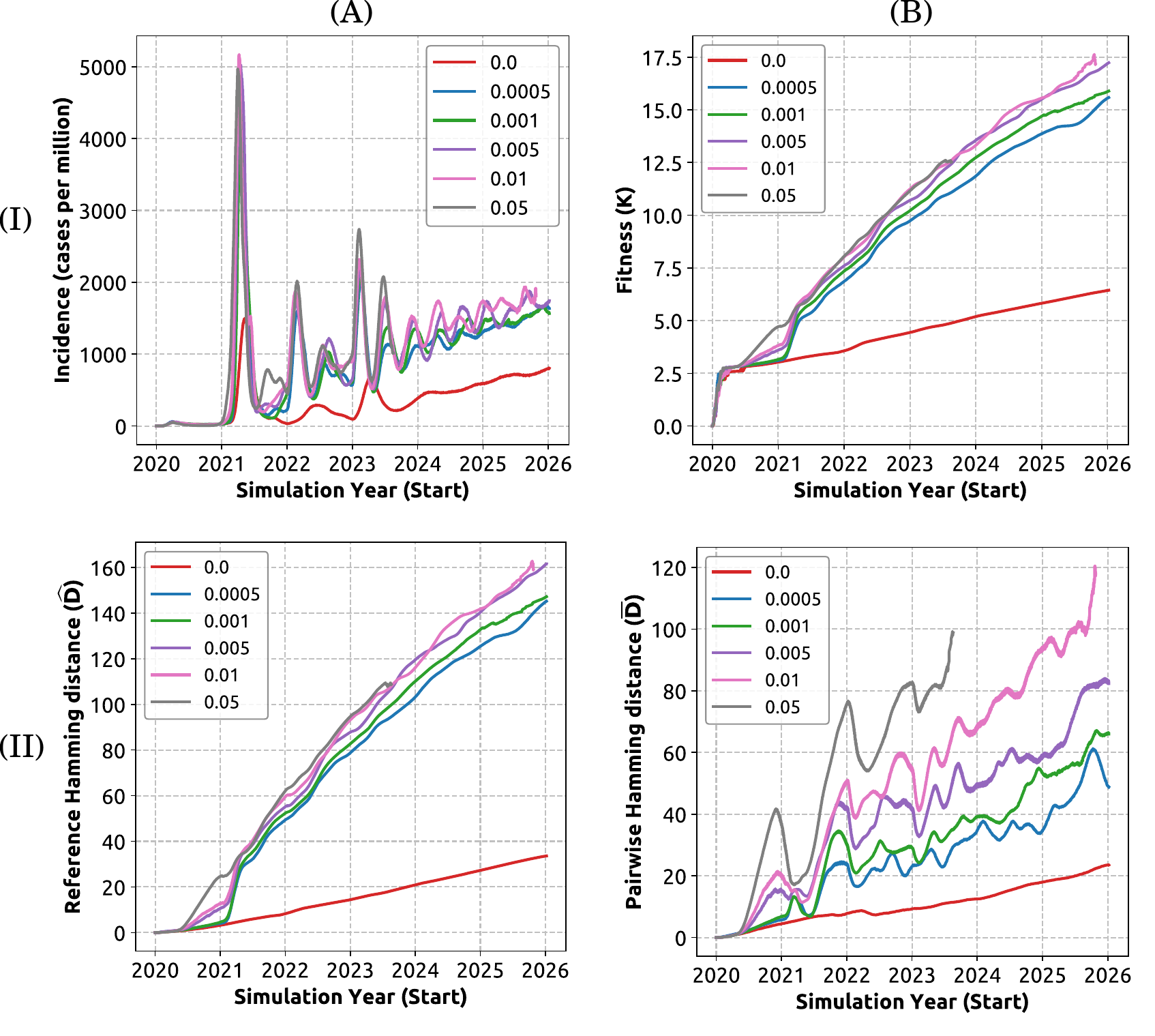}
    \caption{Sensitivity analysis: the fractions of chronic infections. Simulated dynamics of the detected incidence (I-A), average pathogen fitness (I-B), average accumulated mutations $\widehat{D}$ (II-A), and average genomic diversity $\overline{D}$ (II-B) (one realisation only), in a population of 1.7 million. These measures are traced while the fraction of individuals susceptible to chronic infection, with strong positive within-host selective pressure, is varied from 0.0 to 0.05 of the total population.
    }
    \label{sm:fig:sim-fraction_chronic}
\end{figure}

\begin{figure}[ht]
    \centering
    \includegraphics[width=0.85\columnwidth]{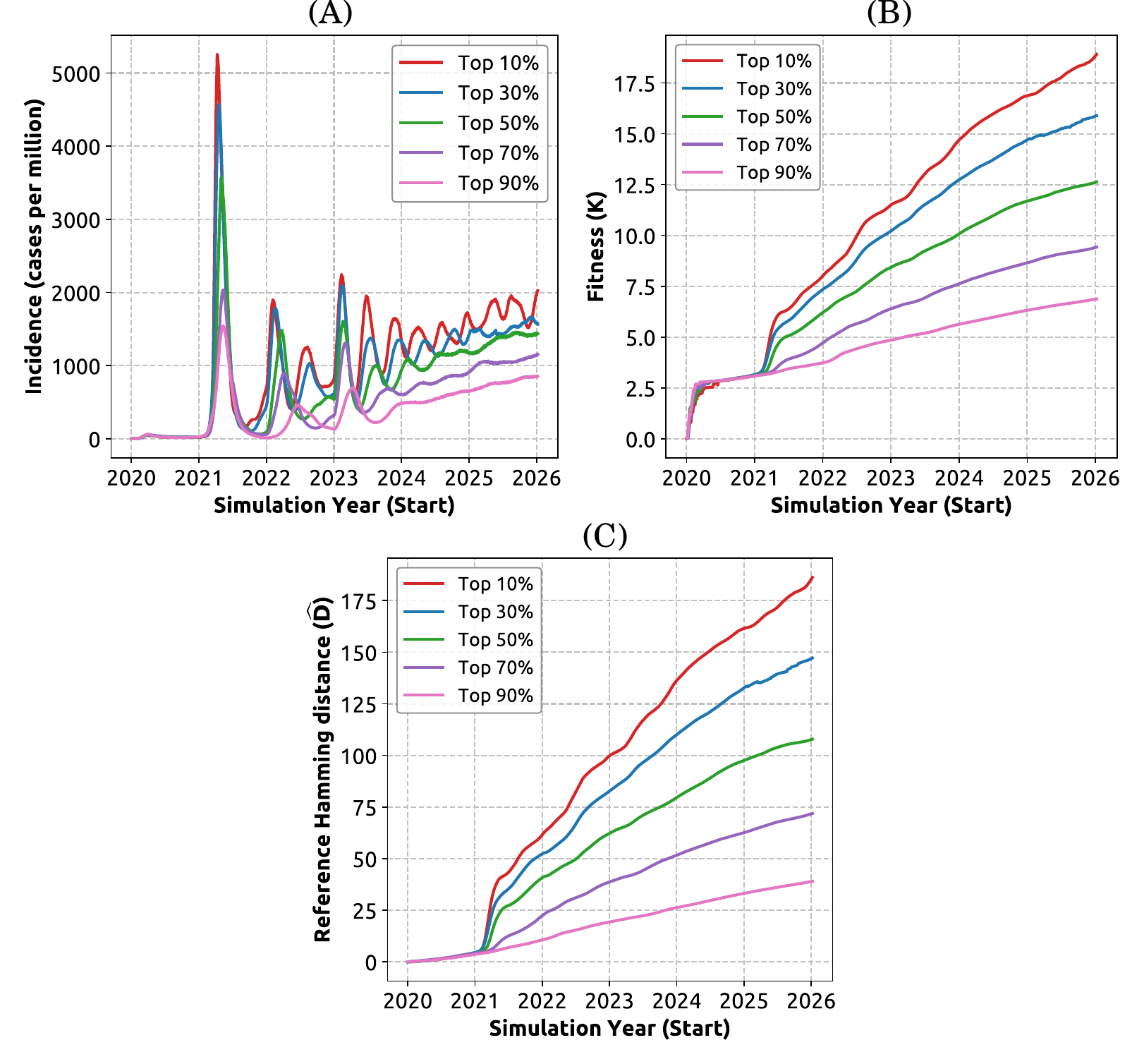}
    \caption{Sensitivity analysis: within-host selective pressure. Simulated dynamics of (A) detected incidence,  (B) average pathogen fitness, and (C) average accumulated mutations $\widehat{D}$ in a population of 1.7 million. The within-host selective pressure for chronically infected hosts, starting from day 60 of infection, is varied by selecting, in each simulation cycle, the mutated genomes from the top 10\% (high selectivity, $X = 10$, $M = 100$) to the top 90\% (low selectivity, $X = 90$, $M = 100$) of the ranked list of mutated genome candidates.}
    \label{sm:fig:sim-Topxx}
\end{figure}

We performed sensitivity analysis by tracing the output variables of interest in response to changes in one input variable while keeping the other inputs specified at default values. In this study, we have four output variables of interest: incidence, transmissibility fitness, accumulated mutations, and genomic diversity. Figs  \ref{sm:fig:sim-fraction_chronic} and \ref{sm:fig:sim-Topxx} show simulated dynamics of the output variables of interest by varying two input parameters: the fraction of chronically infected hosts and the within-host selective pressure in chronically infected hosts, respectively.

\section{Computational complexity and implementation}
\label{sm:sec:comp}
The multi-scale phylodynamic simulator, \phasetrace, builds upon our open-source agent-based epidemiological simulator, AMTRaC-19~\cite{amtract_user_guide_v90},  written in C++. The architecture of \phasetrace includes four layers  (Fig \ref{fig:4-layer-model}), with the immunological and phylogenetic layers being novel additions, which significantly extend the capabilities of AMTRaC-19. \phasetrace is designed to simulate the dynamic interaction between pathogen transmission and evolution over a prolonged timeframe over multiple years (unlike AMTRaC-19 which was typically used to simulate 6-9 months of a pandemic). The new modelling capabilities and simulation requirements  increase the computational complexity, presenting four specific computational challenges: 

\begin{itemize}
    \item \textit{Longer simulation timeframe} (typically, over 6 years or 2,000 simulation days). This is needed to trace the medium-term evolutionary dynamics of pathogens. State-of-the-practice simulators typically employ a short epidemic or pandemic simulation timeframe (e.g., 6-9 months), aiming to examine the short-term impact of public health interventions. 
    \item \textit{Compound and waning immunity}, combining both the immunisation and infection histories. This demands higher memory, continually increasing during the simulation as the agents are likely to accumulate multiple vaccination records and infections during the simulated time frame.
    \item \textit{Complex phylogenetic structure} based on a non-homogeneous genome profile. Each infected agent is assigned a genome profile defined across thousands of nucleotide positions (e.g., 3,090 positions), partitioned between spike and non-spike regions and grouped in terms of antigenicity.  
    Simulation of complex phylodynamics generates a memory-intensive computation task that demands a significant increase in both simulation time and computational resources.
    \item \textit{Simulation across heterogeneous demographics}, with varying population sizes (including very dense populations). Simulating different demographics is needed to investigate conditions for the emergence of variants of concern (VoC), commonly observed in large heterogeneous populations.
\end{itemize}  

\subsection{Performance and scalability}
\begin{figure}[t]
    \centering
    \includegraphics[width=0.95\textwidth]{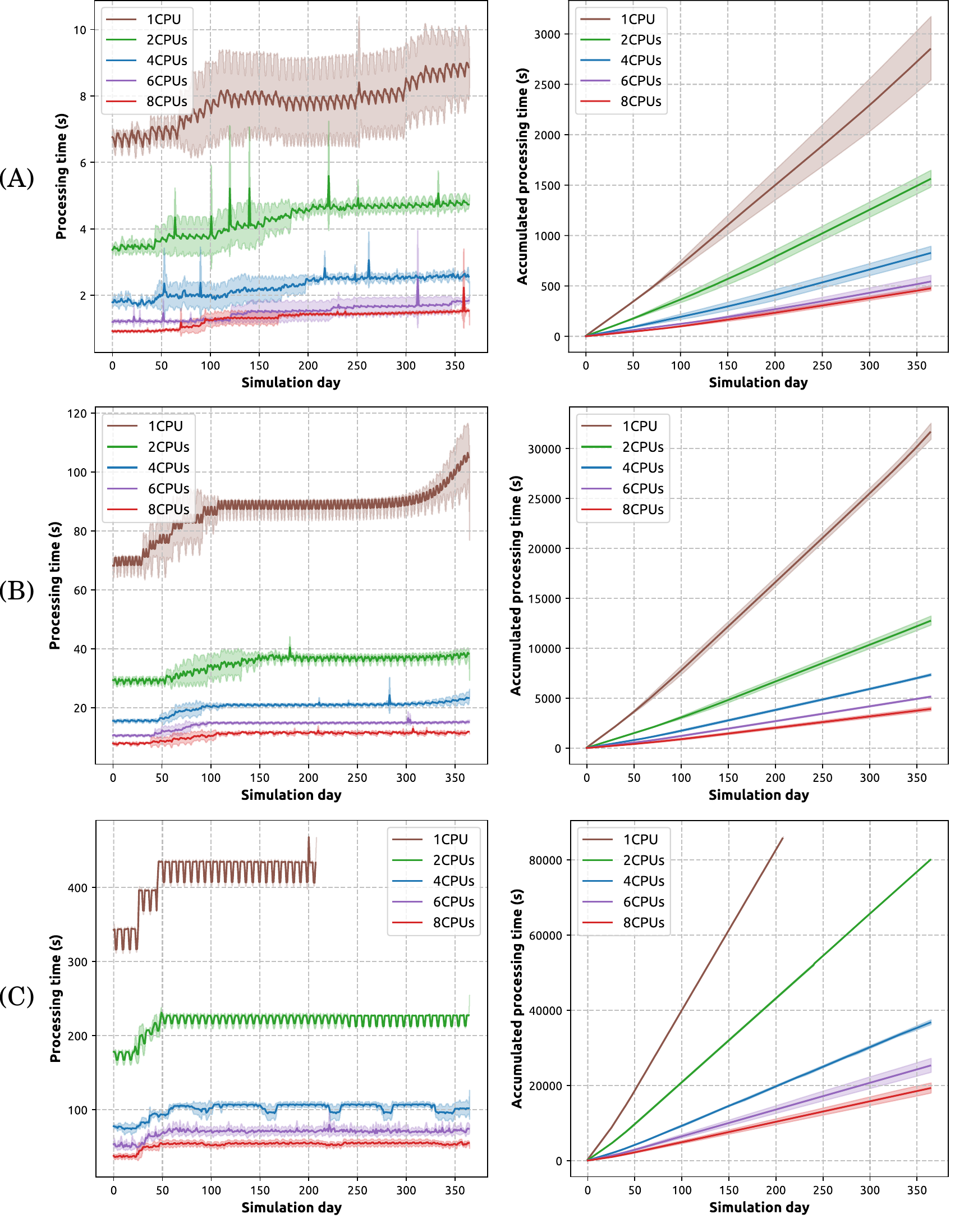}
    \caption{Reduction of the average processing time (seconds) by utilising more CPUs (between 1 and 8). The simulation covers 365 simulation days for three different population sizes: (A) 230K agents, (B) 1.7M agents, and (C) 8M agents. \textit{Left}: the processing time required to simulate a single simulation day. \textit{Right}: the cumulative processing time required to simulate 365 days. Confidence intervals are shown as shaded areas.}
    \label{sm:fig:processing_time_vs_pop}
\end{figure}

To improve computational efficiency, \phasetrace utilises multi-threading processing provided by the C++ OpenMP library. Fig~\ref{sm:fig:processing_time_vs_pop} traces  
the average processing time of simulation runs performed on a high-performance computing cluster, for three different population sizes, ranging from approximately 230,000 agents (Fig~ \ref{sm:fig:processing_time_vs_pop}.A), to approximately 1,700,000 agents (Fig~ \ref{sm:fig:processing_time_vs_pop}.B), and approximately 8,000,000 agents (Fig~ \ref{sm:fig:processing_time_vs_pop}.C). 

\begin{figure}[t]
    \centering
    \includegraphics[width=0.7\textwidth]{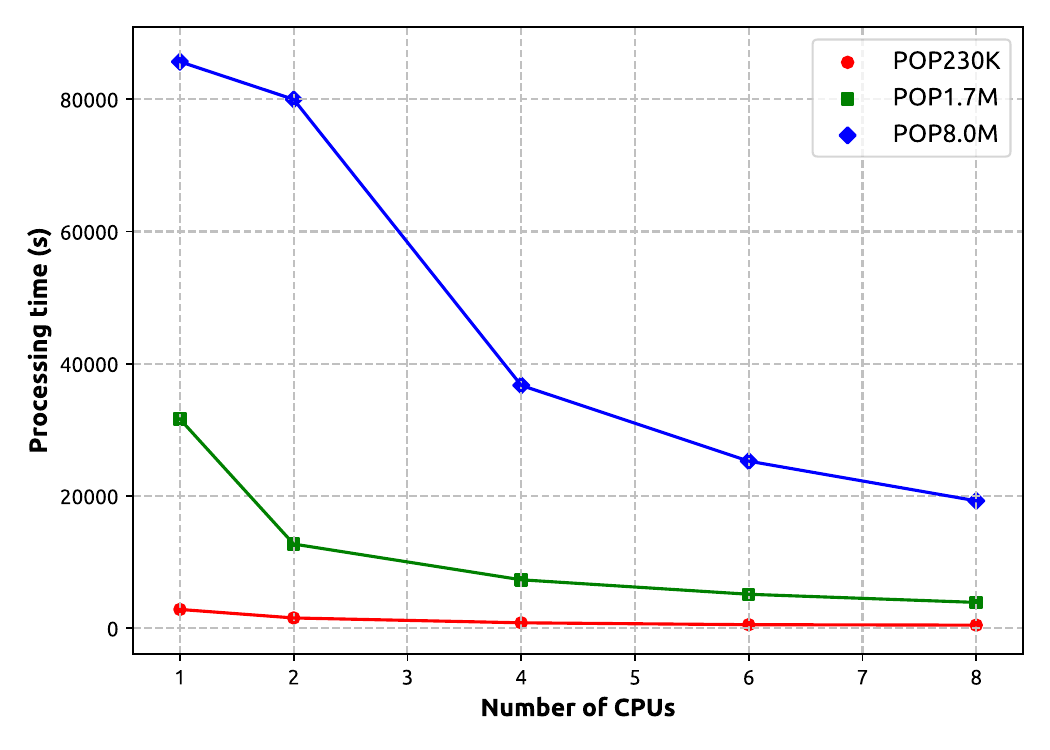}
    \caption{Average cumulative processing time (seconds) for a simulation over 365 days, performed on multiple CPUs in the range between 1 and 8. Population sizes: 230K agents (red), 1.7M agents (green), and 8M agents (blue).}
    \label{sm:fig:processing_time_vs_nCPU}
\end{figure}

The average processing time per simulation day increases during the simulation: this occurs due to the increased memory (see the top row of Fig~ \ref{sm:fig:processing_time_vs_pop}). 
Nevertheless, there is a consistent reduction of the processing time when more CPUs are used per job. We note that, for a given number of CPUs, the average cumulative processing time linearly increases in proportion to the population size. For example, Fig~\ref{sm:fig:processing_time_vs_nCPU} shows that when 8 CPUs are used, a simulation of a 365-day period across a population with 8 million agents completes in approximately 20,000 seconds (over 5.5 hours). 

\subsection{Memory utilisation}

An increasing size of the artificial population requires higher computational resources. This is because the stochastically generated agents are assigned several static demographic attributes, which are preserved during the simulation, and a number of dynamic phylogenetic and immunological attributes, which change during the simulation. These attributes contribute to both fixed memory and dynamic memory utilisation.

Agent attributes that utilise fixed memory include:
\begin{itemize}
    \item {\bf Demographic attributes}: household composition and age group. 
    \item {\bf Working group and school enrollment}: workplace for agents over the age of 18, or school for agents under the age of 18.
    \end{itemize}

Agent attributes that affect dynamic memory utilisation during simulation include:    
    \begin{itemize}
    \item {\bf Vaccination history}: type of the administered vaccine, the total number of vaccination records and the time of vaccination.
    \item {\bf Infection history}:  the number of past infections and their recovery times. 
    \item {\bf Genome profiles}: infected agents are assigned a genome profile transferred from the source of infection, which continues to mutate during the agent's infectivity period.
\end{itemize}

\begin{figure}[t]
    \centering
    \includegraphics[width=0.65\textwidth]{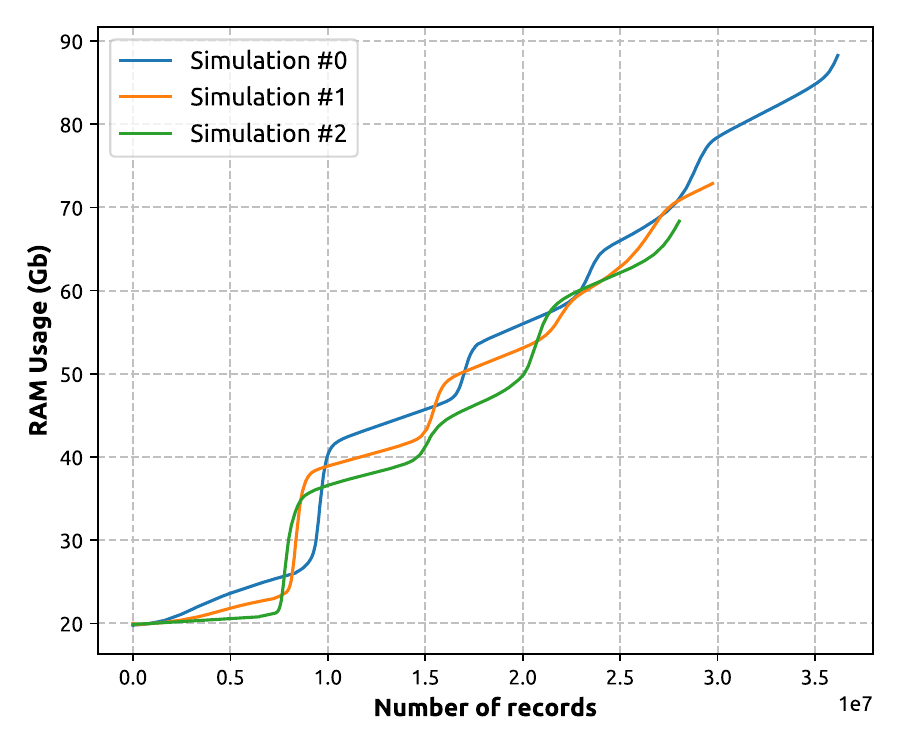}
    \caption{Dependency of the total number of stored immunological records (tens of millions), combining vaccination and infection histories of 8 million agents, and the amount of memory (RAM, Gigabytes) required for a simulation period of 1,200 days. Each profile represents an individual realisation. All realisations are simulated under identical inputs and settings.}
    \label{fig:ram_records}
\end{figure}

As the simulation progresses, the number of stored immunological records grows significantly due to the increased number of infections and administered vaccinations. Fig~\ref{fig:ram_records} shows how the dynamic memory usage (in Gigabytes) increases with the larger number of stored immunological records across the population of 8 million agents. 

\clearpage




\end{document}